\documentclass[10pt]{article}

\usepackage[english]{babel}
\usepackage{latexsym,amsmath,amssymb,amsbsy,amstext,amscd,amsfonts}
\DeclareMathOperator{\arctanh}{arctanh}
\usepackage{graphics,graphicx}
\usepackage{color}
\usepackage{tabularx}
\usepackage{multirow}
\usepackage{booktabs}
\usepackage{epstopdf}
\usepackage{natbib}
\usepackage{ulem}
\usepackage [latin1]{inputenc}
\newtheorem{remark}{Remark}

\date{}

\title{A simplified modelling of hydraulic fractures in elasto-plastic materials\footnote{Preprint submitted to \textit{International Journal of Fracture}.} }

\author{M. Wrobel$^{(1)}\footnote{Corresponding author: wrobel.michal@ucy.ac.cy}$ , P. Papanastasiou$^{(1)}$ and D. Peck$^{(2)}$  
\\
{\it $^{(1)}$\! Department of Civil and Environmental Engineering, University of Cyprus, }
\\ {\it 1678 Nicosia, Cyprus}
\\
{\it $^{(2)}$Department of Mathematics, Aberystwyth University,}
\\ {\it Ceredigion SY23 3BZ, Wales, UK}
 }

\begin{document}

\maketitle

\begin{abstract}
In this paper the problem of a plane strain hydraulic fracture propagating in an elasto-plastic material is analyzed. A new stress redistribution model for the proximity of the fracture tip is formulated and a resulting plasticity-dependent crack propagation condition is introduced. A modified variant of the KGD problem that accounts for the plastic deformations in the near-tip zone only is proposed. It is demonstrated that this model can be a credible substitute for the full elasto-plastic hydraulic fracture problem in the case of moderate plastic deformation. The crack-tip shielding effect introduced by the plastic deformation is quantified.
\end{abstract}

\providecommand{\keywords}[1]
{
  \small	
  \textbf{\textit{Keywords:}} #1
}


\section{Introduction}\label{Introduction}

The phenomenon of hydraulically propelled fracture propagating in a solid material, the so-called hydraulic fracture (HF), is encountered in multiple natural processes. The underlying physical mechanism is also present in technology. It can constitute an undesired side effect of the main technological processes, such as in the case of CO$_2$ sequestration or in induced cracking under hydraulic structures, in which case protective measures against HF need to be taken. On the other hand, intentionally induced hydraulic fracturing can be harnessed to achieve a desired goal, for instance in the stimulation of oil or gas wells. Indeed, since the beginning of this century the fracking technology has revolutionized the exploitation of unconventional hydrocarbon resources.

Efficient and safe utilization of hydraulic fracturing necessitates credible numerical simulation of its underlying mechanisms.  This constitutes a formidable task due to the  complex multiphyscial  nature of the problem. The main computational difficulties stem from: i) strong non-linearity related to the non-local interaction between fluid and solid as well as originating from the non-linear properties of the respective phases, ii) possible singularities of the component physical fields, iii)  moving boundaries, iv) degeneration of the governing equations at the fracture tip, v) nonrecoverable deformation of solid due to the plastic distortion, vi) pronounced multiscale effects, and many others. Since the times when the pioneering works in the field were published a lot of effort has been made to enhance both, the hydraulic fracture models and the computational methods. In this way still more sophisticated descriptions of the HF physics have been implemented and successfully treated. A comprehensive summary of the history and techniques of HF modeling can be found in the paper by \cite{Adachi_2007}. Notably, despite immense progress in the area, still the simplest 1D models, such as PKN \citep{Nordgren}, KGD (plane strain model) \citep{Khristianovic,Geertsma} and the radial (penny shaped) model \citep{Sneddon}, are commonly used. It comes from the fact that, although geometrically simplified, they reflect properly the inherent features of the underlying physical processes. For example, by analyzing the classical 1D models the crack propagation regimes have been identified and categorized \citep{Detournay_2004,Garagash_2009}, the influence of the non-Newtonian fluid rheology has been recognized  \citep{Adachi_Detournay,Wrobel_2020,Wrobel_2021} or power requirements for simultaneous propagation of multiple fractures have been established \citep{Bunger_2013}. Moreover, these simplified HF models can be successfully used in construction and verification of advanced computational algorithms.

One of the most challenging problems when modeling hydraulic fractures is accounting for the plastic deformations of solid. The studies on this aspect of the HF mechanism comprise a relatively small fraction of the overall research in the field. This situation results mainly from the fact that with plasticity it is hardly possible to use any analytical or semi-analytical models that could simplify the problem (e.g. by reducing the order of the problem as it is with elastic solids) and facilitate the computations. The simulations rely here mainly on advanced numerical modeling with the Finite Element Method (FEM) being one of the most popular tools. An efficient algorithm for simulation of hydraulic fractures propagating in elasto-plastic pressure sensitive materials was proposed by \cite{Papanastasiou_algorithm}. The Mohr-Coulomb flow theory of plasticity for solid deformation was combined with the lubrication theory describing fluid flow in the fracture. The cohesive zone model was employed as a crack propagation condition. The algorithm of solution was constructed as a dedicated FEM-based scheme. This simulator was used by \cite{Papanastasiou_2000} to investigate the problem of fracture closure. The shielding effect introduced by plastic deformation was studied in the papers of \cite{Papanastasiou_1997,Papanastasiou_1999a}. It was shown that the near-tip plastic deformation can increase the effective fracture toughness by more than an order of magnitude which results in increased fluid pressure needed to propagate the crack. Consequently, the hydraulic fracture geometry evolves with respect to the elastic variant by reducing the crack length and increasing its aperture. In the paper by \cite{Sarris_2012} the authors analyzed the poroelastoplastic model of the fractured material by means of a FEM-based simulator built in the ABAQUS FEA package with the cohesive zone crack propagation condition (a thorough analysis of the influence of the cohesive zone parameters on the resulting fracture propagation criterion was performed in the publication by \cite{Sarris_2011}). The obtained results  explain the source of discrepancies between the fluid pressures measured in the field operations and those predicted by classical HF models.  \cite{Wang_2015} investigated propagation of a non-planar hydraulic fracture (under plane strain conditions) in a permeable elasto-plastic formation described by Mohr-Coulomb theory of plasticity.  The developed computational algorithm was constructed in the framework of Extended Finite Element Method (XFEM) with the crack propagation condition based on the cohesive zone model. The capabilities of the proposed model were demonstrated for both the near wellbore and far field  scale. This HF simulator was utilized in subsequent studies \citep{Wang_2016} to analyze the multifracturing problem. In the publication by \cite{Liu_2017} a 3D XFEM algorithm was combined with cohesive zone model and a dedicated stabilization method to avoid pressure oscillations on the fracture face. Finally, in the paper by \cite{Zeng_2019} the non-associated Drucker-Prager plastic deformation model was used together with fracture propagation criterion based on J-integral. The solution was obtained by means of XFEM.

From the above description of the recent advances in the area it is evident that very sophisticated models and simulators are constructed to analyze the problem of elasto-plastic hydraulic fracture. Accordingly, the computational cost of obtaining solution is  high. On the other hand, in many situations the extent of plastic yield can be  limited.  Intuitively, especially for long fractures with the plastic process zone centered mainly around the crack tip, the solution should not diverge much from the results obtained in the framework of the linear elastic fracture mechanics (LEFM). If we categorize the plasticity effects into two groups: i) those that affect the rock splitting ahead of the fracture tip and as such contribute to the crack length extension and, ii) the effects of plastic deformation of the bulk of the fractured material, the question arises whether accounting only for the first group can be sufficient for modeling at least in the small yield cases. The question becomes even more vital if we recall that the plastic deformation effects in the near tip zone  can substantially affect fracture development, for example the crack redirection \citep{Wrobel_redirection}.

In the classical fracture mechanics there are a few simplified models that take into account the plastic deformation near the crack tip. Such a model, based on a stress redistribution in the near-tip zone, was introduced by \cite{Irwin}. The results produced by this model were verified numerically, using the example of fracture in an infinite plate loaded by a remote stress, by \cite{Sun_2012}. It was shown that Irwin's solution effectively predicts both, the stress distribution ahead of the fracture tip and the size of the plastic zone, even in those cases where plastic deformations can no longer be considered small. A similar concept to that developed by Irwin was used by \cite{Yao_2011} in the hydraulic fracture problem. Instead of a perfectly plastic yield in the plastic deformation zone (as in the original Irwin's solution) \cite{Yao_2011} assumes a power-law softening behavior of the fractured material. The values of respective softening parameters are taken a priori with the resulting effective fracture toughness being not dependent on the solution.  An interesting approach to the problem of elasto-plastic fracture was formulated in the paper by \cite{Atkinson_1977}. The plastic deformation in the near tip zone was described by two yield strips, the so-called superdislocations, inclined at some angle with respect to the plane of fracture extension. The size and angle of dislocations is to be found as an element of solution for the given loading conditions provided that the zero fracture toughness holds. This concept was further extended for pressure sensitive materials under small scale yielding \citep{Papanastasiou_Atkinson_2000} and large scale yielding \citep{Papanastasiou_Atkinson_2006}. Finally, in the study by \cite{Papanastasiou_CO2} the superdislocation model was combined with a non-zero fracture toughness. Even though some critical comments on the application of the this approach in the hydraulic fracture problem were made in the dissertation by \cite{Wu_phd}, the superdislocation model has not yet been employed in simulation of propagating hydraulic fractures. 

In this paper we aim to determine whether a simplified approach to the problem of hydraulic fracture propagating in an elasto-plastic geomaterial can be justified in the case of small yield. To this end we consider three variants of the HF problem: i) hydrofracturing in an elastic solid, ii) hydraulic fracture in elastic solid but with the plastic deformation effects localized in the near-tip zone only, iii) fully elasto-plastic model of hydraulic fracture. The solution to the first variant of the problem is obtained in the framework of LEFM with the numerical algorithm introduced by \cite{Wrobel_2015} and enhanced further to account for additional features of the HF phenomenon \citep{Perkowska_2016,Wrobel_2017,Wrobel_2018,Wrobel_2020,Wrobel_2021}. As the algorithm proved to be efficient and versatile we modify it to account for the plasticity effects. Due to its modular architecture it is possible to separately amend the subroutines for the crack propagation condition and solid deformation. In this way, for the second variant of the problem the basic algorithm is employed with a new crack propagation condition. The condition based on a stress relaxation model takes into account the plastic deformation effects in the near tip zone. In the third considered variant of the HF problem the original subroutine for computing the fracture aperture from the boundary integral equation of elasticity is substituted by a FEM module of ABAQUS FEA package. The results of computations performed for all three variants of the problem under some typical values of material parameters are compared to assess the legitimacy of the simplified HF plasticity models. An estimation of the plastic deformation effects on the HF process is performed.

The paper is structured as follows. In Section \ref{formulation} the mathematical description of all three variants of the problem is formulated. A new stress relaxation model is proposed and the resulting crack propagation condition that incorporates plastic deformation effects is derived. Section \ref{algorithm} includes description of the numerical scheme used to obtain solutions. In Section \ref{cond_verification} we verify to what degree the underlying assumptions of the employed stress relaxation model are satisfied in the HF problem. This allows us to conclude on the applicability of the new crack propagation condition. Section \ref{results} contains comparative analysis of the results obtained for the respective variants of the HF problem.
The final conclusions are listed in Section \ref{conclusions}.

\section{Problem formulation}
\label{formulation}

\begin{figure}[htb!]
\begin{center}
\includegraphics[scale=0.55]{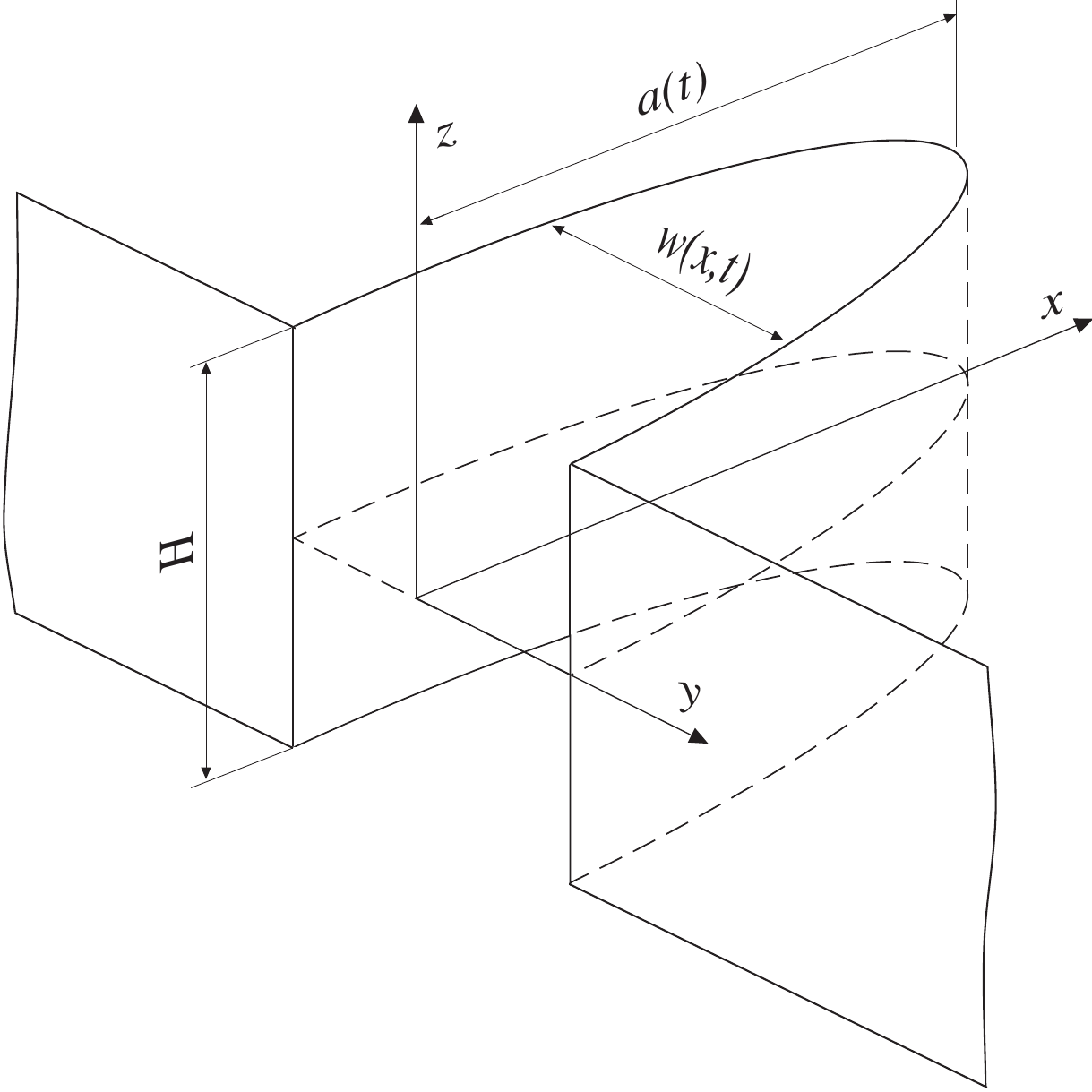}
\caption{The KGD fracture geometry.}
\label{KGD_geom}
\end{center}
\end{figure}

Below we formulate the basic system of equations that governs the analyzed variants of the hydraulic fracture problem. As for the general assumptions, we consider a two-winged plane strain hydraulic fracture of the standard KGD geometry. Due to the problem symmetry we analyze only one of the symmetrical wings - as shown in Figure \ref{KGD_geom}. The fracture half-length is denoted by $a(t)$. Fluid leak-off to the surrounding formation is neglected. Note that, although the last assumption has been commonly accepted in many HF studies (see for example publications by \cite{Papanastasiou_1997,Papanastasiou_2000,Liu_2017}), the pore pressure distribution can substantially influence  the material yielding \citep{Sarris_2012,Wang_2016}. We plan to address the issue of poro-elasto-plastic deformation in our future research.

\subsection{Fluid flow equations}

The mass balance within the fracture is governed by the continuity equation:
\begin{equation}
\label{cont}
\frac{\partial w}{\partial t}+\frac{\partial q}{\partial x}=0, \quad x\in[0,a(t)],
\end{equation}
where $w(x,t)$ [m] is the crack opening, $t$ [s] stands for time and $q(x,t)$ $\left[\frac{\text{m}^2}{\text{s}}\right]$ is the average normalized (by the fracture height $H$) fluid flow rate through the fracture. 

The Newtonian fluid of a constant viscosity $\mu$ [Pa$\cdot$s] is assumed to flow unidirectionally through the crack under a laminar regime. Thus, the fluid flow rate can be expressed by the Poiseulle equation:
\begin{equation}
\label{Poiseulle}
q(x,t)=-\frac{1}{M}w^3\frac{\partial p}{\partial x},
\end{equation}
where $M=12\mu$ and $p(x,t)$ [Pa] is the fluid pressure. The fluid flow rate complies with the following boundary conditions:
\begin{itemize}
\item{the tip boundary condition:
\begin{equation}
\label{BCs_tip}
\quad q(a,t)=0,
\end{equation}}
\item{the influx boundary condition:
\begin{equation}
\label{BC_q}
q(0,t)=q_0(t).
\end{equation}}
\end{itemize}

The average velocity of fluid is computed as:
\begin{equation}
\label{v_def}
v(x,t)=\frac{q}{w}.
\end{equation}
Note that the ratio of $q$ and $w$ becomes indeterminate at the crack tip as both dependent variables go to zero for $x=a$. Thus, rigorous application of the tip asymptotics is needed when one employs the above relation. For a detailed discussion on the advantages and various aspects of using the fluid velocity in computations see the papers by  \cite{Kusmierczyk} and \cite{Wrobel_2015}. 

We assume that fluid fills the whole volume of the fracture, i.e. there is no lag between the fluid front and the crack tip. This implies that the following Stefan-type condition has to be satisfied:
\begin{equation}
\label{SE}
v(a,t)=\frac{\text{d}a}{\text{d}t}.
\end{equation}
From the above assumptions it follows that the fluid velocity is bounded in any point in space and time. Equation \eqref{SE} is used in the employed computational scheme to trace the fracture front. Description of the front tracing mechanism for different formulations of the hydraulic fracture problem can be found in the publication by \cite{Wrobel_2015}.

\subsection{Solid mechanics equations}
\label{sol_mech}

This group of equations describes deformation of the fractured material under the applied hydraulic pressure. As mentioned in the introduction two models of solid material are considered.  

The first one assumes a linear elastic behavior of the rock formation. In this case one can employ the standard boundary integral equation of elasticity to obtain relation between the crack opening and the fluid pressure \citep{Adachi_Detournay}:
\begin{equation}
\label{elasticity_1}
p(x,t)=  \frac{E}{2\pi(1-\nu^2)}\int_{0}^{a(t)}\frac{\partial w(s,t)}{\partial s} \frac{s\, ds}{x^2 - s^2}, \quad 0\le x<a(t),
\end{equation}
with the inverse operator defined as:
\begin{equation}
\label{inverse_KGD}
w(x,t)=\frac{4(1-\nu)}{\pi E}\int_0^{a(t)} p(s,t) \ln \left|\frac{\sqrt{a^2(t)-x^2}+\sqrt{a^2(t)-s^2}}{\sqrt{a^2(t)-x^2}-\sqrt{a^2(t)-s^2}}\right|ds.
\end{equation}
In the above formulae $E$ [Pa] is the Young's modulus and $\nu$ denotes the Poisson's ratio. In order to secure convergence of the singular integral in equation \eqref{elasticity_1} the following condition needs to be satisfied \citep{Wrobel_2015}:
\begin{equation}
\label{w_sym}
\frac{\partial w}{\partial x}(0,t)=0.
\end{equation}
Naturally, the zero opening boundary condition at the crack tip holds:
\begin{equation}
\label{w_tip}
w(a(t),t)=0.
\end{equation}
The system of equation \eqref{inverse_KGD} -- \eqref{w_tip} enables one to compute the crack opening profile, $w(x,t$), provided that the fluid pressure distribution, $p(x,t)$, is given. If however the stress distribution in the whole $(x,y)$ plane for an arbitrary fluid pressure is of interest, the full 2D numerical solution needs to be obtained.

	The second considered model of the fractured material assumes elasto-plastic behaviour. We do not detail here a complete system of equations for the elasto-plastic problem as in our solver we directly employ the ABAQUS FEA package to obtain the stress and displacement fields (the interested reader is directed to the software documentation of ABAQUS FEA \citep{ABAQUS}). The elastic deformation of the material is described by linear elasticity. The plastic deformation model is based on the Mohr-Coulomb flow theory which in the ABAQUS software includes: i) the classical Mohr-Coulomb yield criterion, ii) the smooth flow potential proposed by \cite{Menetrey_1995} instead of the classical  hexagonal pyramid (the flow potential forms a hyperbola in the meridional plane and a piecewise elliptic shape in the deviatoric stress plane). The ABAQUS module is used in computations to obtain the crack opening profile, $w(x,t)$, for the predefined fluid pressure distribution, $p(x,t)$. Here, the boundary conditions \eqref{w_sym}-\eqref{w_tip} also hold. 
	
	Among the three variants of the HF problem considered in this paper and mentioned in the introduction it is only the third one that uses the elasto-plastic model to describe the deformation of the bulk of fractured material.

\subsection{Crack propagation condition}\label{Crack_propagation_cond}

In line with the main objective of this paper we aim at quantifying the shielding mechanism of plastic deformation. To this end we `decouple' the plasticity effect embedded in the process of rock splitting from that related to the plastic deformation of the of bulk fractured material. Below we derive a dedicated crack propagation condition which takes into account the plastic deformation  in the plane of fracture extension. It is included in the so-called effective fracture toughness. Such an approach allows us to combine this crack propagation condition with the classical KGD model of hydraulic fracture that assumes only elastic deformation of the fractured material.  In this way we can assess whether accounting for the plasticity solely via the crack propagation condition can produce a sufficiently good approximation of the elastic-plastic solution at least in the case of small scale yielding. Simultaneously, the new condition can be combined with the full elasto-plastic deformation model of the bulk material to account for both plastic effects throughout the fracture length. Thus, we can estimate  to what extent the aforementioned two mechanisms contribute to the final solution.

\subsubsection{Stress relaxation model}
\label{stress_relaxation}
In our analysis we assume a priori that the plastic yield scale is small and contained in the near-tip region. For such conditions Irwin proposed the stress redistribution model that accounts for the plastic deformation of the fractured material in the near-tip zone in the case of Mode I fracture \citep{Irwin}. The model assumes that the original LEFM problem for a remote loading can be replaced by an equivalent elastic-plastic problem where inside the plastic yield area the stress equals the yield strength of the material, $\sigma_\text{F}$, and outside the plastic zone the stress distribution follows the standard $K$ asymptote.  Numerical results presented by \cite{Sun_2012} imply that this approach can be effective and accurate even in the cases of relatively large extent of plastic deformation. On the other hand, in many publications (see the paper by \cite{Dyskin_1997} and references therein) the importance of including also the non-singular terms in the fracture analysis is emphasized.
When trying to directly adopt the concept of the Irwin's stress relaxation model to the conditions of hydraulic fracture we arrived at conclusion that the original way of approximating the elastic solution near the crack tip by a single term of $K$ asymptote  may not be sufficient for accurate prediction of the plastic zone size and stress distribution. Thus, we have amended the Irwin's approach in the following way.

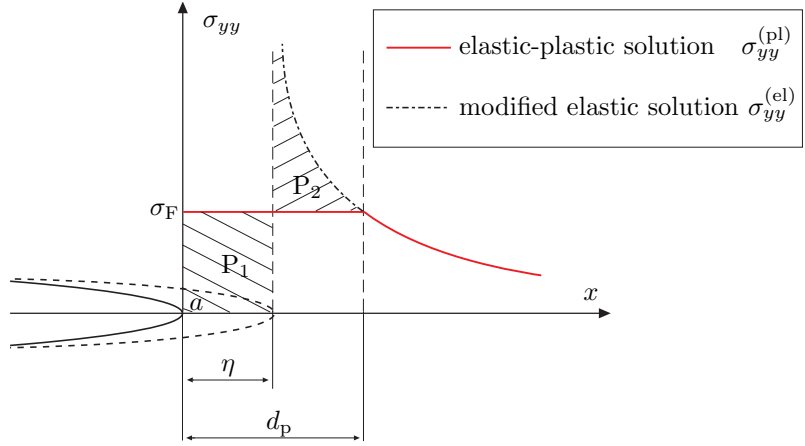
\begin{figure}[htb!]
\begin{center}
\includegraphics[scale=1.2]{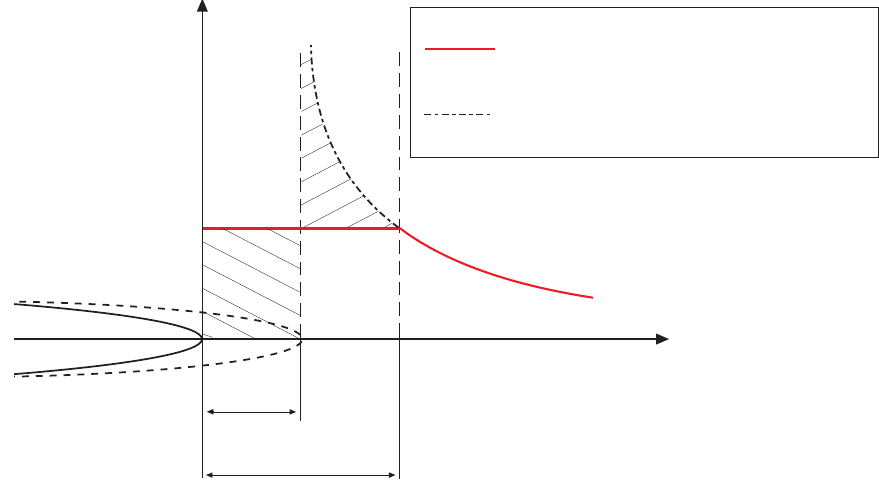}
\put(-83,55){$x$}
\put(-227,158){$\sigma_{yy}$}
\put(-247,87){$\sigma_\text{F}$}
\put(-220,65){P$_1$}
\put(-193,95){P$_2$}
\put(-220,28){$\eta$}
\put(-232,51){$a$}
\put(-203,6){$d_\text{p}$}
\put(-130,148){$\text{elastic-plastic solution} \quad \sigma_{yy}^\text{(pl)}$}
\put(-130,125){$\text{modified elastic solution }  \sigma_{yy}^\text{(el)}$}
\caption{The near tip zone and stress distributions in the stress relaxation model.}
\label{tip}
\end{center}
\end{figure}

It is assumed that the original problem for the crack of half-length $a$ can be replaced by a modified elastic solution 
where the (fictitious) crack tip is located at $x=a+\eta$ - see Figure \ref{tip}. This modified solution  
is defined such that it is equivalent to the elastic-plastic solution (marked in the figure by a red line). The elastic-plastic solution coincides with the elastic solution outside the plastic deformation zone whose extent is described by $d_\text{p}$. Inside the plastic deformation zone the material is assumed to  be perfectly plastic. The stress distributions pertaining to the respective solutions (i.e. the modified elastic solution and elastic-plastic solution) are assumed to produce the same overall load in the plane of fracture extension.

In our analysis we adopt the solution given by \cite{Inglis} for an elliptic cavity in an infinite plane subjected to the uniaxial tension. However, the original formula is modified to account for internal pressure, $p$, applied to the crack faces instead of remote loading and to include confining stress\footnote{The original formula provided by Inglis for the remote load yields: $\sigma_{yy}=\frac{p}{2}\left[2+\frac{(m+1)^2}{s^2-m}+(m^2-1)(m-1)\frac{3s^2-m}{(s^2-m)^3}\right]$.}, $\sigma_{yy}^\text{c}$:
\begin{equation}
\label{Inglis_stress}
\sigma_{yy}=\frac{p}{2}\left[\frac{1+2(1+m)-m^2}{s^2-m}-(m^2-1)(m-1)\frac{3s^2-m}{(s^2-m)^3}\right]+\sigma_{yy}^\text{c}\frac{x}{\sqrt{x^2-a^2}}, \quad x>a,
\end{equation}
where:
\[
s=\frac{x}{2B}+\sqrt{\left(\frac{x}{2B}\right)^2-m}, \quad B=\frac{1}{2}(a+b), \quad  m=\frac{a-b}{a+b}.
\]
In the above formulae $a$ and $b$ are the semi-major and semi-minor axes of the cavity, respectively. 

Equation \eqref{Inglis_stress} defines an exact solution for stress distribution in the plane of fracture extension, provided that the applied fluid pressure, $p$, is constant along the crack faces. In the proposed model we account for the possible non-uniform distribution of pressure by introducing an effective fluid pressure, $p^\text{eff}$, computed as:
\begin{equation}
\label{p_eff}
p^\text{eff}=\frac{\int_0^a p \text{d}x}{a}.
\end{equation}  
Moreover, the semi-minor axis of the cavity, $b$, is defined from the standard LEFM solution for an elliptic fracture loaded by an internal pressure $p^\text{eff}$: 
\begin{equation}
\label{b_def}
b=\frac{2(1-\nu^2)}{E}a p^\text{eff}.
\end{equation}
As a consequence of the above assumption: $b=b\left(p^\text{eff}\right)$, $m=m\left(p^\text{eff}\right)$ and $s=s\left(p^\text{eff}\right)$. Note that also the non-singular terms of the stress field around the fracture tip are included in the model which eliminates the basic deficiency of the Irwin's solution (a single-term approximation of $\sigma_{yy}$). The legitimacy of the proposed approach will be demonstrated in Section \ref{cond_verification}. We would like emphasize here that formula \eqref{Inglis_stress} is very convenient in numerical implementation of the HF problem as it provides an explicit algebraic relation between $p$ and $\sigma_{yy}$ without a need to integrate the singular fluid pressure (as it is in the standard integral definition of $\sigma_{yy}$ - see e.g. the paper by \cite{Dyskin_1997}), which in turn contributes to stability of the iterative computations.  


We assume that in the modified elastic solution $\sigma_{yy}^\text{(el)}$ follows the distribution \eqref{Inglis_stress} but translated with respect to the fracture tip  by the value of $\eta$, as shown in Figure \ref{tip} (i.e. it complies with the fictitious crack tip located at $x=a+\eta$) . The elastic-plastic solution is defined as:
\begin{equation}
\label{TP_def1}
\sigma_{yy}^\text{(pl)}=
  \begin{cases}
		\sigma_\text{F}       & \quad \text{for } \quad x\leq a+d_\text{p},\\
		
    \sigma_{yy}^\text{(el)}  & \quad \text{for } \quad x>a+d_\text{p}.
  \end{cases}
\end{equation}

The values of the plastic zone size, $d_\text{p}$, and the location of the fictitious crack tip, $\eta$, are elements of solution to be found from two conditions:
\begin{itemize}
\item{the stress continuity at the elastic-plastic boundary:
\begin{equation}
\label{sig_cont}
\sigma_{yy}^\text{(el)}\big|_{x=a+d_\text{p}}=\sigma_\text{F},
\end{equation}}
\item{the load balance condition:
\begin{equation}
\label{bal_cond}
\sigma_\text{F}d_\text{p}=\int_{a+\eta}^{a+d_\text{p}}\sigma_{yy}^\text{(el)}dx.
\end{equation}}
\end{itemize}
Equation \eqref{bal_cond} ensures that for both solutions, the modified elastic solution and the elastic-plastic solution, the external loading carried in the plane of fracture extension is the same. In graphical interpretation this condition reduces to the hatched areas $P_1$ and $P_2$ being equal (compare Figure \ref{tip}). Unfortunately, unlike in the Irwin approach, the system of equations \eqref{sig_cont}--\eqref{bal_cond} cannot be solved analytically. On the other hand, numerical solution of this system can be easily obtained.

For the sake of completeness we would like to mention that in the Irwin's stress relaxation model the respective solution components are found analytically as:
\begin{equation}
\label{d_p}
d_\text{p}=\frac{1}{\pi}\left(\frac{K_I}{\sigma_\text{F}}\right)^2, \quad \eta=\frac{1}{2\pi}\left(\frac{K_I}{\sigma_\text{F}}\right)^2,
\end{equation}
where $K_I$ is the mode I stress intensity factor.

\subsubsection{Plane strain fracture - a modified stress intensity factor}
\label{KGD}

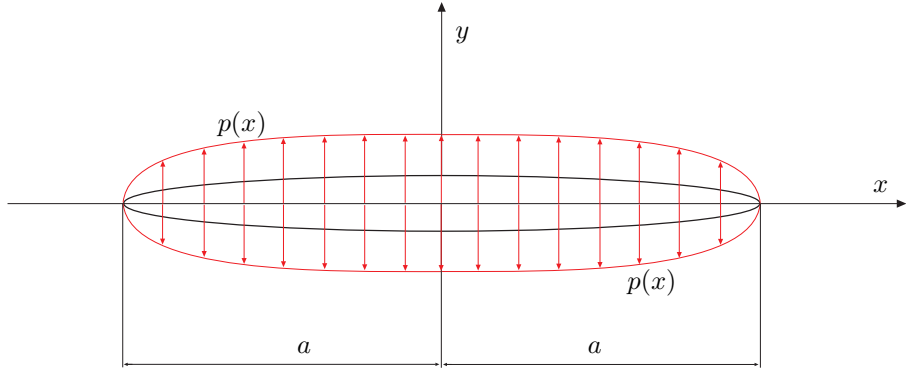
\begin{figure}[htb!]
\begin{center}
\includegraphics[scale=0.8]{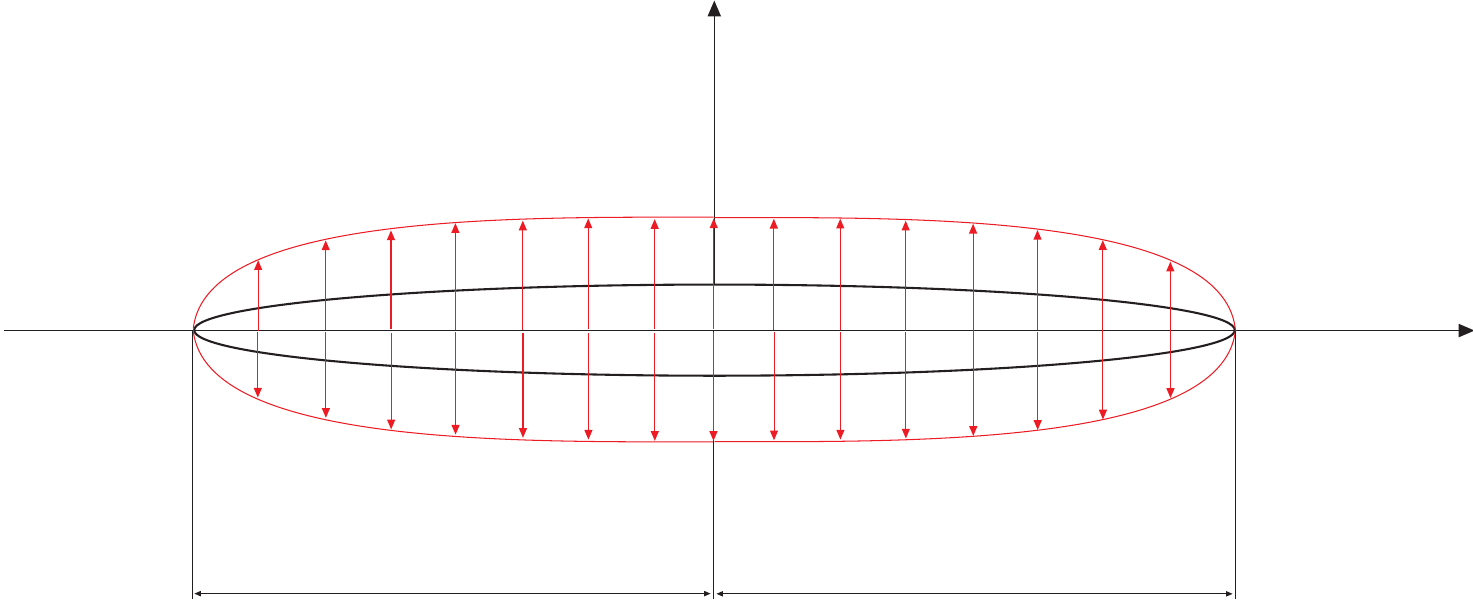}
\put(-12,67){$x$}
\put(-170,125){$y$}
\put(-260,90){$p(x)$}
\put(-105,30){$p(x)$}
\put(-230,6){$a$}
\put(-120,6){$a$}
\caption{Plane strain fracture of length $2a$ loaded by internal pressure $p(x)$. }
\label{SIF}
\end{center}
\end{figure}

The standard crack propagation condition for a fracture in elastic medium is based on the Energy Release Rate (ERR) criterion. It reduces to the following relation:
\begin{equation}
\label{K_cond}
K_I=K_{I\text{c}},
\end{equation}
where $K_{I\text{c}}$ is the material fracture toughness. For a two winged fracture of length $2a$ loaded by internal fluid pressure $p(x)$ (see Figure \ref{SIF}) the Mode I stress intensity factor can be expressed as:
\begin{equation}
\label{K_I}
K_I=2\sqrt{\frac{a}{\pi}}\int_0^a\frac{p(x)}{\sqrt{a^2-x^2}}\text{d}x.
\end{equation}

\begin{figure}[htb!]
\begin{center}
\includegraphics[scale=0.8]{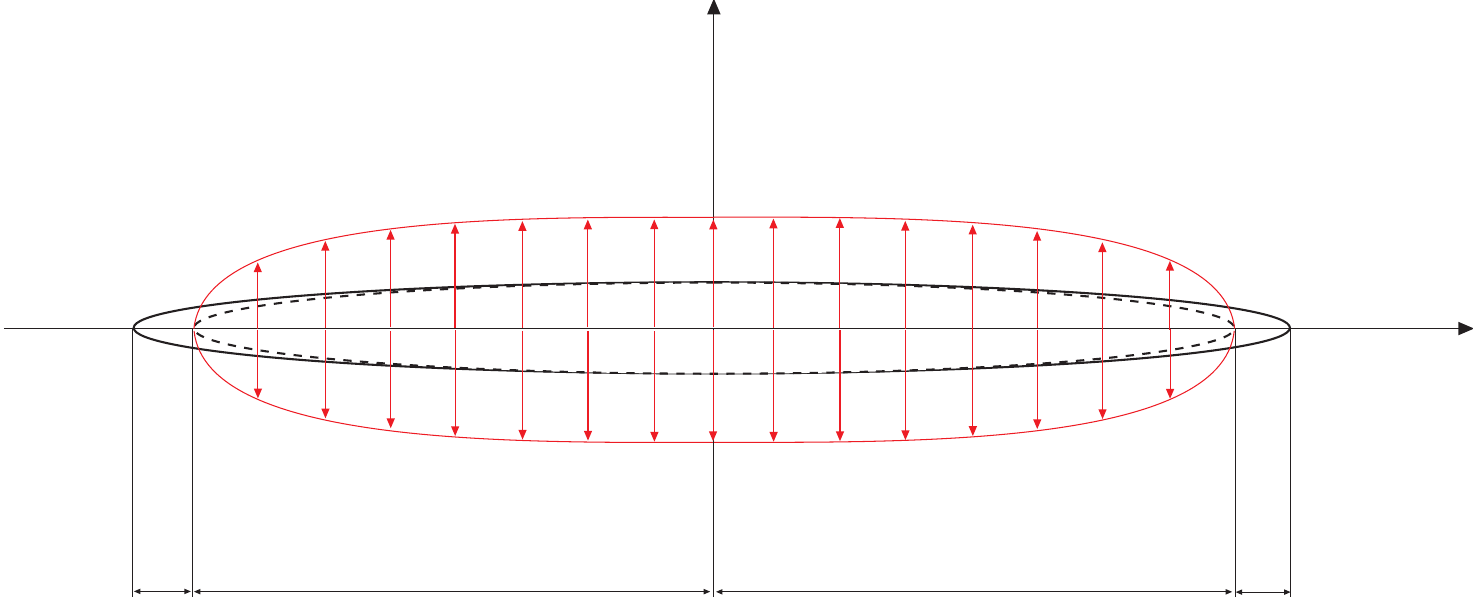}
\put(-12,67){$x$}
\put(-170,125){$y$}
\put(-260,90){$p(x)$}
\put(-105,30){$p(x)$}
\put(-230,6){$a$}
\put(-120,6){$a$}
\put(-306,6){$\eta$}
\put(-50,6){$\eta$}
\caption{Plane strain fracture of length $2(a+\eta)$ loaded by internal pressure.}
\label{SIF_1}
\end{center}
\end{figure}
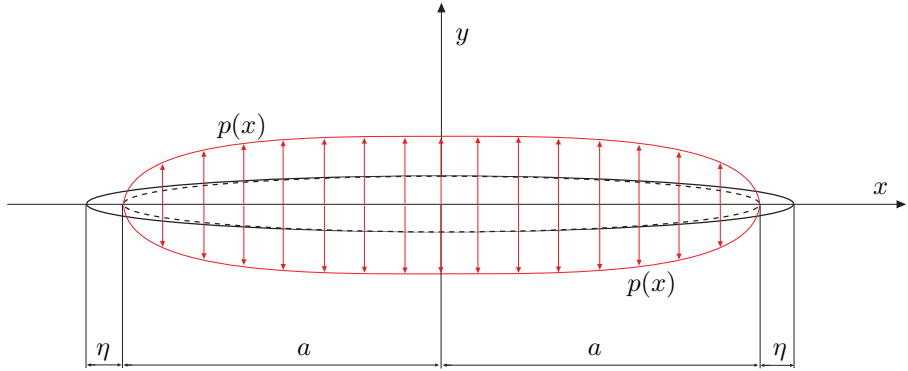

Let us assume that for the elasto-plastic problem the stress redistribution presented above holds and the equivalent LEFM problem for a fracture of an effective length $2(a+\eta)$ is considered - see Figure \ref{SIF_1}. The fluid pressure remains the same as in the previous case (i. e. there is no additional fluid pressure within the region $x\in [a,a+\eta]$). Under these conditions the Mode I SIF can be computed as:
\begin{equation}
\label{K_Ia}
K_I^{(\text{p})}=2\sqrt{\frac{a+\eta}{\pi}}\int_0^a\frac{p(x)}{\sqrt{(a+\eta)^2-x^2}}\text{d}x,
\end{equation}
where $\eta$ is obtained from the system \eqref{sig_cont}--\eqref{bal_cond}.

It is clear when comparing \eqref{K_I} and \eqref{K_Ia} that, for $\eta>0$ and $p>0$:
\begin{equation}
\label{KIp_est}
K_I^{(\text{p})}<K_I.
\end{equation}

The physical explanation of this fact is that, with the stress redistribution pertaining to the plastic yielding, the stress concentration in the near-tip zone is alleviated. Thus, when considering two cases: i) the standard one (fully elastic) from Figure \ref{SIF}, ii) the one that accounts for the plastic deformation -  Figure \ref{SIF_1}, for the same values of loading, $p(x)$, and fracture toughness, $K_{I\text{c}}$, it is the first variant that brings us closer to fulfillment of the crack propagation condition \eqref{K_cond}. In other words, plastic yielding reduces the actual value of the stress intensity factor and increases the material resistance to fracture.
In this case the crack propagation condition can be reformulated as:
\begin{equation}
\label{K_Ip}
K_I^{(\text{p})}=K_{I\text{c}}.
\end{equation}

Let us assume that in the problem whose geometry is described by the variant I (standard elastic plane strain crack), the crack propagation condition \eqref{K_Ip} holds. Then, we can rewrite this condition as:
\begin{equation}
\label{cond_equiv}
\alpha K_I=K_{I\text{c}},
\end{equation}
where:
\begin{equation}
\label{alfa_def}
\alpha=\frac{K_I^{(\text{p})}}{K_I}<1.
\end{equation}

In this way condition \eqref{K_cond} can be replaced by the following one:
\begin{equation}
\label{K_eff}
K_I=K_{I\text{c}}^\text{eff},
\end{equation}
where:
\begin{equation}
\label{K_I_eff}
K_{I\text{c}}^\text{eff}=\frac{K_{I\text{c}}}{\alpha}.
\end{equation}
Clearly, the following estimation can be deduced from the above relations:
\begin{equation}
\label{K_I_est}
K_{I\text{c}}^\text{eff}>K_{I\text{c}}.
\end{equation}
When solving the HF problem, coefficient $\alpha$ is computed in each iteration according to \eqref{alfa_def} with the respective definitions of $K_I$ and $K_I^{(\text{p})}$ given by \eqref{K_I} and \eqref{K_Ia}. The values of $\eta$ and $d_\text{p}$ are found from the solution of the system \eqref{sig_cont} -- \eqref{bal_cond}. Note that as the fracture length increases the influence of the plastic deformation in the near tip zone is expected to be less pronounced, given that the ratio $\eta/a$ decreases.

\section{Numerical algorithm}
\label{algorithm}

The computational scheme employed to simulate the problem of hydraulic fracture is based on the universal algorithm originally introduced in the paper by \cite{Wrobel_2015} and further developed by \cite{Perkowska_2016,Wrobel_2017,Wrobel_2020,Wrobel_2021} with the respective version for the radial (penny shaped) model  described in the publications by \cite{Peck_2018_1,Peck_2018_2}. The algorithm utilizes two dependent variables: the crack opening, $w$ and the fluid velocity $v$. A modular structure of the scheme  facilitates its modifications to account for different rock deformation and fluid flow models as well as the crack propagation conditions. In general, the iterative algorithm of solution assumes consecutive application of two basic modules:
\begin{itemize}
\item{the so-called `$v$ module' - here the continuity equation \eqref{cont} is solved to obtain the fluid velocity inside the fracture;}
\item{the so-called `$w$ module' - in this block of the algorithm the solid deformation under the applied hydraulic pressure is computed (in the case of elastic model of solid equation \eqref{inverse_KGD} is applied).}
\end{itemize}
Additional subroutines are used for fracture front tracing (on the basis of equation \eqref{SE}) and integration of the fluid pressure derivative (determined according to equation \eqref{Poiseulle}) to obtain $p(x,t)$. The computed distribution of the fluid pressure is used in accordance with the newly introduced stress relaxation model and formula \eqref{alfa_def} to produce the toughness scaling coefficient, $\alpha$. As mentioned previously, the value of $\eta$, needed for the stress intensity factor \eqref{K_Ia}, is obtained by solving the system \eqref{sig_cont} -- \eqref{bal_cond}. In this way $\eta$ constitutes a sought component of solution. The information on solution tip asymptotics (including the crack propagation condition) is extensively used in the simulations, primarily to determine the position of the fracture front and to cancel out the main singular terms in the governing equations.
The computations are implemented in the Matlab environment. For a detailed description of the algorithm we direct the prospective reader to the cited publications. 

The solution for the classical KGD model (the first analyzed variant of the HF problem) is obtained by a direct application of the algorithm proposed by \cite{Wrobel_2015}. In the second variant of the HF problem, where the elastic deformation of the solid  is combined with the plasticity affected  crack propagation condition, we modify the computational scheme by replacing the standard LEFM condition, $K_I=K_{I\text{c}}$, with the condition \eqref{K_eff}.

However, when analyzing the case where the elasto-plastic material model is used for the whole bulk of the fractured rock (the third variant of the HF problem), the respective subroutine for the solid deformation from the original algorithm needs to be replaced by a dedicated FEM based scheme. Due to the modular structure of the algorithm it is sufficient to replace only the module that computes the crack opening, $w(x,t)$, with the remaining elements of the numerical scheme being the same as in the previous variant. Naturally, this time the complete 2D fields of stresses and displacements in the $(x,y)$ plane need to be analyzed. A corresponding finite element model for the solid deformation sub-problem is built in the ABAQUS FEA package. The interaction between ABAQUS and Matlab is provided by the \textit{Abaqus2Matlab} interface \citep{Abaqus2Matlab}. 

\begin{figure}[htb!]
\begin{center}
\includegraphics[scale=0.5]{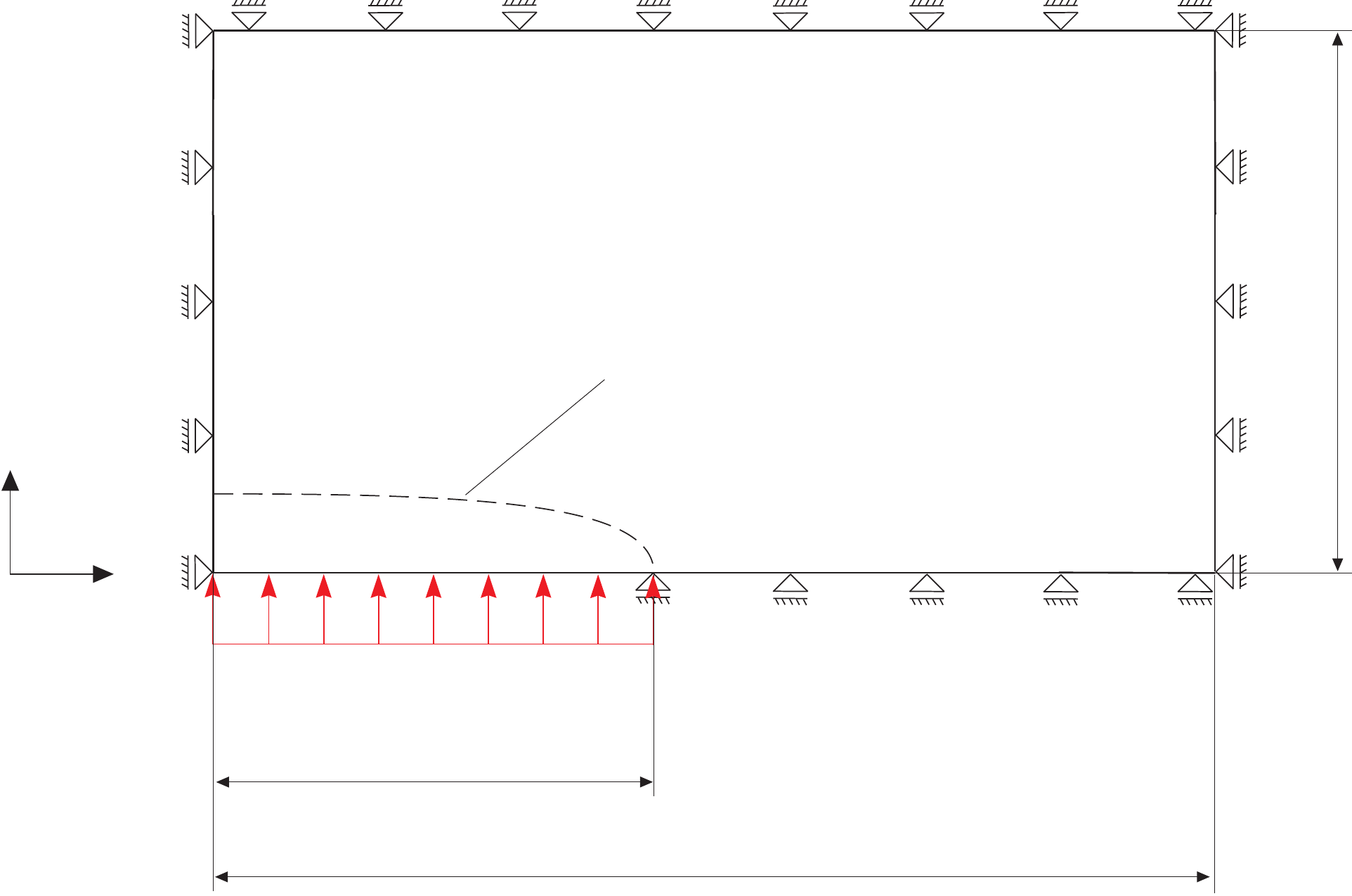}
\put(-260,70){$\tilde x$}
\put(-272,80){$\tilde y$}
\put(-12,115){\rotatebox{90}{100}}
\put(-200,40){$p(\tilde x,t)$}
\put(-190,25){1}
\put(-140,6){101}
\put(-215,110){$\text{resulting fracture profile} \ \frac{1}{2}w(\tilde x,t)$}
\caption{A pattern geometry of the FEM problem in the normalized variables: $\tilde x=x/a$, $\tilde y =y/a$.}
\label{FEM_geometry}
\end{center}
\end{figure}

\begin{figure}[htb!]
\begin{center}
\includegraphics[scale=0.45]{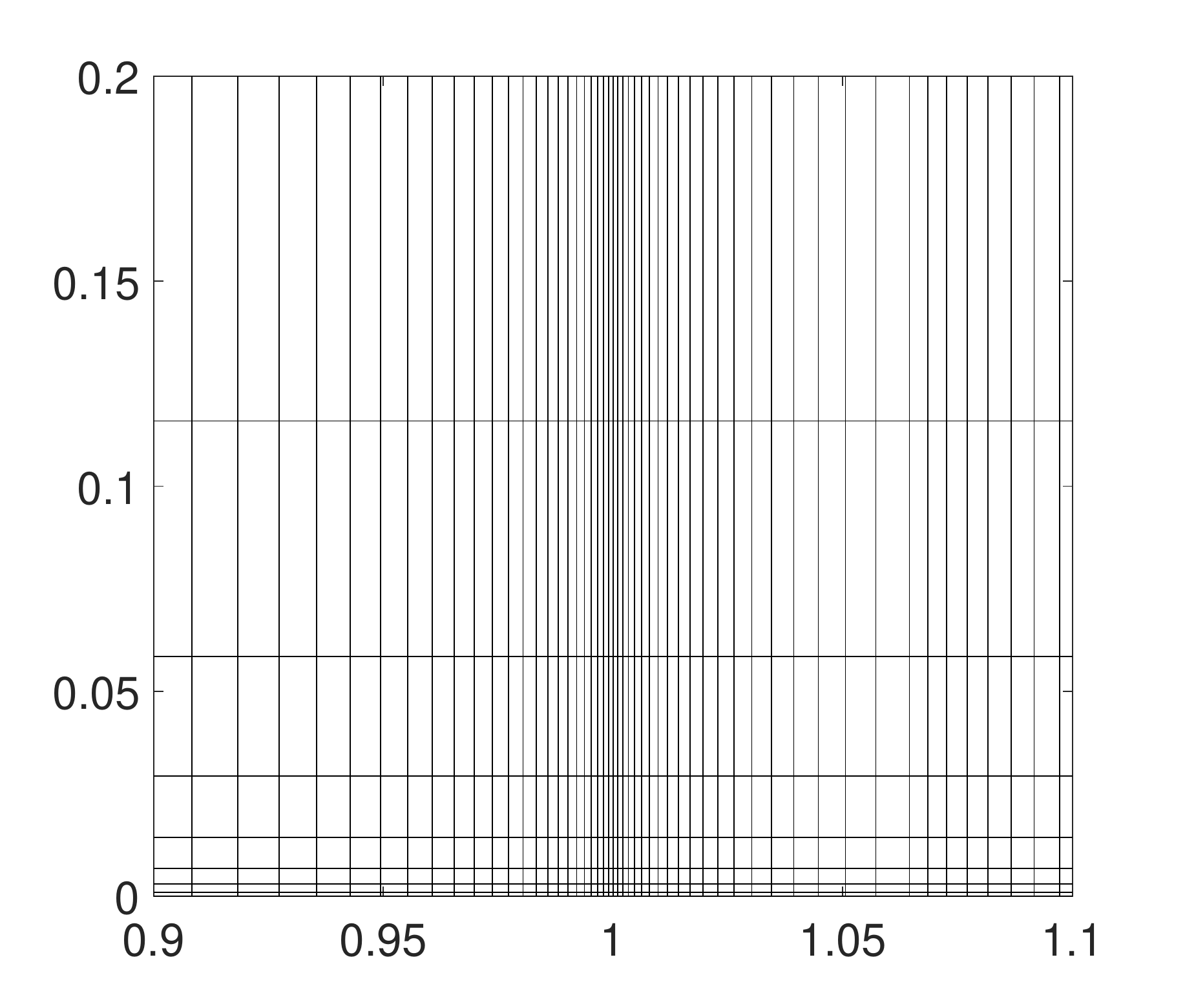}
\put(-118,0){$\tilde x$}
\put(-240,103){$\tilde y$}
\caption{FEM mesh pattern in the proximity of the crack tip. The normalized spatial variables are used: $\tilde x=x/a$, $\tilde y =y/a$.}
\label{mesh}
\end{center}
\end{figure}

A sketch of the FEM problem geometry is depicted in Figure \ref{FEM_geometry}. The domain is shown in the scaled spatial variables: $\tilde x=x/a$, $\tilde y=y/a$. Due to symmetry, only one quadrant of the original domain is considered with the respective symmetry boundary conditions imposed at $\tilde x=0$ and $\tilde y=0$ (for $\tilde x \geq 1$). In numerical computations this pattern geometry is re-scaled at every time step by the value of the crack length $a(t)$. In this way, the ratio of the crack length to the external dimensions of the domain is kept constant throughout the whole computational process. Moreover, as the mesh pattern is retained the same during computations, the technical implementation of the domain rescaling reduces to a simple multiplication of the nodal coordinates in the ABAQUS input file by a constant value. Thus, rescaling of the FEM domain at each iteration does not additionally detract from the efficiency of computations   as there is no need for a repeated use of mesh generator.

In the original version of the algorithm (designed for the elastic deformation of solid) the computations are carried out over a normalized spatial interval ($\tilde x=x/a$) with the mesh density increased near both ends of the fracture. In the FEM-based variant of the scheme the spatial meshes in the respective modules of the algorithm are different. Spatial mesh in the `$v$ module' remains the same as in the original scheme, while the crack discretization in the `$w$ module' results directly from the employed mesh of finite elements (note that the computational domain in this block of the algorithm is no longer normalized by the crack length). For this reason, the outputs from the respective algorithm modules need to be interpolated over the proper mesh before being reintroduced to the next block.

 The computations are carried out with the eight-node bi-quadratic plane strain elements (CPE8R). The mesh consists of 1751 finite elements, with the density increasing near the fracture tip. The mesh pattern is depicted in Figure \ref{mesh}. This pattern has been selected in order to facilitate the control of the mesh density over the fracture area and interpolation of the results between the respective modules of the algorithm. As a consequence, the aspect ratio of the finite elements adjacent to the plane $x=a$ becomes very high as moving away from the fracture plane. On the other hand, the elements located away from the crack plane do not undergo large deformations. Thus, in our algorithm the aforementioned drawback does not negatively affect the accuracy and stability of computations, which  has been confirmed in numerical simulations. A comprehensive analysis of the algorithm performance can be found in the preprint by \cite{Wrobel_FEM} where also the solution sensitivity to the density of the finite element meshing was investigated.  It has been established that the for the FEM mesh of 1751 elements the relative accuracy of computations for both $w$ and $v$ is of the order 10$^{-3}$. For a detailed description of the computational algorithm and verification of its performance we refer the prospective reader to the recalled preprint.


\section{Verification of the stress relaxation model}
\label{cond_verification}

In this section we verify  the underlying assumptions of the stress relaxation model introduced in Subsection \ref{stress_relaxation}. This investigation is going to provide us with an indirect verification of the legitimacy of the newly introduced crack propagation condition. Note that full evidence of the condition applicability to the HF problems would require comparison with other computational models or/and experimental data. First we check the quality of approximation of the stress component $\sigma_{yy}$ in the plane of fracture extension provided by the formula \eqref{Inglis_stress} for the LEFM problem when non-uniform fluid pressure is applied. Then a quantitative analysis of the elasto-plastic solution based on the relaxation model is performed.

In order to investigate the applicability of formula \eqref{Inglis_stress} in those cases where fluid pressure is non-uniform along the crack length we utilize two benchmark solutions. The first one is taken from the paper by \cite{Dyskin_1997} and involves an example of a fracture loaded by fluid pressure applied only to a part of the crack faces:
\begin{equation}
\label{p_step_bench}
p(x)=
  \begin{cases}
		\bar p       & \quad \text{for } \quad |x|\leq a-d,\\	
    0  & \quad \text{for } \quad |x|>a-d,
  \end{cases}
\end{equation}
where $d<a$. The analytical solution  in terms of $\sigma_{yy}$ yields:
\begin{equation}
\label{sigma_step_bench}
\sigma_{yy}(x)=\bar p\left[ \frac{2x\arcsin(1-d/a)}{\pi \sqrt{x^2-a^2}}-\frac{2}{\pi}\arctan\left(\frac{(a-d)\sqrt{x^2-a^2}}{x\sqrt{d(2a-d)}}\right)\right].
\end{equation}
Note that this benchmark example does not describe any realistic scenario of the HF problem. On the other hand, it constitutes quite challenging test for the proposed stress approximation formula and simultaneously provides fully analytical solution for $\sigma_{yy}$.

In the second analyzed benchmark example we assume that the fluid pressure is expressed as:
\begin{equation}
\label{p_bench}
p(\tilde x)=\sum_{i=0}^3  k_i \Pi_i(\tilde x),
\end{equation}
where:
\[
\tilde x=x/a, \quad E'=\frac{E}{1-\nu^2}, \quad \Pi_0(\tilde x)=\frac{E'}{4  a}, \quad \Pi_1(\tilde x)=\frac{E'}{\pi a}\left[1-\tilde x \arctanh(\tilde x) \right],
\]
\begin{equation}
\label{Pi_def}
\end{equation}
\[
\Pi_2(\tilde x)=\frac{E'}{4 a}\left[1-2\tilde x^2+\frac{3}{2}\left(1-4\tilde x \sqrt{1-\tilde x^2}\arcsin(\tilde x) +4\ln(2)\tilde x^2-\ln(4)\right) \right], \quad \Pi_3(\tilde x)=\frac{E'}{a}\left(1-\pi/2\tilde x\right).
\]
Respective coefficients $k_i$ are taken to ensure physically realistic behaviour of the solution. The above representation was used in the paper of \cite{Wrobel_2015} (and subsequent publications of the author) to construct an analytical benchmark solution for the hydraulic fracture problem. The tip asymptotics of $p$ complies here with the so-called toughness dominated regime of crack propagation where a logarithmic singularity holds at the the fracture front. Moreover, formula \eqref{p_bench} enables analytical integration of the boundary integral equation of elasticity \eqref{inverse_KGD} to obtain a closed form analytical solution for the crack opening. Unfortunately, to the best of authors' knowledge, no closed form analytical solution for $\sigma_{yy}$ can be derived here. Nevertheless, numerical computation of stress function in the fracture plane is not problematic, and the numerical solution will be considered here as a reference (benchmark) function. Note that this benchmark, unlike the one described by equations \eqref{p_step_bench} -- \eqref{sigma_step_bench}, reflects properly the qualitative behavior of the fluid pressure in the HF problem.

\begin{remark}
\label{remark_sc}
The term related to the confining stress, $\sigma_{yy}^\text{c}$, in formula \eqref{Inglis_stress} constitutes an exact analytical solution to the problem of a fracture subjected to the remote load. Thus, it does not introduce any error of approximation to the relation \eqref{Inglis_stress}.  For this reason it is expected that with growing confining stress magnitude the quality of approximation \eqref{Inglis_stress} increases. In other words, for two load cases with the same magnitude and distribution of the net fluid pressure it is the one with higher absolute value of $\sigma_{yy}^\text{c}$ which should provide better resemblance of the exact LEFM  solution.
\end{remark}

\begin{figure}[htb!]
\begin{center}
\includegraphics[scale=0.4]{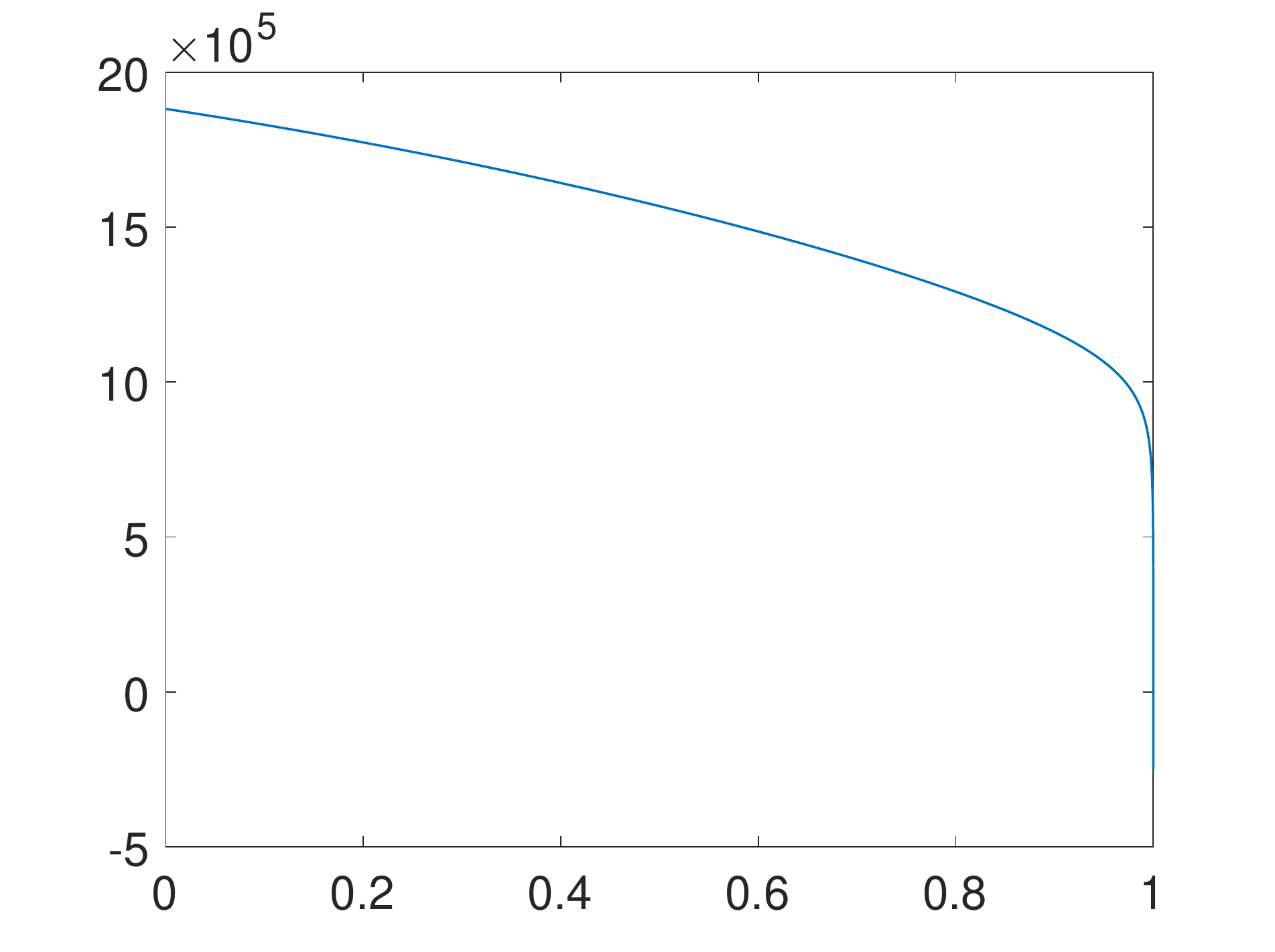}
\put(-110,-5){$x$}
\put(-215,80){$p$}
\caption{The distribution of the fluid pressure, $p$ [Pa], computed according to relation \eqref{p_bench}. }
\label{p_HF_bench}
\end{center}
\end{figure}

In this part of our analysis we use the following values of the respective benchmarks' parameters:
\begin{itemize}
\item{benchmark \eqref{p_step_bench}: $\bar p=2$ MPa, $a=1$ m, $d=0.3$ m;}
\item{benchmark \eqref{p_bench}: $k_0=3.54\cdot 10^{-4}$, $k_1=4.1\cdot 10^{-5}$, $k_2=5\cdot 10^{-6}$, $k_3=2\cdot 10^{-5}$, $E=14.5$ GPa, $\nu=0.24$, $a=1$ m. The resulting distribution of the fluid pressure function is depicted in Figure \ref{p_HF_bench}. The logarithmic singularity of the fluid pressure produces negative values of $p$ at the crack tip.}
\end{itemize}

\begin{figure}[htb!]
\begin{center}
\includegraphics[scale=0.40]{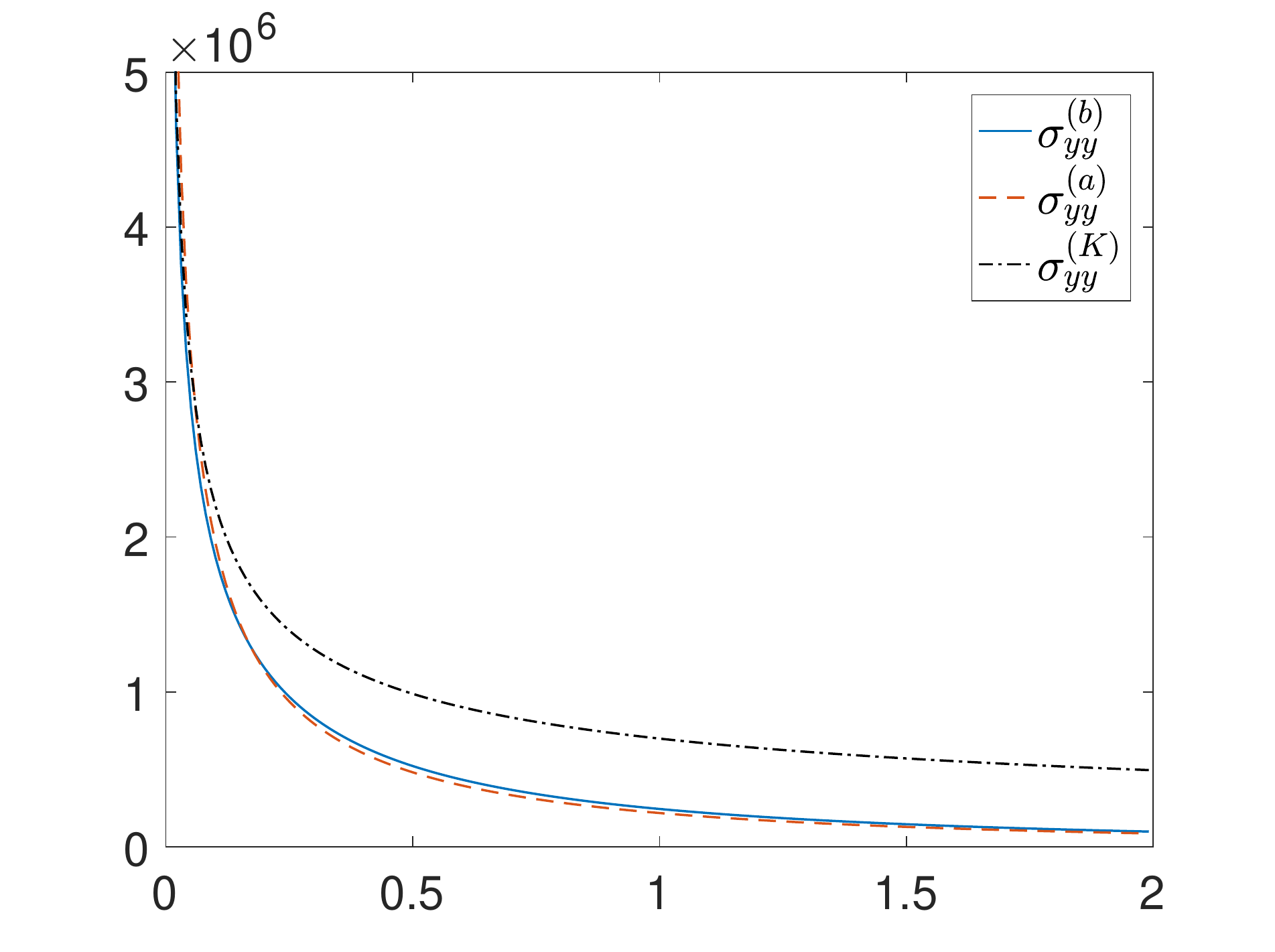}
\hspace{0mm}
\includegraphics[scale=0.40]{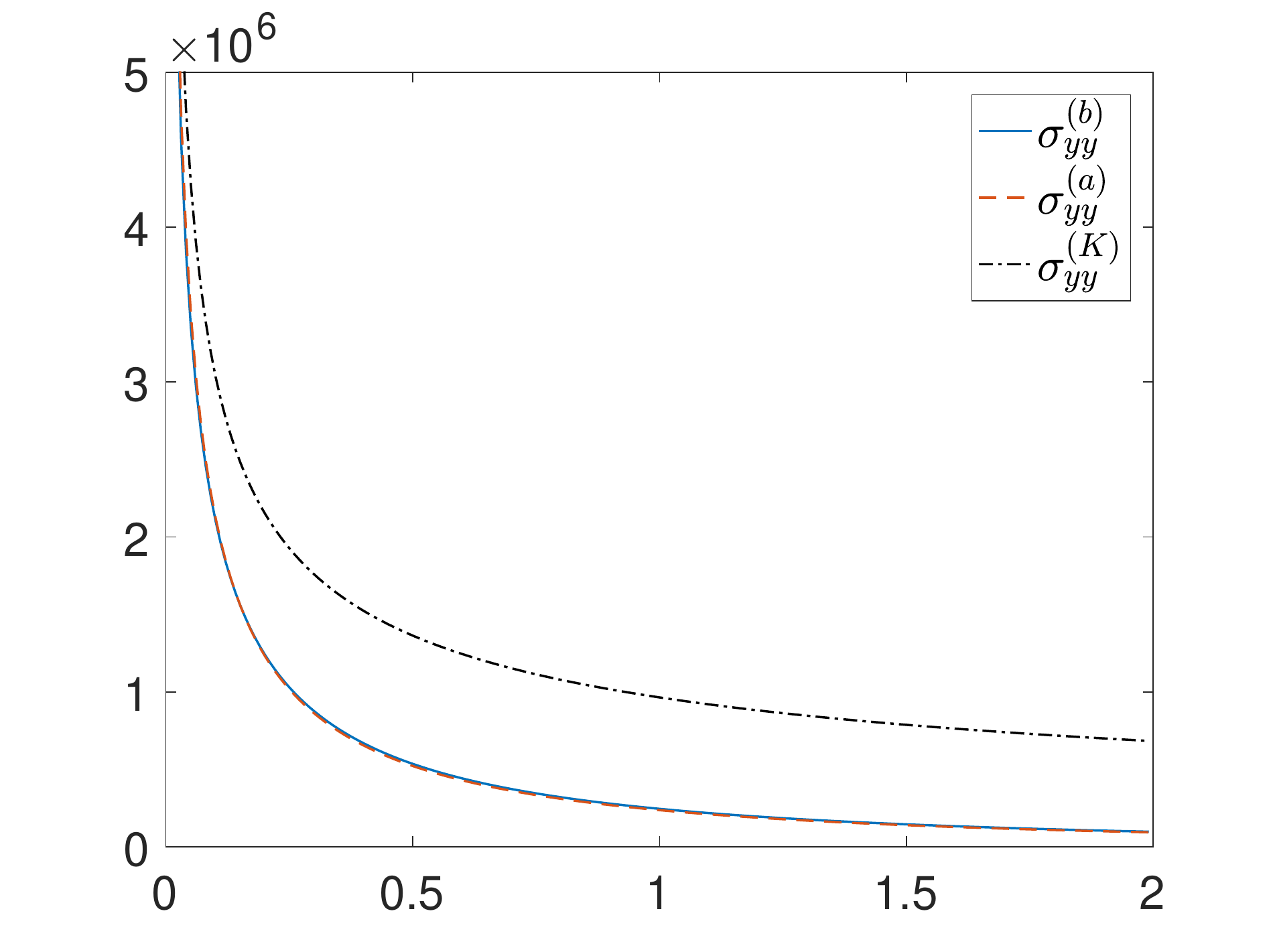}
\put(-335,-5){$r$}
\put(-110,-5){$r$}
\put(-450,155){$\textbf{a)}$}
\put(-225,155){$\textbf{b)}$}
\put(-450,85){$\sigma_{yy}$}
\put(-220,85){$\sigma_{yy}$}
\caption{Stress distribution ahead of the crack tip, $\sigma_{yy}$ [Pa], for: a) the benchmark example \eqref{p_step_bench} -- \eqref{sigma_step_bench}, b) the  benchmark example \eqref{p_bench} -- \eqref{Pi_def}. Respective notations stand for: $\sigma_{yy}^{(\text{b})}$ - the benchmark stress distribution,  $\sigma_{yy}^{(\text{a})}$ - stress values approximated by formula \eqref{Inglis_stress}, $\sigma_{yy}^{(K)}$ - the leading term approximation of the stress ($K$ - asymptote).}
\label{sigma_comparison}
\end{center}
\end{figure}
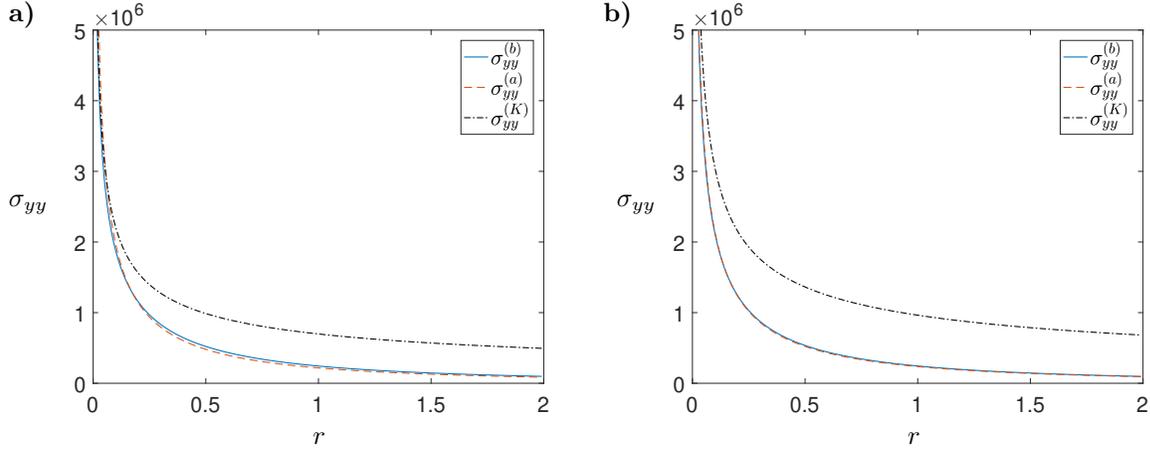

The results obtained by employing relation \eqref{Inglis_stress} are shown in Figure \ref{sigma_comparison}  as functions of variable $r=x-a$ which describes the distance from the crack tip. Note that this test verifies only the quality of approximating the elastic solution. No elasto-plastic problem is considered at this stage. Also the confining stress is taken zero here as, in line with  Remark \ref{remark_sc}, such a variant constitutes a more challenging case. Respective notations in the legend refer to: $\sigma_{yy}^{(\text{b})}$ - the benchmark stress distribution,  $\sigma_{yy}^{(\text{a})}$ - stress values approximated by formula \eqref{Inglis_stress}, $\sigma_{yy}^{(K)}$ - the leading (singular) term approximation of the stress ($K$ - asymptote) defined as:
\begin{equation}
\label{K_stress}
\sigma_{yy}^{(K)}=\frac{K_I}{\sqrt{2\pi(x-a)}},
\end{equation}
where for the Mode I SIF, $K_I$, an exact benchmark value is taken. 

As can be seen in Figure \ref{sigma_comparison}a) the stress approximation for the first benchmark is very good, yet a slight difference  between $\sigma_{yy}^{(\text{b})}$ and $\sigma_{yy}^{(\text{a})}$ is observed. On the other hand, in the case of the second benchmark example, respective graphs for $\sigma_{yy}^{(\text{b})}$ and $\sigma_{yy}^{(\text{a})}$ are virtually indistinguishable. In both cases the simple $K$ - asymptote approximation does not provide reasonable estimation of the stress except in the immediate vicinity of the crack tip.

\begin{figure}[htb!]
\begin{center}
\includegraphics[scale=0.40]{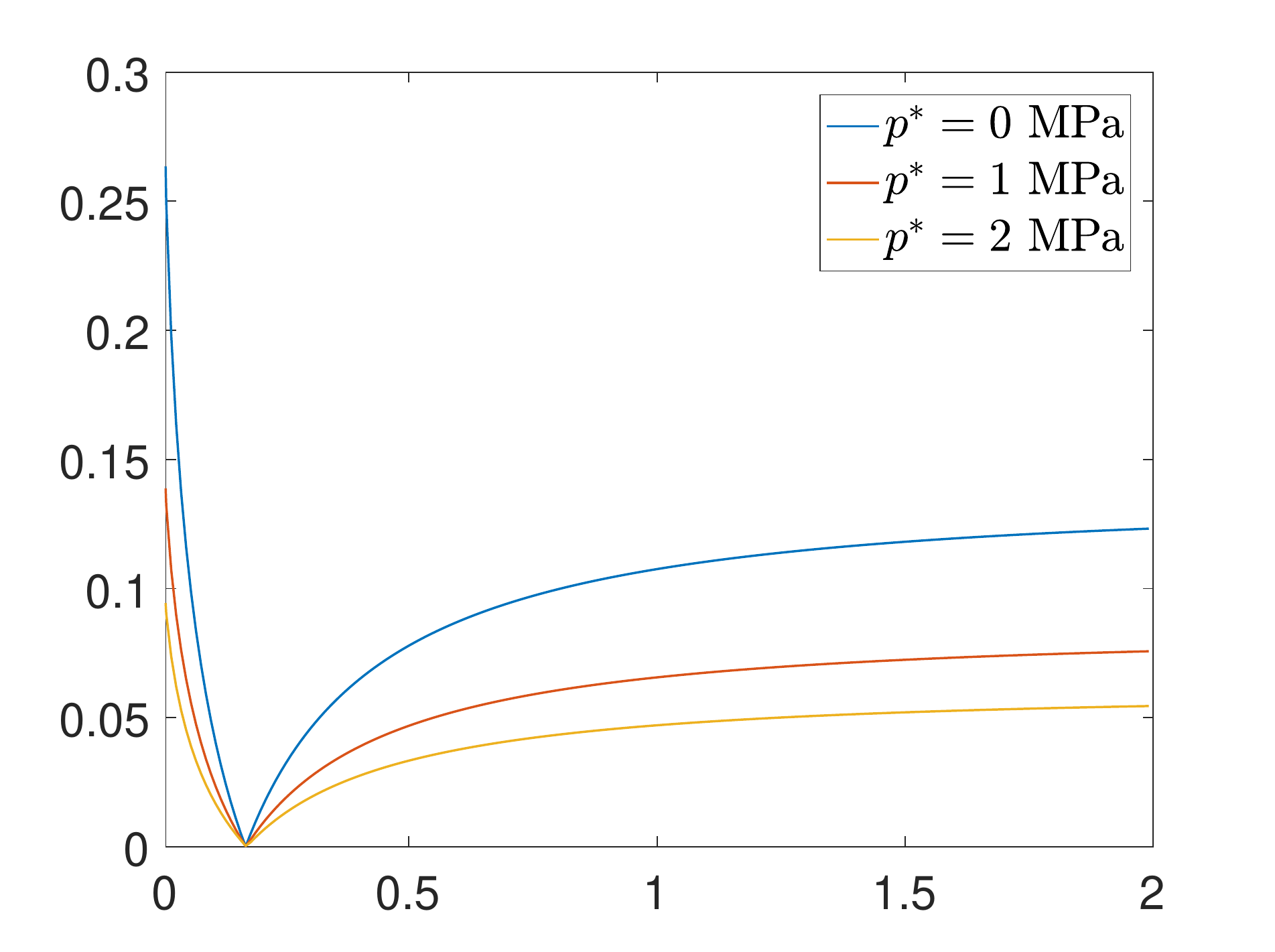}
\hspace{0mm}
\includegraphics[scale=0.40]{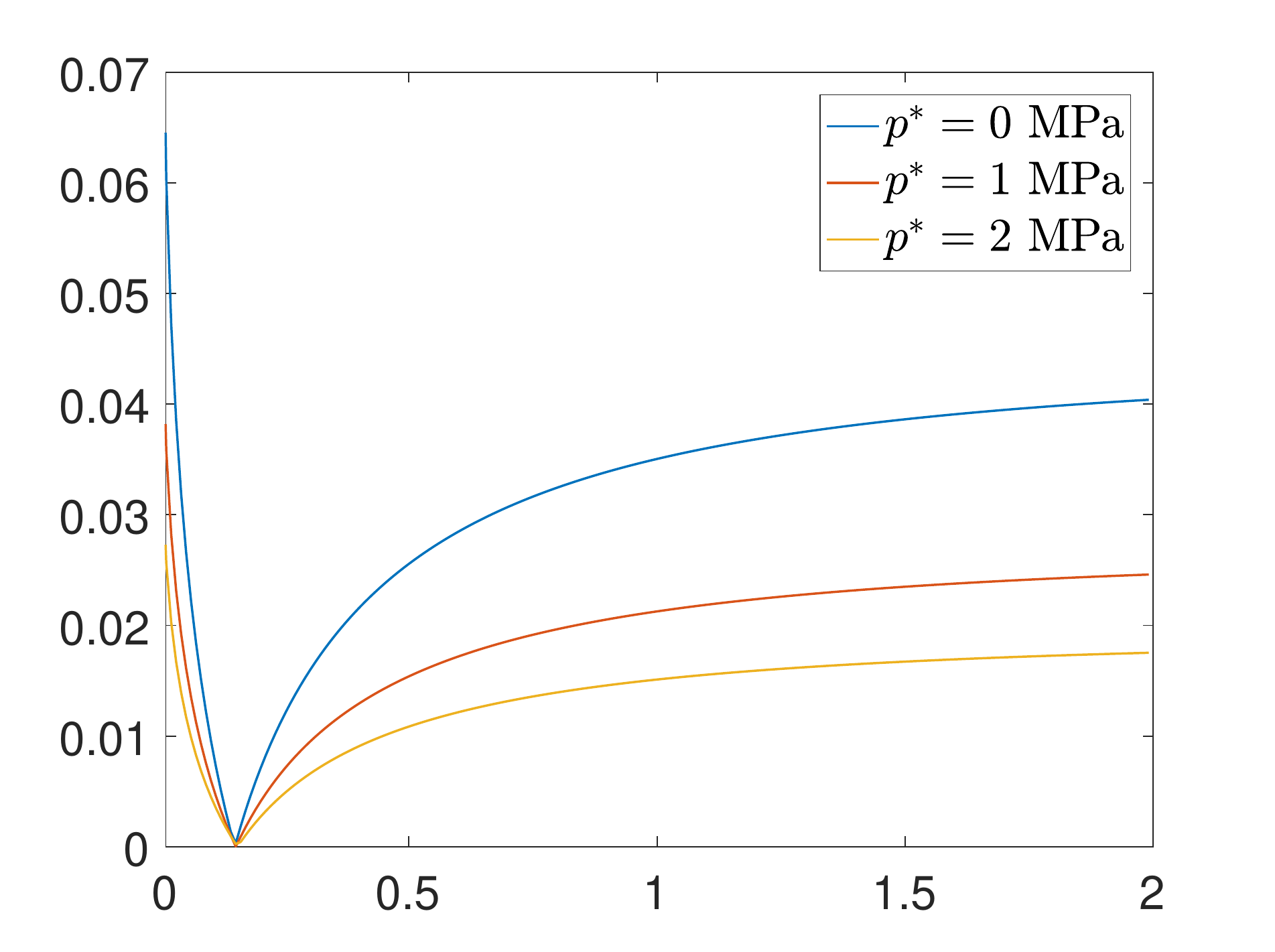}
\put(-335,-5){$r$}
\put(-110,-5){$r$}
\put(-450,155){$\textbf{a)}$}
\put(-225,155){$\textbf{b)}$}
\put(-460,85){$\delta \sigma_{yy}$}
\put(-230,85){$\delta \sigma_{yy}$}
\caption{The relative errors of approximation of $\sigma_{yy}$ for: a) the benchmark example \eqref{p_step_bench} -- \eqref{sigma_step_bench}, b) the benchmark example \eqref{p_bench} -- \eqref{Pi_def}. The value of $p^*$ [MPa] defines an additional constant pressure term employed in the respective variant of the problem.}
\label{delta_sigma_comparison}
\end{center}
\end{figure}
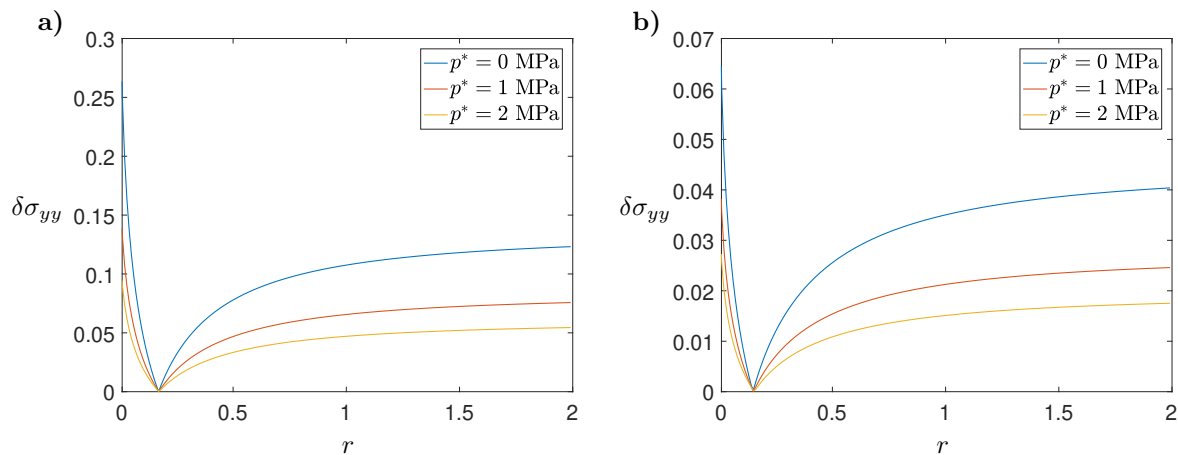

It was mentioned in Remark \ref{remark_sc} that increasing confining stress should contribute to improvement of quality of approximation by formula \eqref{Inglis_stress} due to the constant term of the absolute pressure being increased. In the next test we estimate the influence of the constant component of $p$ on the quality of stress approximation.  To this end we use the benchmark examples considered previously by adding a constant increment $p^*$ to each of them. Two different increment magnitudes are analyzed: $p^*=1$ MPa and $p^*=2$ MPa. The results in terms of the relative error of stress approximation, $\delta \sigma_{yy}$, are depicted in Figure \ref{delta_sigma_comparison} (the case of $p^*=0$ MPa refers directly to the results presented in Figure \ref{sigma_comparison}). As anticipated, the relative error of approximation decreases with growing  $p^*$ (note that with growing compressive confining stress the absolute value of fluid pressure needs to increase accordingly to sustain the crack opening). For the first benchmark, the average value of $\delta \sigma_{yy}$ drops from around 10$\%$ for $p^*=0$ MPa to 4$\%$ for $p^*=2$ MPa. In the case of the second benchmark the average error is reduces from little above 3$\%$ for $p^*=0$ MPa to 1$\%$ for $p^*=2$ MPa.

The above tests have proven that the assumed stress approximation for the elastic solution in the form of relation \eqref{Inglis_stress} is very satisfactory even in the case of non-uniform pressure distribution. In the next step we investigate quantitatively the stress relaxation model  introduced in Subsection \ref{stress_relaxation}. To this end we simulate with FEM a plane strain problem for a stationary elliptic fracture located in an infinite medium and loaded by a predefined constant internal fluid pressure, $p$. The resulting FEM solutions obtained for the elasto-plastic material are considered here as the reference solutions. The Mohr-Coulomb plasticity model is employed. The FEM results are compared with the newly introduced stress relaxation model and a simplified version based on the $K$ stress asymptote  \eqref{K_stress} (i. e. the original Irwin's model). In our analysis (and the figures' legends) we use the following nomenclature: i) the results obtained with the stress relaxation from Subsection \ref{stress_relaxation} will be called the `$\sigma$ relaxation model', ii) the results corresponding to the simplified $K$ stress asymptote variant will be named as the `$K$ relaxation model'. Note that the FEM computations do not involve here a complete HF problem but only the solid deformation sub-problem for predefined crack length and fluid pressure (thus only the module for computation of the crack opening of the main algorithm from Section \ref{algorithm} is employed here).

\begin{figure}[htb!]
\begin{center}
\includegraphics[scale=0.40]{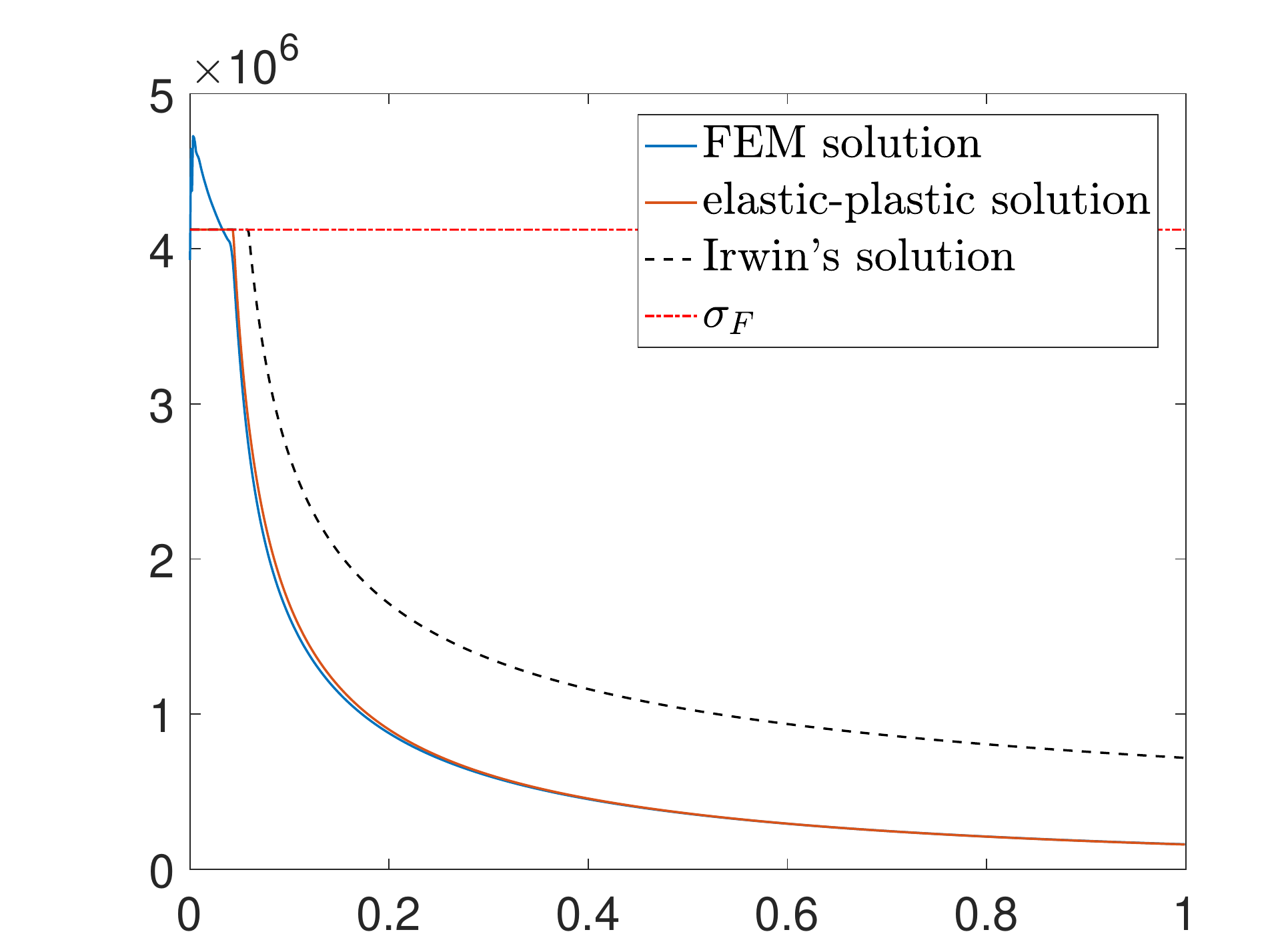}
\hspace{0mm}
\includegraphics[scale=0.40]{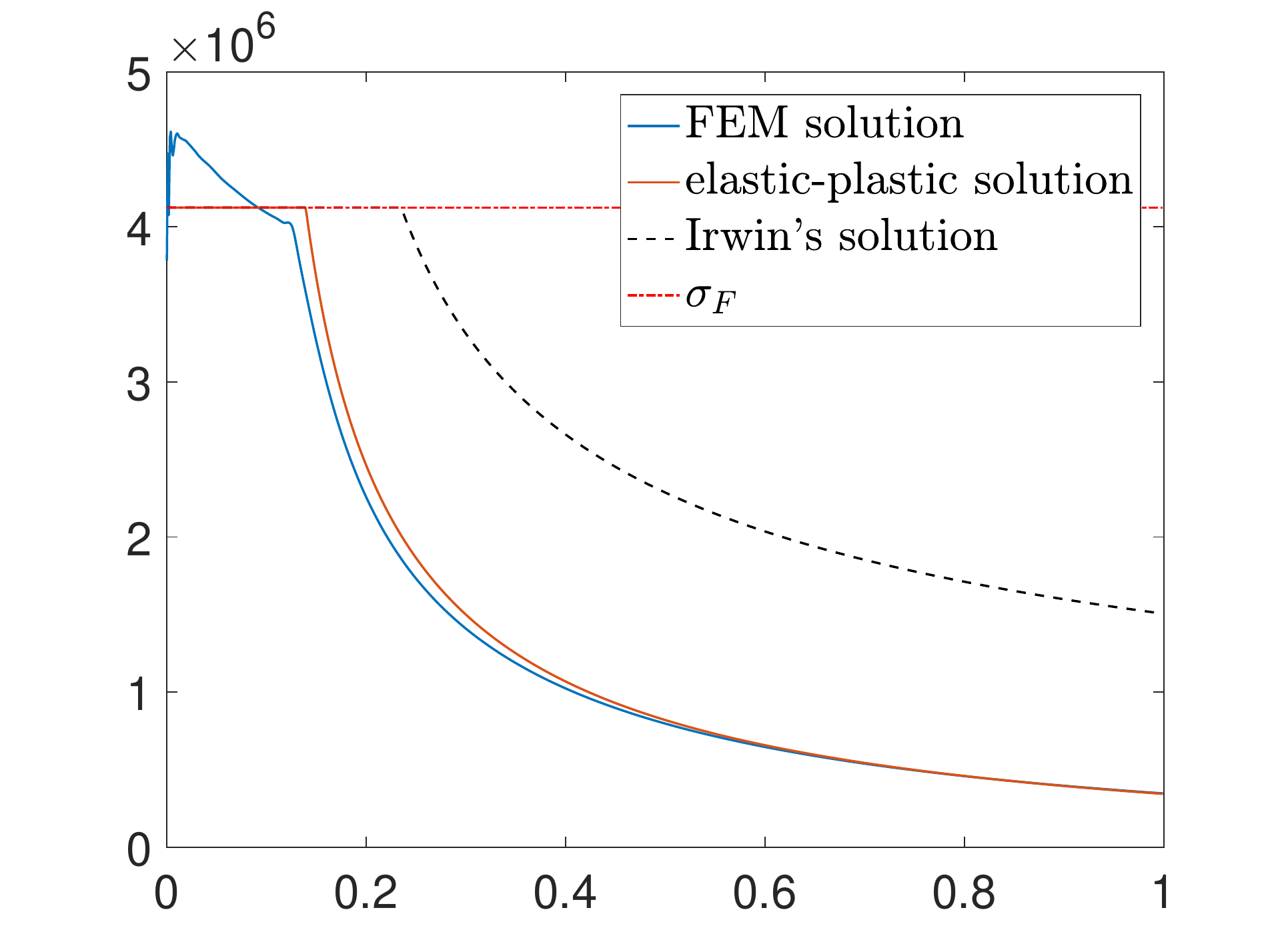}
\put(-335,-5){$r$}
\put(-110,-5){$r$}
\put(-450,155){$\textbf{a)}$}
\put(-225,155){$\textbf{b)}$}
\put(-450,85){$\sigma_{yy}$}
\put(-220,85){$\sigma_{yy}$}
\caption{Stress distribution ahead of the crack tip, $\sigma_{yy}$ [Pa], for: a) fluid pressure $p=10^6$ Pa, material cohesion $c=3 \cdot 10^6$ Pa, b) fluid pressure $p=2\cdot 10^6$ Pa, material cohesion $c=3 \cdot 10^6$ Pa. In both cases the assumed material friction angle is $\varphi=30^\circ$ and the crack half-length is $a=1$ m.}
\label{sigma_y_1}
\end{center}
\end{figure}

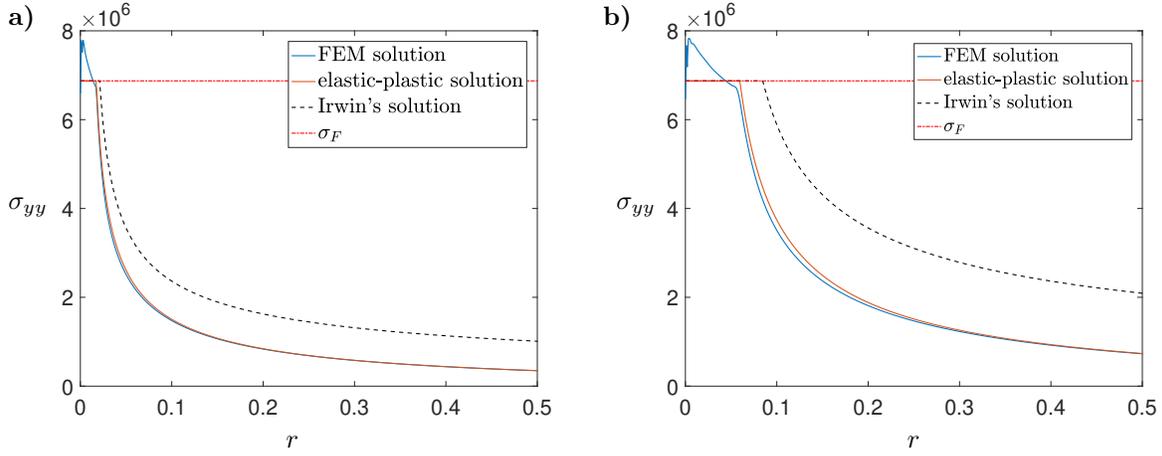
\begin{figure}[htb!]
\begin{center}
\includegraphics[scale=0.4]{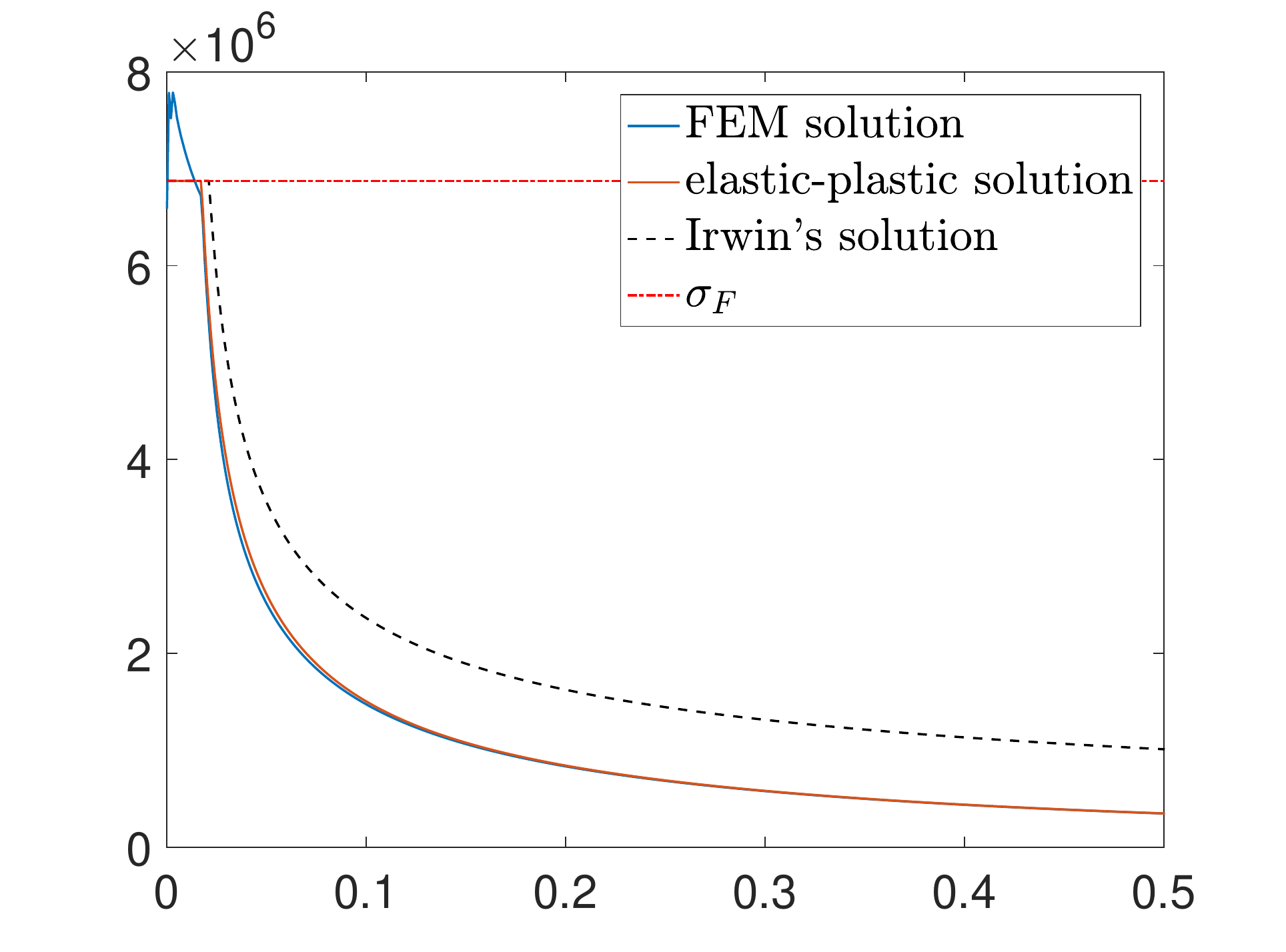}
\hspace{0mm}
\includegraphics[scale=0.4]{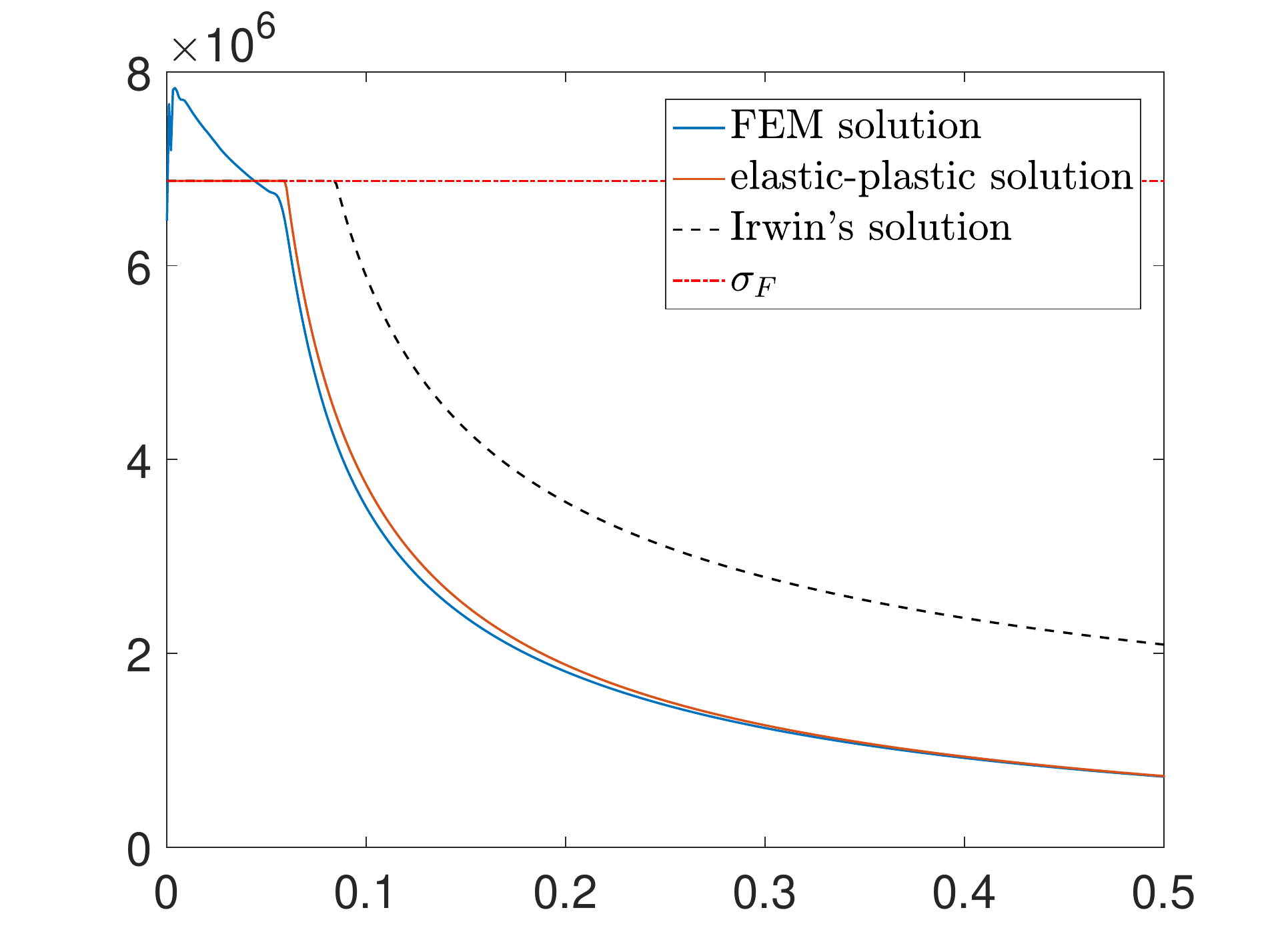}
\put(-345,-5){$r$}
\put(-110,-5){$r$}
\put(-450,155){$\textbf{a)}$}
\put(-225,155){$\textbf{b)}$}
\put(-450,85){$\sigma_{yy}$}
\put(-220,85){$\sigma_{yy}$}
\caption{Stress distribution ahead of the crack tip, $\sigma_{yy}$ [Pa], for: a) fluid pressure $p=10^6$ Pa, material cohesion $c=5 \cdot 10^6$ Pa, b) fluid pressure $p=2\cdot 10^6$ Pa, material cohesion $c=5 \cdot 10^6$ Pa. In both cases the assumed material friction angle is $\varphi=30^\circ$ and the crack half-length is $a=1$ m.}
\label{sigma_y_2}
\end{center}
\end{figure}

Simulations are performed for two different values of the fluid pressure, $p$: 1 MPa and 2 MPa. As for the elastic properties of material we assume: $E=14.5$ GPa and $\nu=0.24$. The plastic behaviour is analyzed for two variants of Mohr-Coulomb parameters: i) $c=3$ MPa, $\varphi=\psi=30^\circ$, and ii) $c=5$ MPa, $\varphi=\psi=30^\circ$, where $c$ denotes cohesion, $\varphi$ stands for the angle of friction and $\psi$ is the dilation angle. The confining stress is neglected for the time being so that the relative measure of the stress approximation error can be used ($\sigma_{yy}$ does not change its sign).

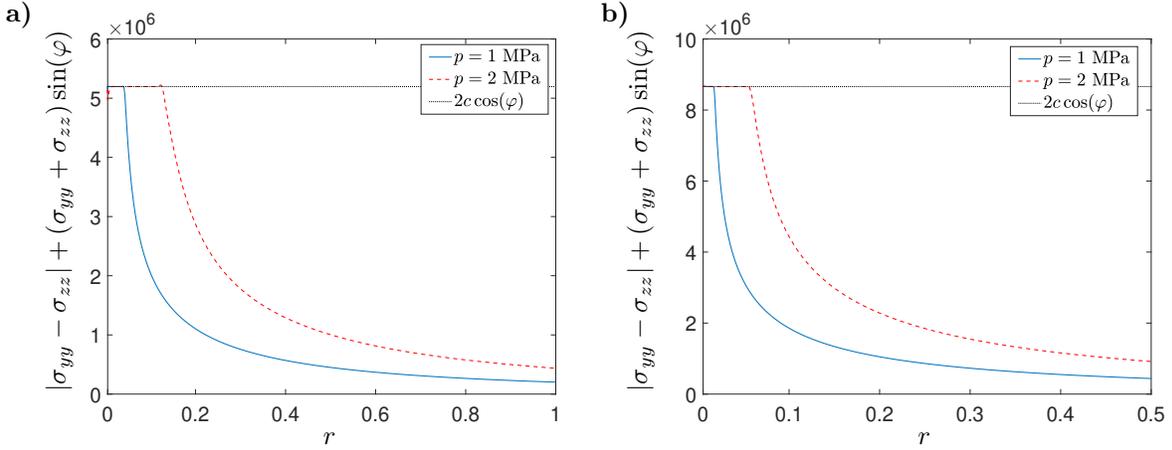
\begin{figure}[htb!]
\begin{center}
\includegraphics[scale=0.38]{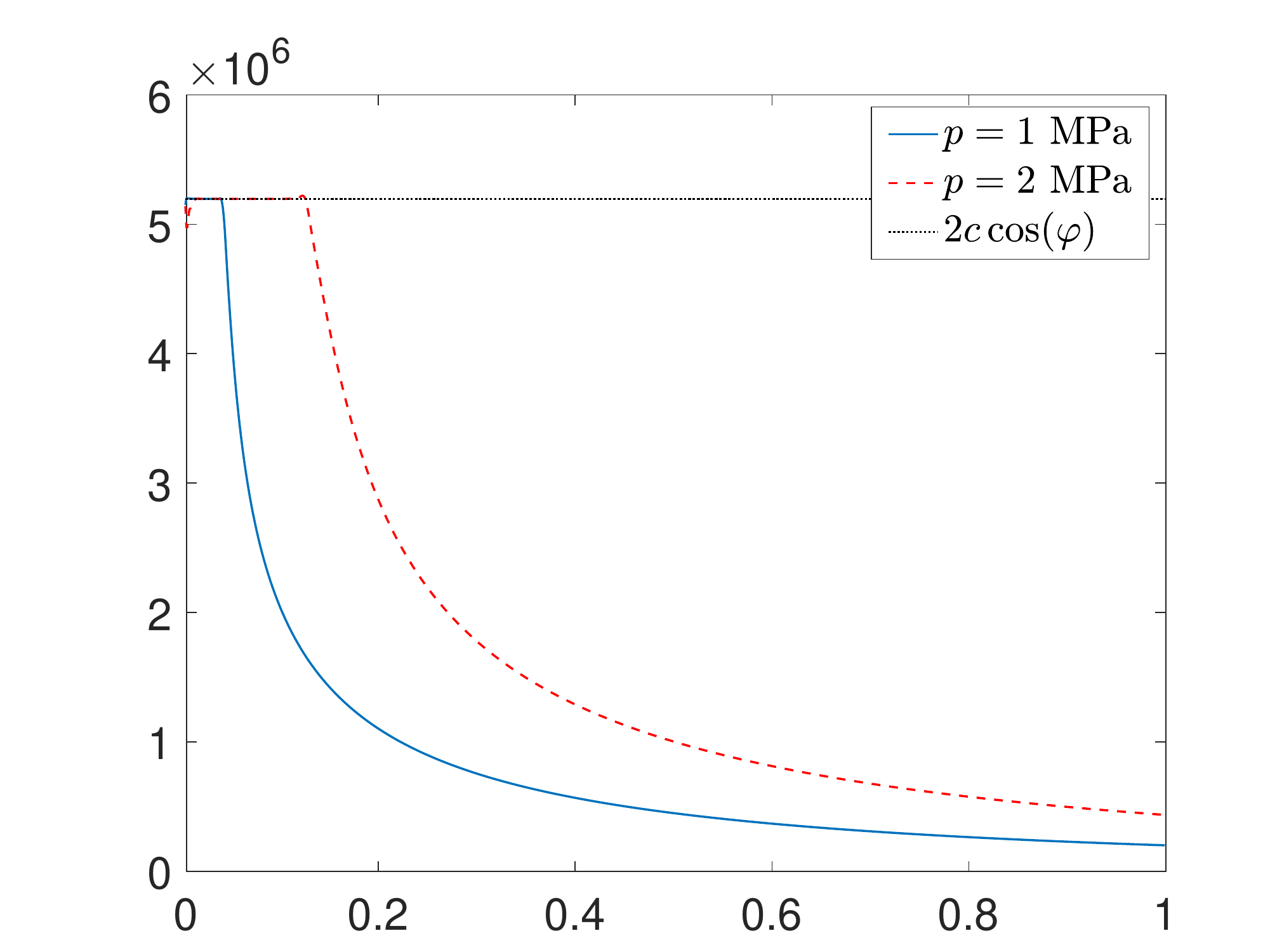}
\hspace{0mm}
\includegraphics[scale=0.38]{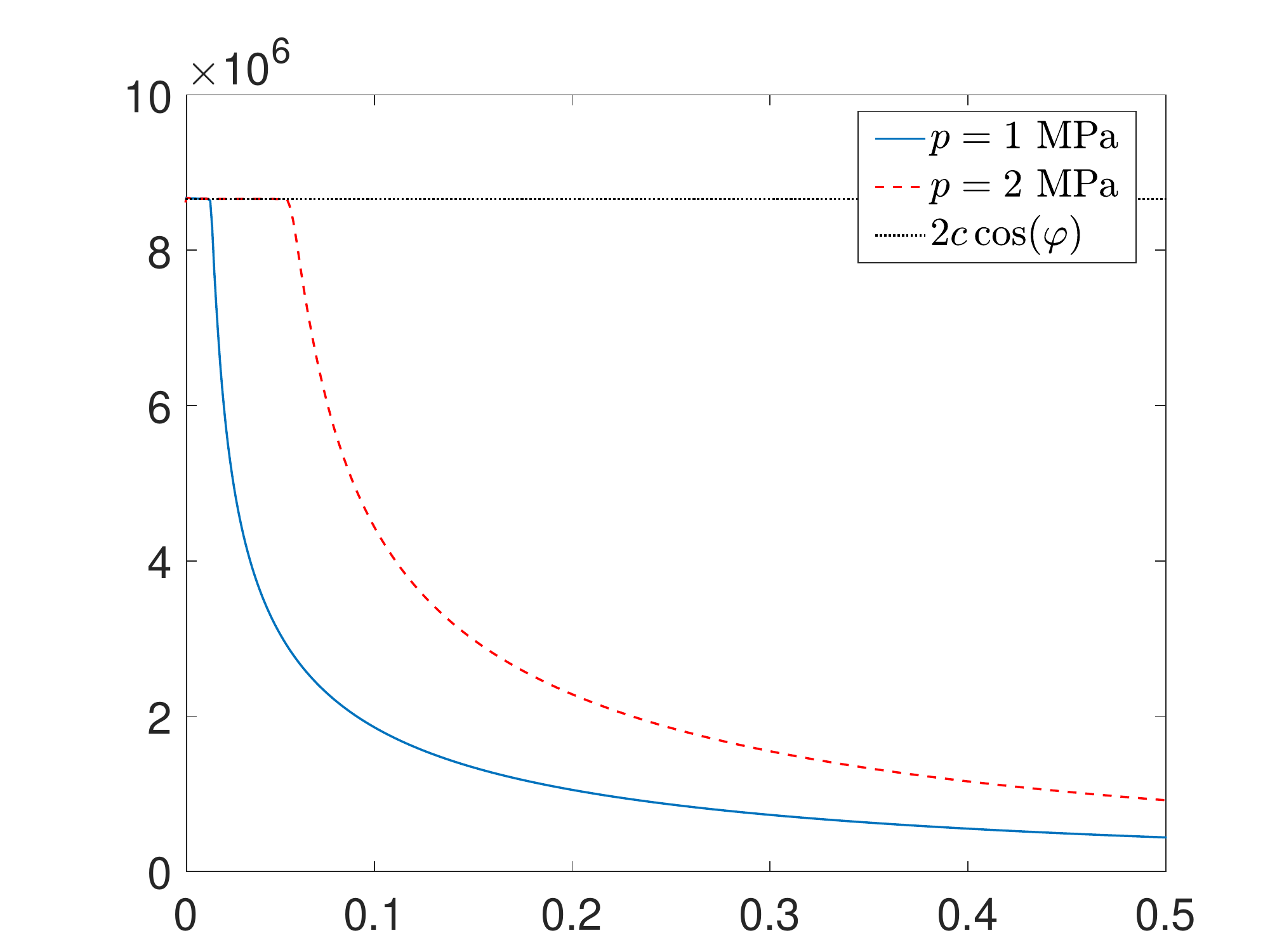}
\put(-330,-5){$r$}
\put(-105,-5){$r$}
\put(-450,155){$\textbf{a)}$}
\put(-225,155){$\textbf{b)}$}
\put(-435,15){\rotatebox{90}{$|\sigma_{yy}-\sigma_{zz}|+(\sigma_{yy}+\sigma_{zz})\sin(\varphi)$}}
\put(-215,15){\rotatebox{90}{$|\sigma_{yy}-\sigma_{zz}|+(\sigma_{yy}+\sigma_{zz})\sin(\varphi)$}}
\caption{Fulfillment of the Mohr-Coulomb yield criterion by the numerical FEM solution for: a) material cohesion $c=3 \cdot 10^6$ Pa - this variant corresponds to the data from Figure \ref{sigma_y_1}, b) material cohesion $c=5 \cdot 10^6$ Pa - this variant corresponds to the data from Figure \ref{sigma_y_2}. In both cases the assumed material friction angle is $\varphi=30^\circ$ and the crack half-length is $a=1$ m.}
\label{FEM_yield}
\end{center}
\end{figure}

The values of $\sigma_{yy}$ are depicted in Figures \ref{sigma_y_1} -- \ref{sigma_y_2} for: i) numerical FEM  solution, ii) the $\sigma$ relaxation model, $\sigma_{yy}^\text{(pl)}$, computed according to formula \eqref{TP_def1}, iii) the $K$ relaxation model. It shows that the $\sigma$ relaxation model provides  good approximation of the FEM solution even in the case in which  the size of plastic deformation zone amounts to over 10 $\%$ of the crack half-length ($p=2$ MPa, $c=3$ MPa - Fig. \ref{sigma_y_1}b)). It can be also seen that both, the level of yield stress and the extent of the plastic deformation zone, are in surprisingly good agreement with those produced by the FEM model. It is only inside the plastic deformation zone that respective results differ noticeably as the  FEM solution does not return a constant value of stress, even though it satisfies the Mohr-Coulomb yield condition.  The level to which the FEM solution fulfills the yield criterion, defined under the analyzed conditions as:
\[
|\sigma_{yy}-\sigma_{zz}|+(\sigma_{yy}+\sigma_{zz})\sin(\varphi)=2c\cos(\varphi),
\]
is depicted in Figure \ref{FEM_yield}. 
On the other hand, the $K$ relaxation model (i. e. classical Irwin's stress relaxation model) overestimates the size of the plastic zone and more importantly, it does not provide good approximation of the stress outside the plastic zone in any case. Clearly, a simple $K$ asymptote approximation is not sufficient for the assumed loading and material parameters. The values of the size of the plastic deformation zone, $d_\text{p}$, and the fictitious extension of the crack, $\eta$, are collated in Table \ref{tabela_d_eta}.

\begin{figure}[htb!]
\begin{center}
\includegraphics[scale=0.4]{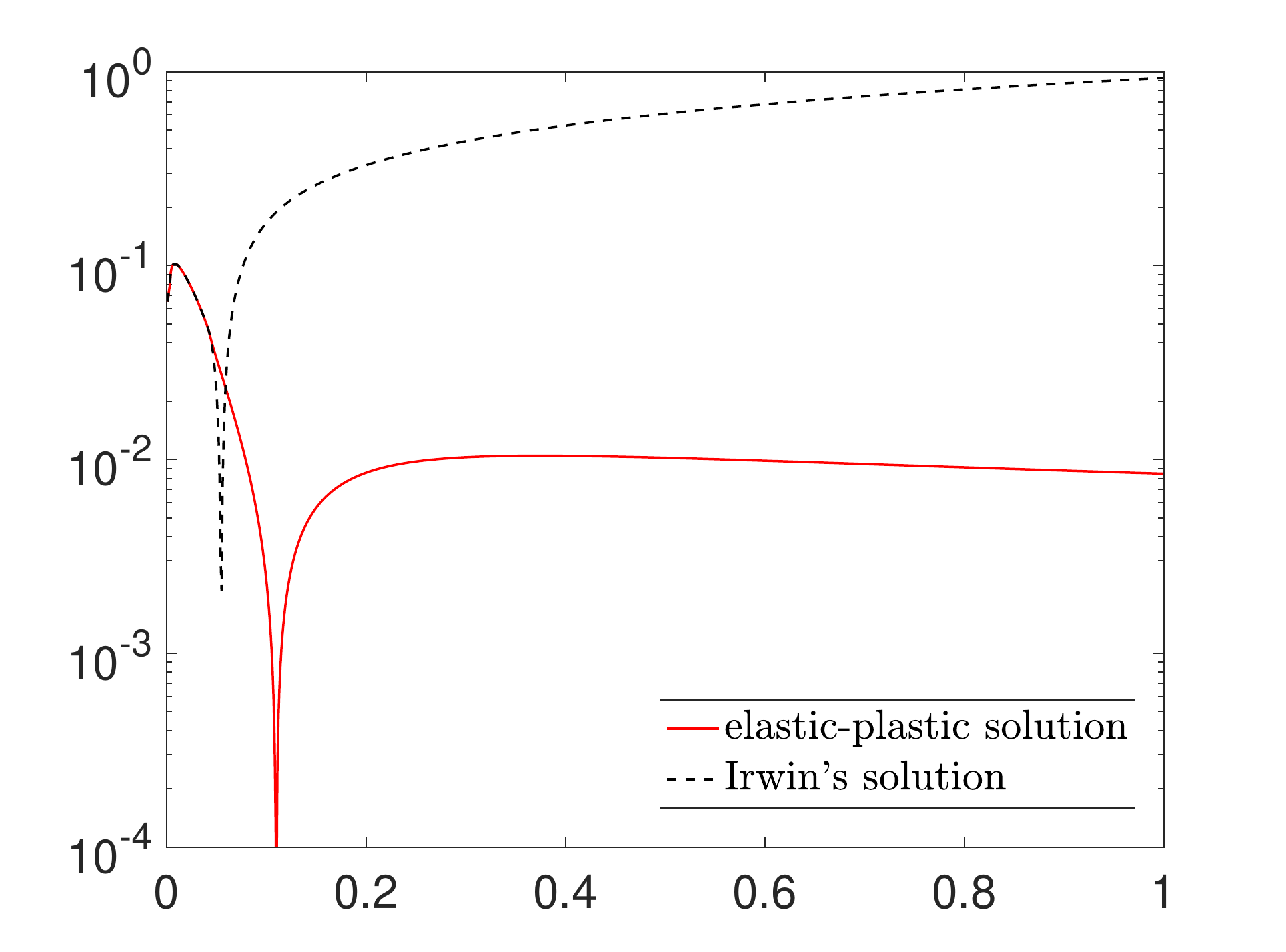}
\hspace{0mm}
\includegraphics[scale=0.4]{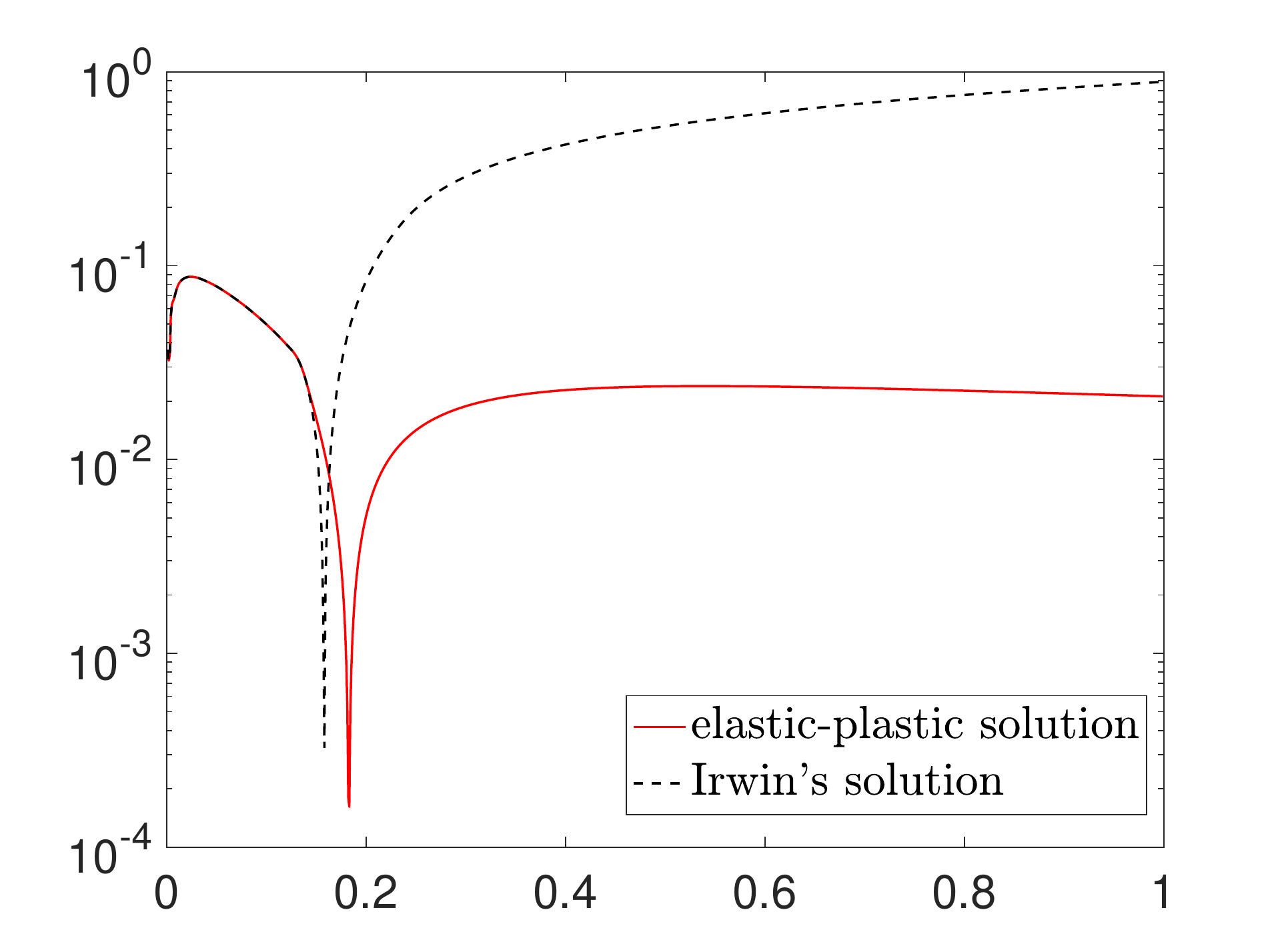}
\put(-345,-5){$r$}
\put(-110,-5){$r$}
\put(-450,155){$\textbf{a)}$}
\put(-225,155){$\textbf{b)}$}
\put(-455,85){$\delta S$}
\put(-225,85){$\delta S$}
\caption{The relative error of stress equivalence in the plane of fracture extension for: a) fluid pressure $p=10^6$ Pa, material cohesion $c=3 \cdot 10^6$ Pa, b) fluid pressure $p=2\cdot 10^6$ Pa, material cohesion $c=3 \cdot 10^6$ Pa. In both cases the assumed material friction angle is $\varphi=30^\circ$ and the crack half-length is $a=1$ m.}
\label{delta_S_1}
\end{center}
\end{figure}

\begin{figure}[htb!]
\begin{center}
\includegraphics[scale=0.4]{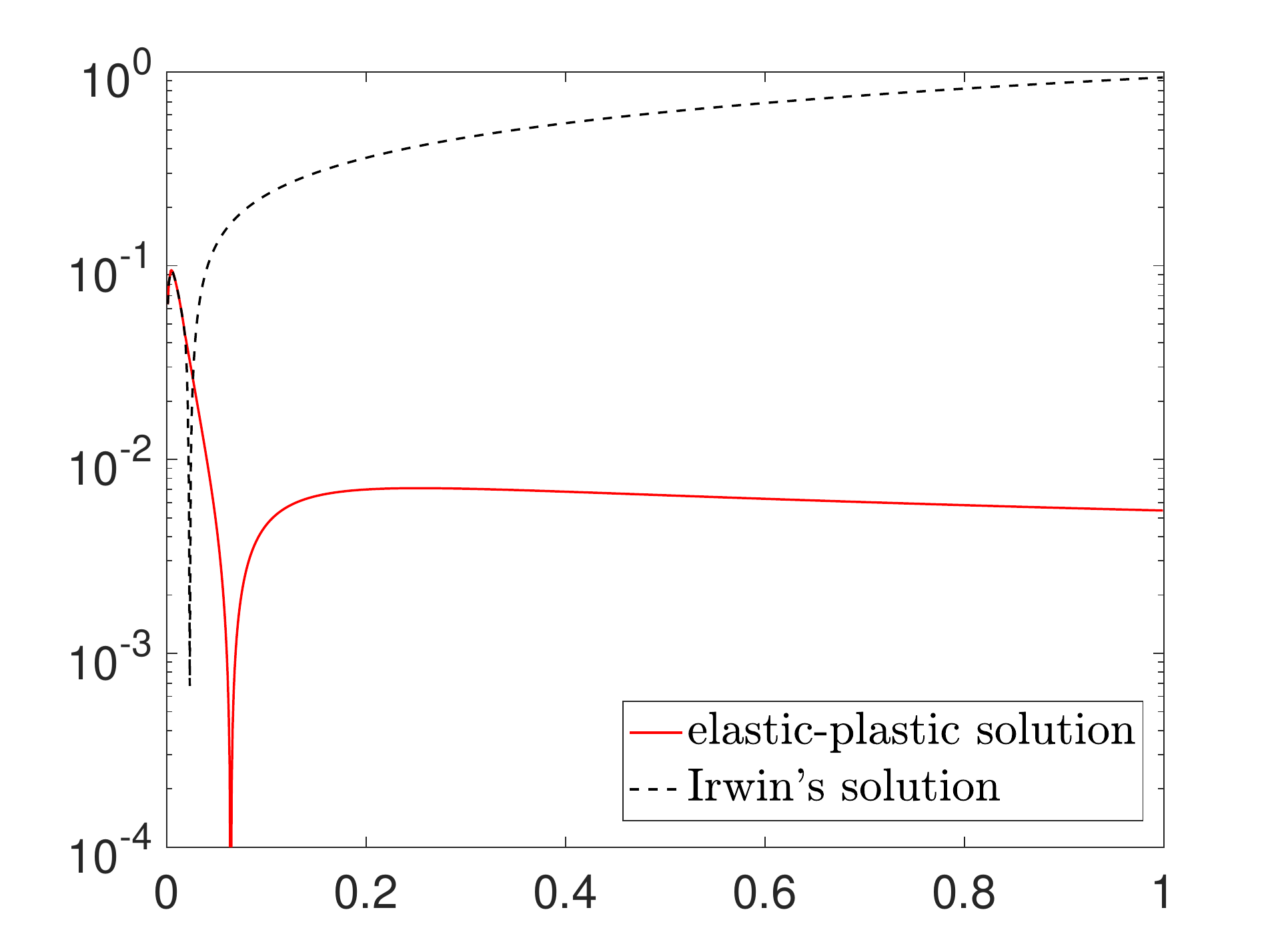}
\hspace{0mm}
\includegraphics[scale=0.4]{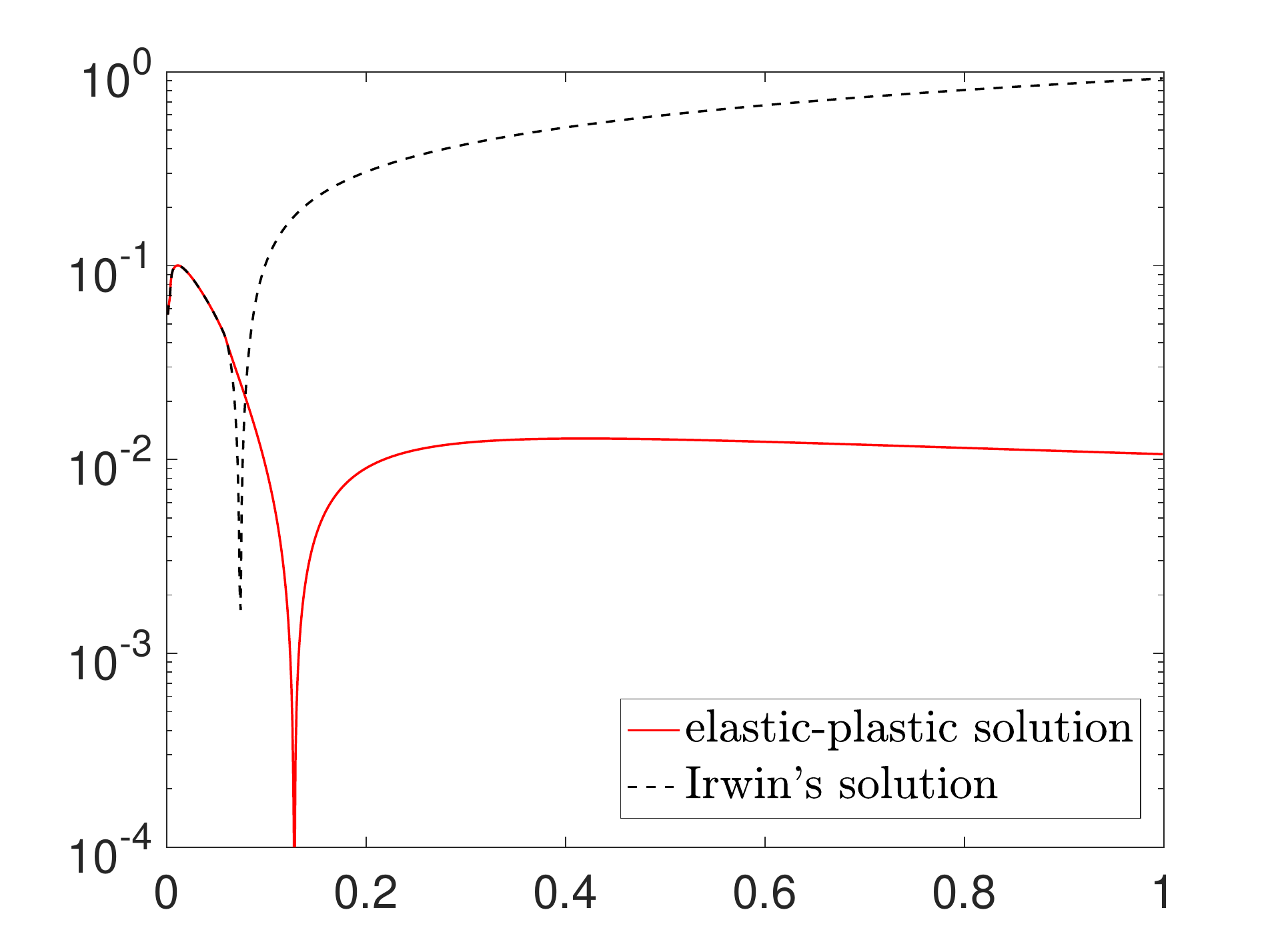}
\put(-345,-5){$r$}
\put(-110,-5){$r$}
\put(-450,155){$\textbf{a)}$}
\put(-225,155){$\textbf{b)}$}
\put(-455,85){$\delta S$}
\put(-225,85){$\delta S$}
\caption{The relative error of stress equivalence  in the plane of fracture extension for: a) fluid pressure $p=10^6$ Pa, material cohesion $c=5 \cdot 10^6$ Pa, b) fluid pressure $p=2\cdot 10^6$ Pa, material cohesion $c=5 \cdot 10^6$ Pa. In both cases the assumed material friction angle is $\varphi=30^\circ$ and the crack half-length is $a=1$ m.}
\label{delta_S_2}
\end{center}
\end{figure}
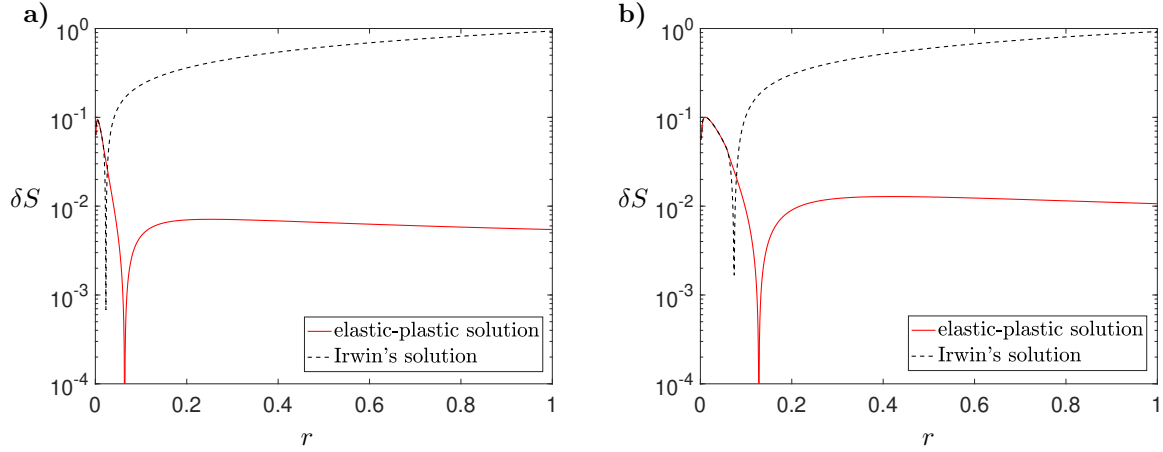

Let us now investigate to what degree the stress distribution in the plane of fracture extension computed according to the elastic-plastic solution $\sigma_{yy}^\text{(pl)}$ can be considered in a non-local sense as an equivalent  for the stress obtained with the FEM solution. In this way one can verify whether the external loads carried over the plane of crack propagation are comparable to each other in both cases. To this end we introduce the following dimensionless parameter:
\begin{equation}
\label{delta_S}
\delta S(r)=\Big| \frac{\int_a^r \sigma_{yy}^\text{(FEM)}dr-\int_a^r \sigma_{yy}dr}{\int_a^r \sigma_{yy}^\text{(FEM)}dr}\Big|,
\end{equation}
where $\sigma_{yy}^\text{(FEM)}$ is the stress distribution computed by the FEM, while  $\sigma_{yy}$ refers  either to the $\sigma$ relaxation model or the $K$ relaxation model. Thus, $\delta S$ constitutes a relative measure of stress equivalence (with respect to the FEM solution)  in the plane of fracture extension. The graphs of $\delta S(r)$ for the respective values of load and material parameters are shown in Figures \ref{delta_S_1}--\ref{delta_S_2}.

As can be seen in the figures, even though the elastic-plastic solution based on the $\sigma$ relaxation model diverges locally from the FEM results in the plastic deformation zone, it still provides very similar results in terms of the overall load carried across the zone itself and the whole plane of  fracture propagation. Supplementary data given in Table \ref{tabela_delta_S} shows that there is less than 5$\%$ difference between the respective stresses integrated over the entire radius of the plastic deformation zone, and this difference declines swiftly as the upper integration limit moves towards larger values of $r$. This trend holds for all the considered variants of load and material parameters. On the other hand, when using the $K$ relaxation model one can observe an essential divergence from the FEM results with the relative difference growing bigger with increasing $r$. In  particular, when analyzing the integrals over $r \in [0,a]$ we see that this model returns around 90$\%$ higher values of the overall load than the FEM results in all the considered cases.

\begin{table}[]
\begin{center}
\begin{tabular}{c|c|c|c|c|}
\cline{2-5}
\multicolumn{1}{l|}{\multirow{2}{*}{}} & \multicolumn{2}{c|}{$p=1$ MPa}                                & \multicolumn{2}{c|}{$p=2$ MPa}                                \\ \cline{2-5} 
\multicolumn{1}{l|}{}                  & $c=3$ MPa                       & $c=5$ MPa                       &$c=3$ MPa                       &$c=5$ MPa                       \\ \hline \hline
\multicolumn{1}{|c||}{$\sigma_\text{F}$}       & \multicolumn{1}{l|}{4.12 MPa} & \multicolumn{1}{l|}{6.87 MPa} & \multicolumn{1}{l|}{4.12 MPa} & \multicolumn{1}{l|}{6.87 MPa} \\ \hline
\multicolumn{5}{|c|}{$\sigma$ relaxation model}                                                                                                                 \\ \hline
\multicolumn{1}{|c||}{$\eta$}           & 0.0238  m                   & 0.0092      m               & 0.0813      m              & 0.0331      m                \\ \hline
\multicolumn{1}{|c||}{$d_\text{p}$}             & 0.0435     m                & 0.0174      m               & 0.1393        m              & 0.0595       m               \\ \hline
\multicolumn{5}{|c|}{$K$ relaxation model}                                                                                                                         \\ \hline
\multicolumn{1}{|c||}{$\eta$}           & 0.0294     m                 & 0.0106       m               & 0.1176      m                & 0.0424      m                \\ \hline
\multicolumn{1}{|c||}{$d_\text{p}$}             & 0.0588    m                  & 0.0212         m             & 0.2352   m                   & 0.0848       m               \\ \hline
\end{tabular}
\caption{The crack length extension values, $\eta$, and the sizes of the plastic deformation zone, $d_p$, for the case of uniform pressure distribution.}
\label{tabela_d_eta}
\end{center}
\end{table}

\begin{table}[]
\begin{center}
\begin{tabular}{c|c|c|c|c|}
\cline{2-5}
                                & \multicolumn{2}{c|}{$p=1$ MPa} & \multicolumn{2}{c|}{$p=2$  MPa} \\ \cline{2-5} 
                                & $c=3$ MPa         & $c=5$ MPa        & $c=3$ MPa        &$ c=5$ MPa         \\ \hline \hline
\multicolumn{1}{|c||}{$r$ {[}m{]}} & \multicolumn{4}{c|}{$\delta S(r)$}                             \\ \hline 
\multicolumn{5}{|c|}{$\sigma$ relaxation model}                                                \\ \hline
\multicolumn{1}{|c||}{$d_p$}      & 0.0436        & 0.0476       & 0.0258       & 0.0414        \\ \hline
\multicolumn{1}{|c||}{0.5}       & 0.0102        & 0.0065       & 0.0239       & 0.0013        \\ \hline
\multicolumn{1}{|c||}{1}         & 0.0084        & 0.0055       & 0.0212       & 0.0011       \\ \hline
\multicolumn{5}{|c|}{$K$ relaxation model}                                                        \\ \hline
\multicolumn{1}{|c||}{$d_p$}      & 0.0598        & 0.0183       & 0.1640       & 0.0463        \\ \hline
\multicolumn{1}{|c||}{0.5}       & 0.6070        & 0.6193       & 0.5239       & 0.5979        \\ \hline
\multicolumn{1}{|c||}{1}         & 0.9290        & 0.9354       & 0.8873       & 0.9241        \\ \hline
\end{tabular}
\caption{The relative measure of stress equivalence  in the plane of fracture extension, $\delta S$, for the case of uniform pressure distribution. The values of $d_p$ for respective variants correspond to the figures from Table \ref{tabela_d_eta}.}
\label{tabela_delta_S}
\end{center}
\end{table}

\begin{table}[]
\begin{center}
\begin{tabular}{c|c|c|}
\cline{2-3}
                                        & $c=3$ MPa & $c=5$ MPa \\ \hline \hline
\multicolumn{1}{|c||}{$\sigma_\text{F}$} & 4.12 MPa  & 6.87 MPa  \\ \hline 
\multicolumn{3}{|c|}{$\sigma$ relaxation model}                 \\ \hline
\multicolumn{1}{|c||}{$\eta$}            & 0.0509 m  & 0.0203 m  \\ \hline
\multicolumn{1}{|c||}{$d_\text{p}$}      & 0.0895 m  & 0.0373 m  \\ \hline
\multicolumn{3}{|c|}{$K$ relaxation model}                      \\ \hline
\multicolumn{1}{|c||}{$\eta$}            & 0.0546 m  & 0.0197 m  \\ \hline
\multicolumn{1}{|c||}{$d_\text{p}$}      & 0.1092 m  & 0.0393 m  \\ \hline
\end{tabular}
\caption{The crack length extension values, $\eta$, and the sizes of the plastic deformation zone, $d_p$, for the case of non-uniform pressure distribution (fluid pressure is described by relation \eqref{p_bench}).}
\label{d_p_eta_HF_p}
\end{center}
\end{table}

\begin{table}[]
\begin{center}
\begin{tabular}{c|c|c|}
\cline{2-3}
                                   & $c=3$ MPa        & $c=5$ MPa       \\ \hline \hline
\multicolumn{1}{|c||}{$r$ {[}m{]}}  & \multicolumn{2}{c|}{$\delta S(r)$} \\ \hline
\multicolumn{3}{|c|}{$\sigma$ relaxation model}                         \\ \hline
\multicolumn{1}{|c||}{$d_\text{p}$} & 0.0312           & 0.0390          \\ \hline
\multicolumn{1}{|c||}{0.5}          & 0.0357           & 0.0310          \\ \hline
\multicolumn{1}{|c||}{1}            & 0.0269           & 0.0214          \\ \hline
\multicolumn{3}{|c|}{$K$ relaxation model}                              \\ \hline
\multicolumn{1}{|c||}{$d_\text{p}$} & 0.0214           & 0.0284          \\ \hline
\multicolumn{1}{|c||}{0.5}          & 0.4558           & 0.4767          \\ \hline
\multicolumn{1}{|c||}{1}            & 0.7442           & 0.7524          \\ \hline
\end{tabular}
\caption{The relative measure of stress equivalence  in the plane of fracture extension, $\delta S$, for the case of non-uniform pressure distribution  (fluid pressure is described by relation \eqref{p_bench}). The values of $d_p$ for respective variants correspond to the figures from Table \ref{d_p_eta_HF_p}.}
\label{tabela_delta_S_p_HF}
\end{center}
\end{table}

\begin{figure}[htb!]
\begin{center}
\includegraphics[scale=0.4]{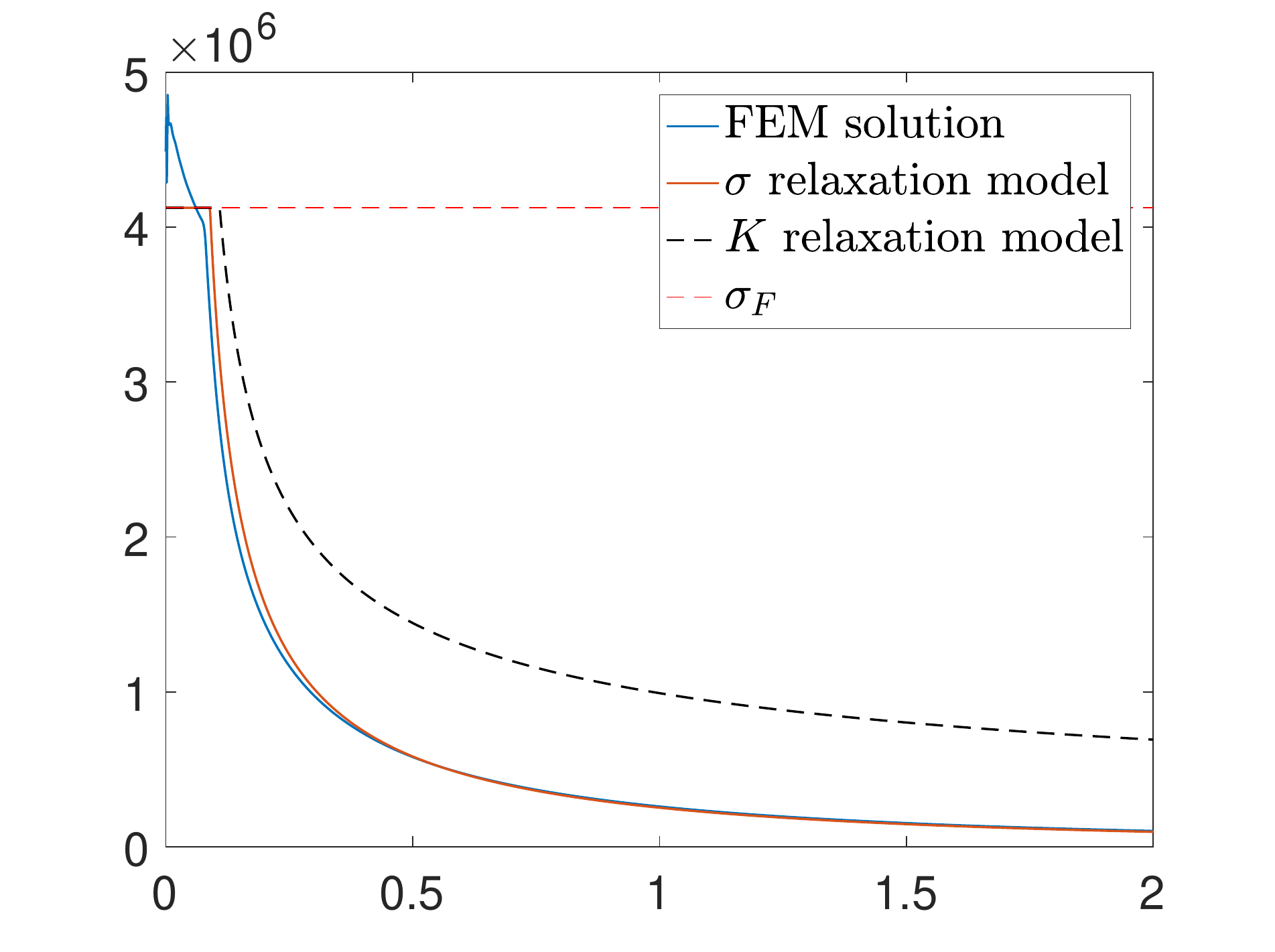}
\hspace{0mm}
\includegraphics[scale=0.4]{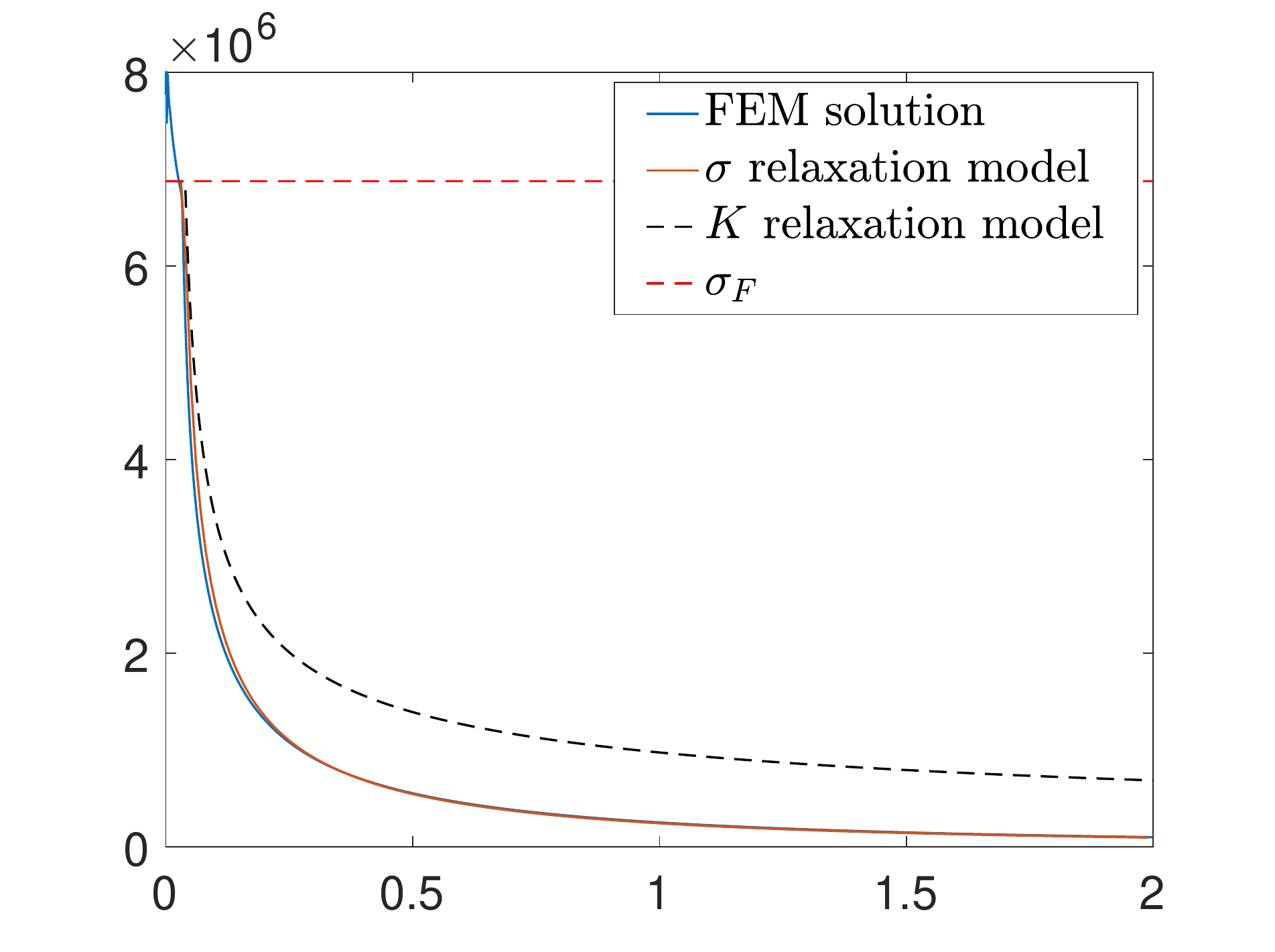}
\put(-345,-5){$r$}
\put(-110,-5){$r$}
\put(-450,155){$\textbf{a)}$}
\put(-225,155){$\textbf{b)}$}
\put(-450,85){$\sigma_{yy}$}
\put(-220,85){$\sigma_{yy}$}
\caption{Stress distribution ahead of the crack tip, $\sigma_{yy}$ [Pa], under the fluid pressure described by relation \eqref{p_bench} for: a) material cohesion $c=3 \cdot 10^6$ Pa, b) material cohesion $c=5 \cdot 10^6$ Pa. In both cases the assumed material friction angle is $\varphi=30^\circ$ and the crack half-length is $a=1$ m.}
\label{sigma_HF_zero_conf}
\end{center}
\end{figure}

\begin{figure}[htb!]
\begin{center}
\includegraphics[scale=0.4]{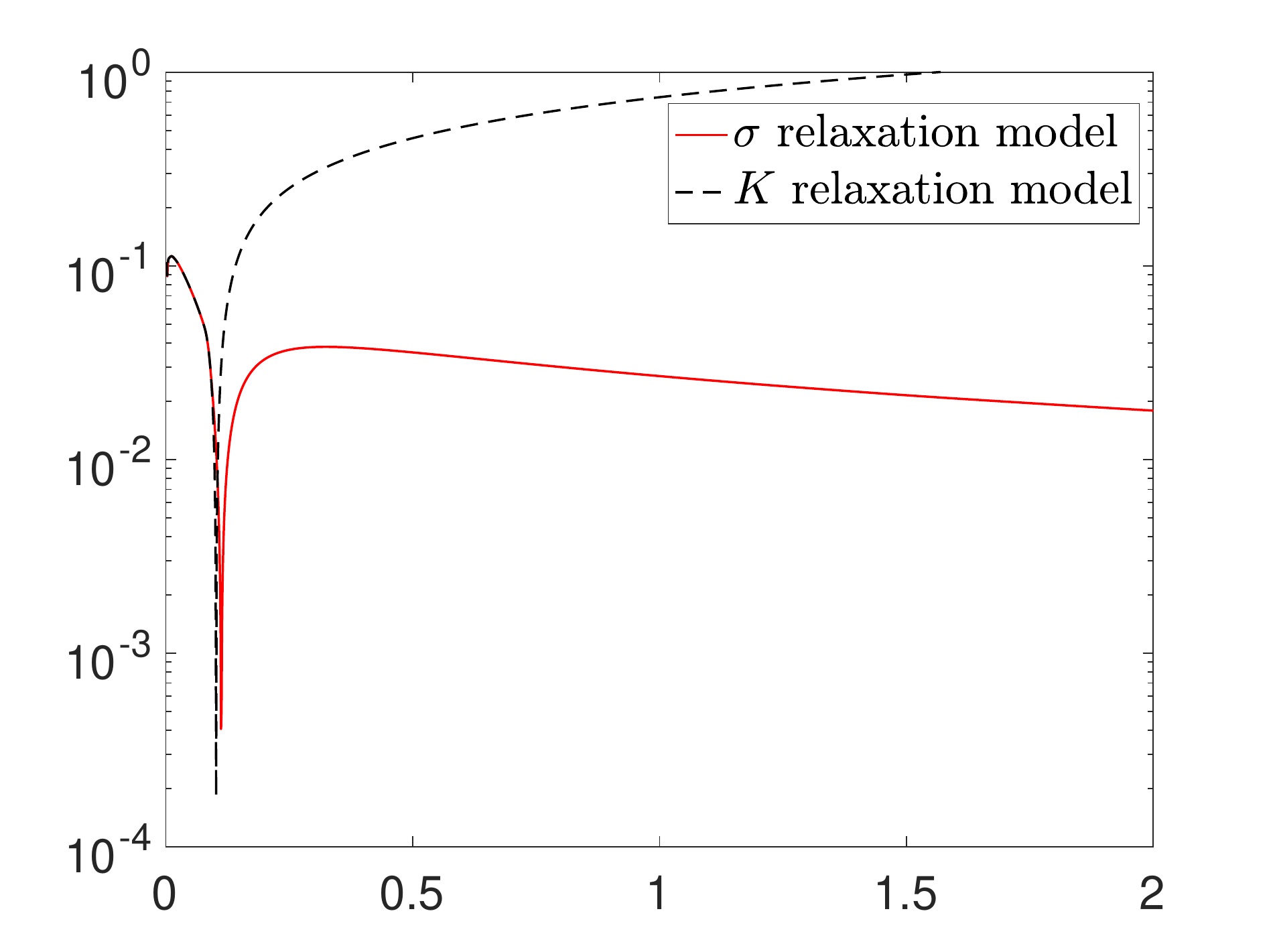}
\hspace{0mm}
\includegraphics[scale=0.4]{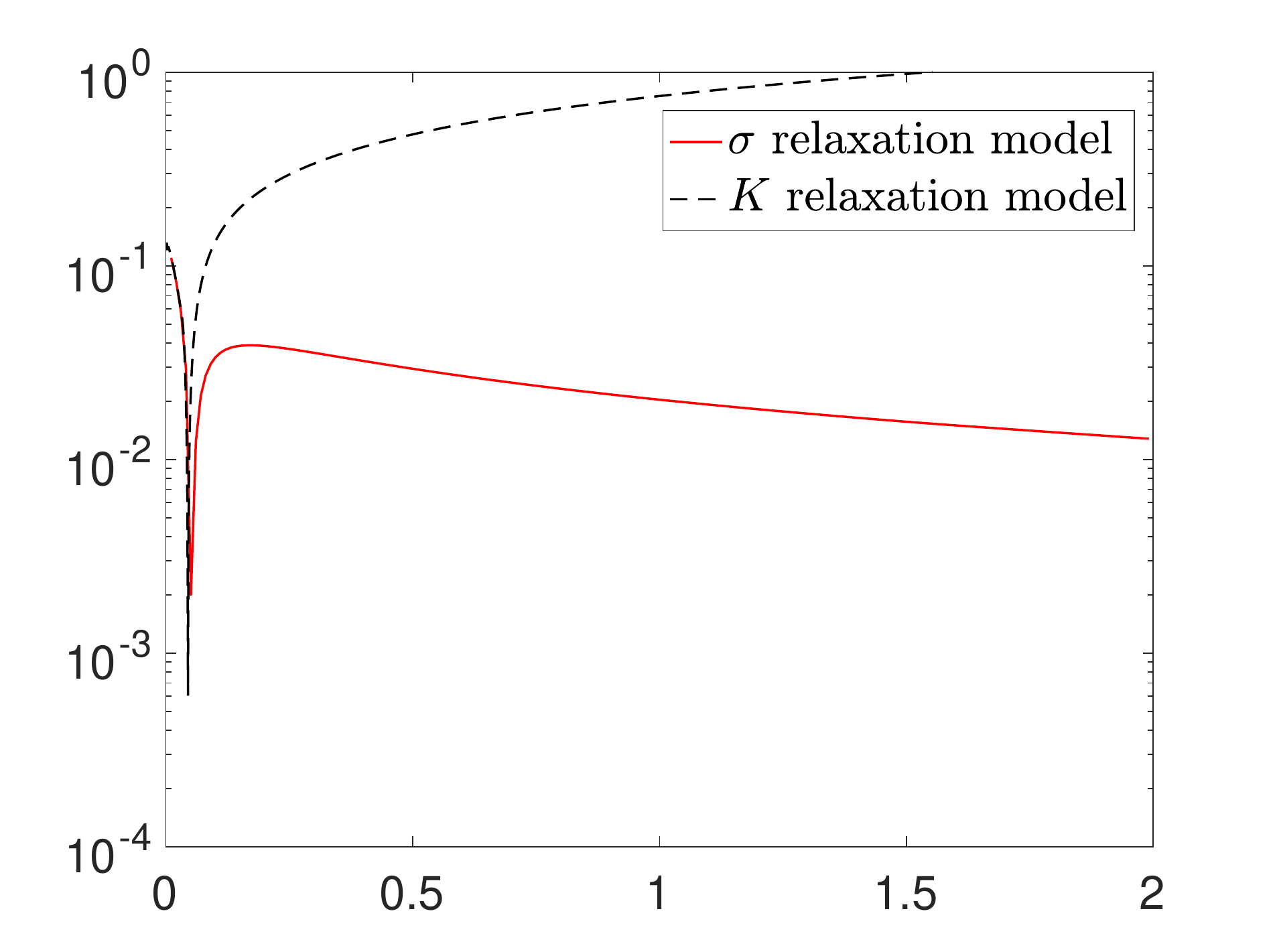}
\put(-345,-5){$r$}
\put(-110,-5){$r$}
\put(-450,155){$\textbf{a)}$}
\put(-225,155){$\textbf{b)}$}
\put(-455,85){$\delta S$}
\put(-225,85){$\delta S$}
\caption{The relative error of stress equivalence  in the plane of fracture extension under the fluid pressure described by relation \eqref{p_bench} for: a) material cohesion $c=3 \cdot 10^6$ Pa, b) material cohesion $c=5 \cdot 10^6$ Pa. In both cases the assumed material friction angle is $\varphi=30^\circ$ and the crack half-length is $a=1$ m.}
\label{delta_S_HF_zero_conf}
\end{center}
\end{figure}

As the above results were obtained for a constant fluid pressure let us check now if the observed trends remain similar when non-uniform pressure is applied. To this end we employ $p$ distribution described by benchmark \eqref{p_bench}--\eqref{Pi_def}. The material constants are the same as in the previous example. Respective stress distributions are depicted in Figure \ref{sigma_HF_zero_conf} while the relative error of stress equivalence is shown in Figure \ref{delta_S_HF_zero_conf}. The corresponding tabular data is collated in Table \ref{d_p_eta_HF_p} ($\eta$ and $d_\text{p}$) and Table \ref{tabela_delta_S_p_HF} ($\delta S(r)$).

The presented data confirms that the basic tendencies identified previously for the constant fluid pressure hold also for the non-uniform $p$. Moreover, even though the size of the plastic deformation zone obtained with the $K$ relaxation model yields here results that are relatively consistent with those produced with the $\sigma$ relaxation model, it is the latter variant which provides much better stress approximation in the plane of fracture extension.

\begin{figure}[htb!]
\begin{center}
\includegraphics[scale=0.4]{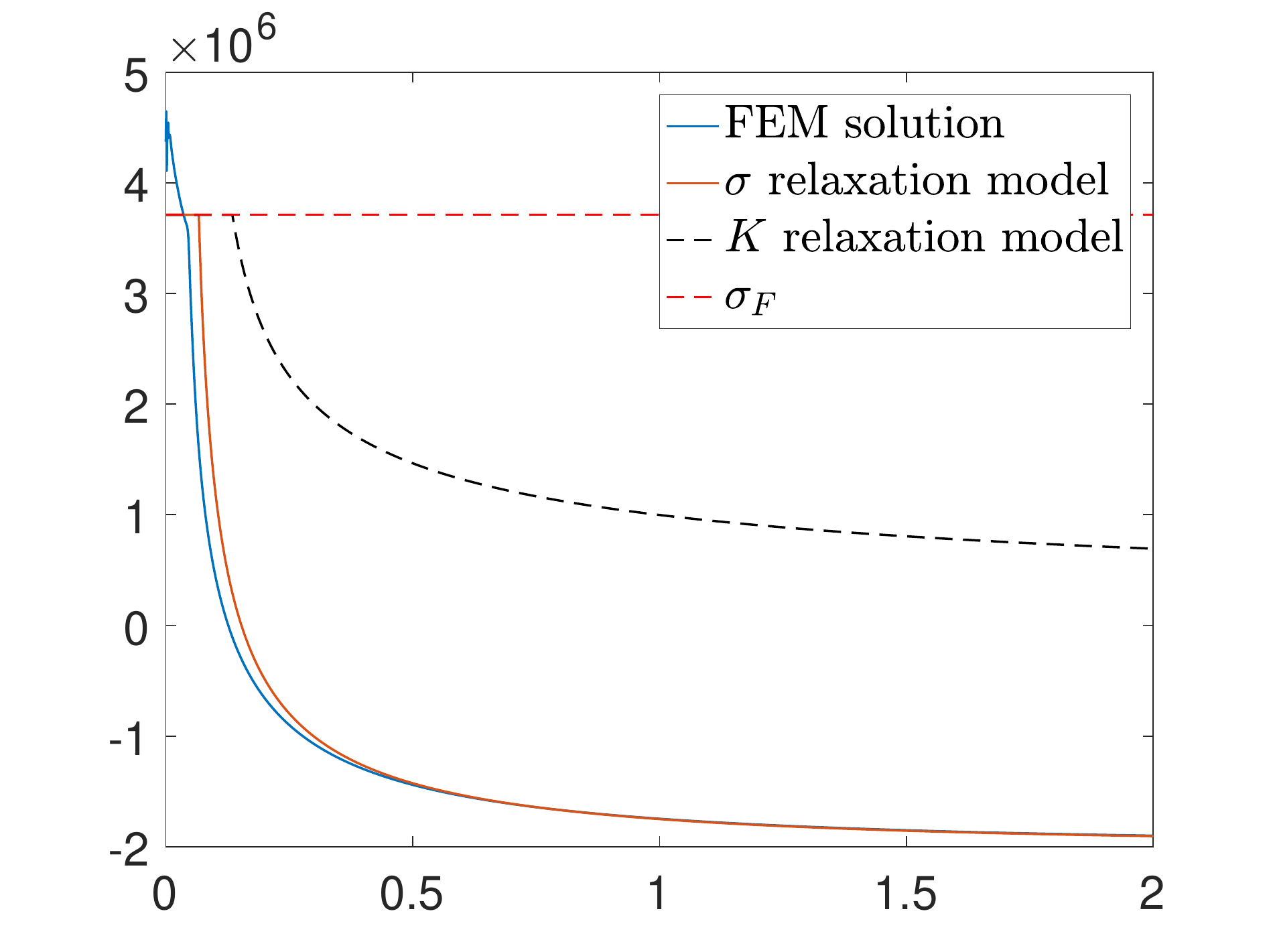}
\hspace{0mm}
\includegraphics[scale=0.4]{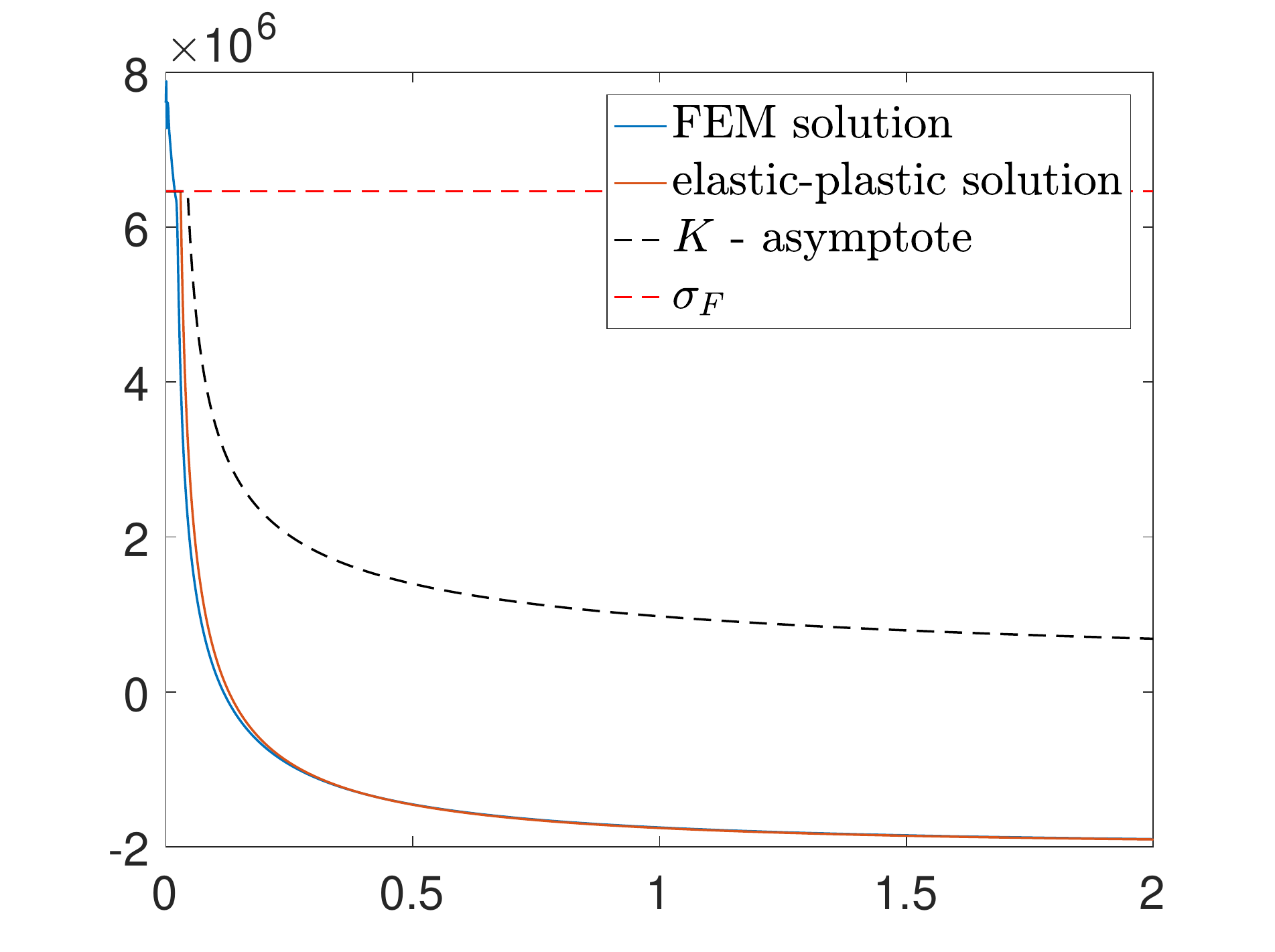}
\put(-345,-5){$r$}
\put(-110,-5){$r$}
\put(-450,155){$\textbf{a)}$}
\put(-225,155){$\textbf{b)}$}
\put(-450,85){$\sigma_{yy}$}
\put(-220,85){$\sigma_{yy}$}
\caption{Stress distribution ahead of the crack tip, $\sigma_{yy}$ [Pa], under the net fluid pressure described by relation \eqref{p_bench} for: a) material cohesion $c=3 \cdot 10^6$ Pa, b) material cohesion $c=5 \cdot 10^6$ Pa. The confining stress components are: $\sigma_{xx}^\text{c}=\sigma_{yy}^\text{c}=\sigma_{zz}^\text{c}=-2$ MPa. In both cases the assumed material friction angle is $\varphi=30^\circ$ and the crack half-length is $a=1$ m.}
\label{sigma_HF_conf_hydr}
\end{center}
\end{figure}

\begin{figure}[htb!]
\begin{center}
\includegraphics[scale=0.4]{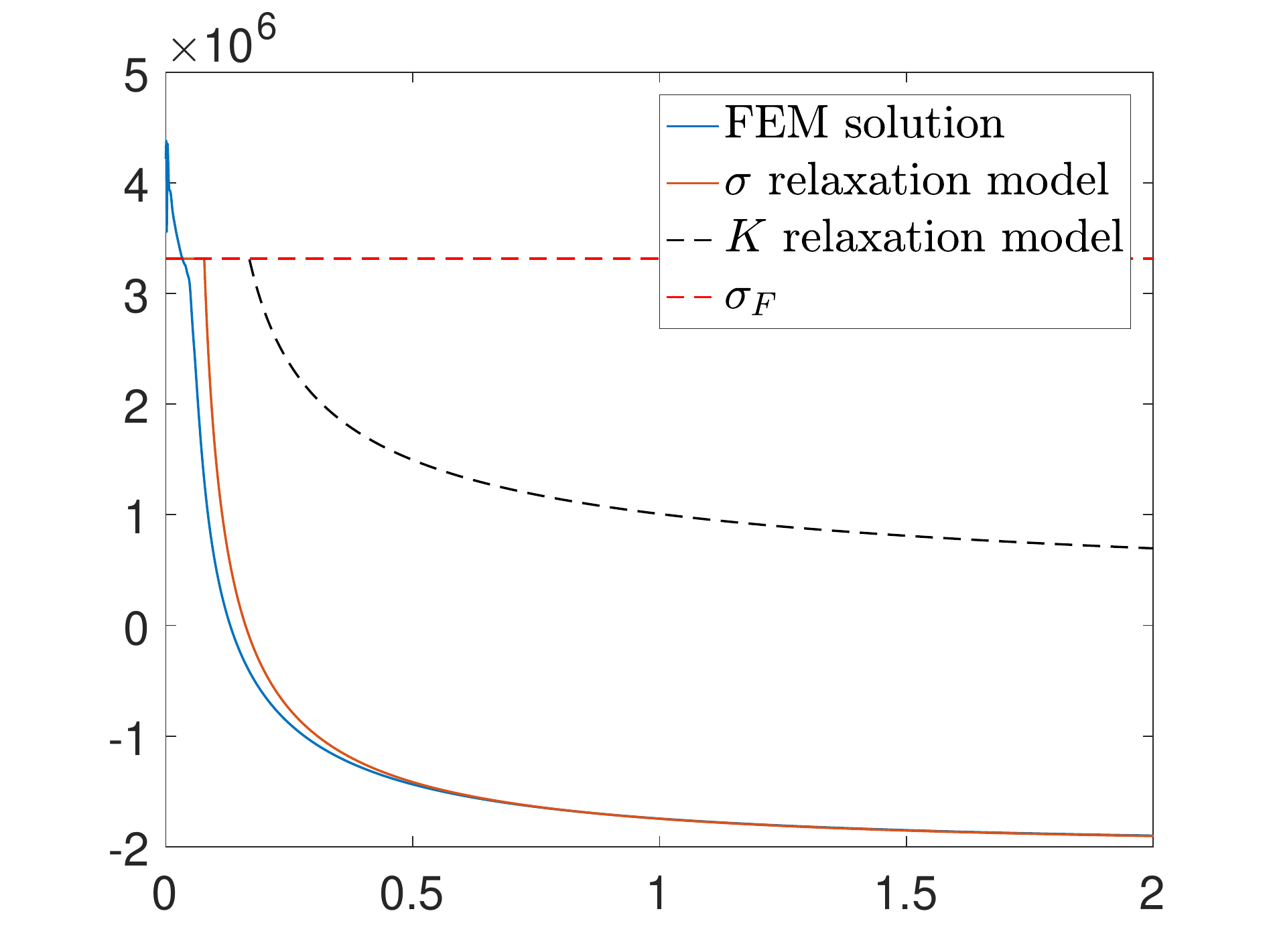}
\hspace{0mm}
\includegraphics[scale=0.4]{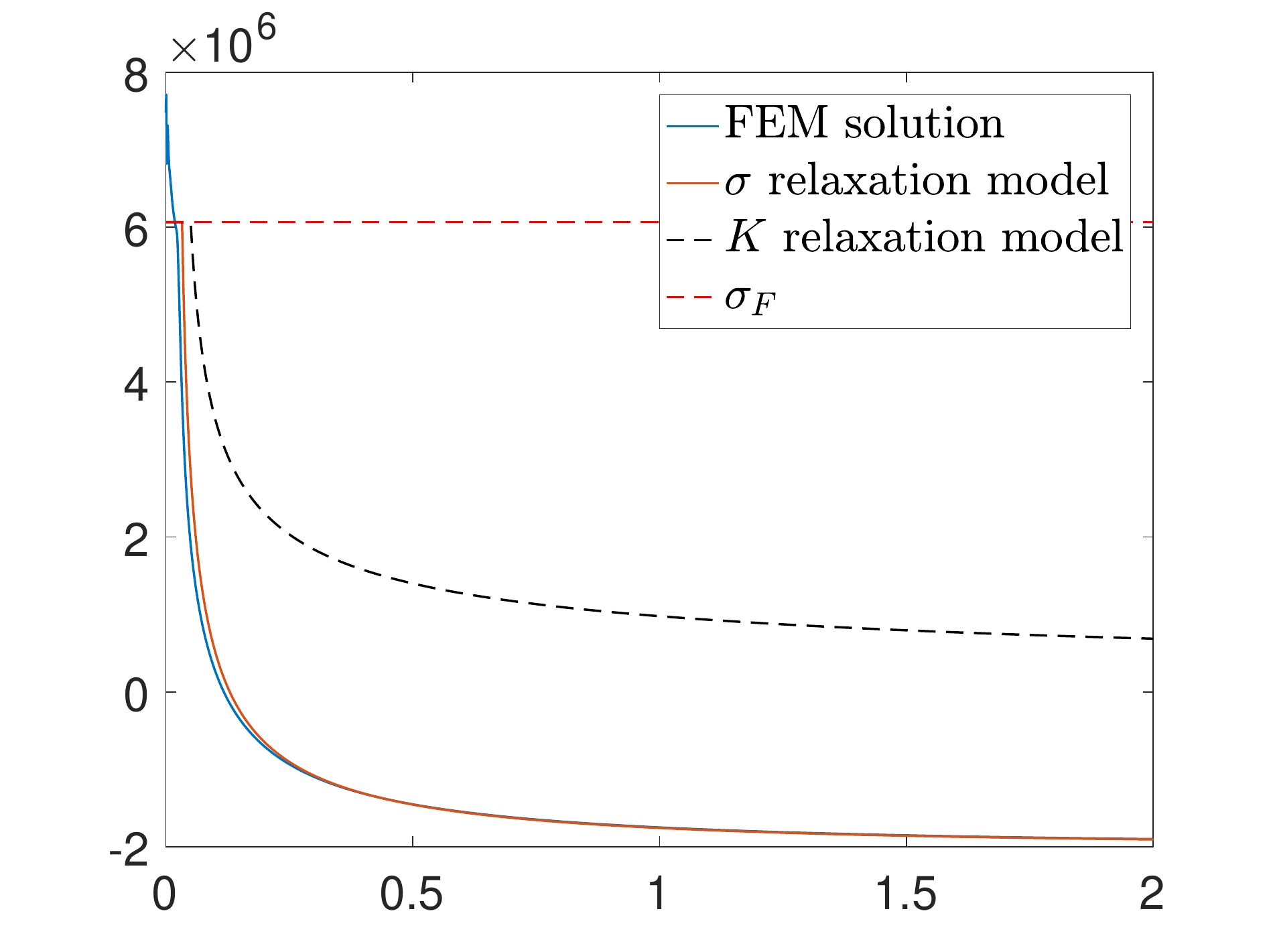}
\put(-345,-5){$r$}
\put(-110,-5){$r$}
\put(-450,155){$\textbf{a)}$}
\put(-225,155){$\textbf{b)}$}
\put(-450,85){$\sigma_{yy}$}
\put(-220,85){$\sigma_{yy}$}
\caption{Stress distribution ahead of the crack tip, $\sigma_{yy}$ [Pa],  under the net fluid pressure described by relation \eqref{p_bench} for: a) material cohesion $c=3 \cdot 10^6$ Pa, b) material cohesion $c=5 \cdot 10^6$ Pa. The confining stress components are: $\sigma_{xx}^\text{c}=-4$ MPa, $\sigma_{yy}^\text{c}=-2$ MPa, $\sigma_{zz}^\text{c}=-3$ MPa. In both cases the assumed material friction angle is $\varphi=30^\circ$ and the crack half-length is $a=1$ m.}
\label{sigma_HF_conf_non_un}
\end{center}
\end{figure}

In the last test in this section we analyze the case where the non-uniform distribution of the fluid pressure is combined with the confining stress imposed on the solid. The net pressure is defined in accordance with the benchmark \eqref{p_bench} -- \eqref{Pi_def}, while the absolute value of $p$ is obtained by adding the $\sigma_{yy}^\text{c}$ component of the confining stress. Two configurations of the confining stress are analyzed: i) $\sigma_{xx}^\text{c}=\sigma_{yy}^\text{c}=\sigma_{xx}^\text{c}=-2$ MPa, ii) $\sigma_{xx}^\text{c}=-4$ MPa, $\sigma_{yy}^\text{c}=-2$ MPa, $\sigma_{zz}^\text{c}=-3$ MPa. Respective results of computations are depicted in Figure \ref{sigma_HF_conf_hydr} and Figure \ref{sigma_HF_conf_non_un}, whereas the information on $\eta$ and $d_\text{p}$ can be found in Table \ref{tabela_d_eta_p_non_un}. This time we do not provide the relative error of stress equivalence $\delta S(r)$ as, due to the presence of the confining stress, $\sigma_{yy}$ changes its sign over the plane of crack propagation (and thus computation of  $\delta S(r)$ includes division by zero at some point).

\begin{table}[]
\begin{center}
\begin{tabular}{c|c|c|c|c|}
\cline{2-5}
\multirow{2}{*}{}                       & \multicolumn{2}{c|}{$\sigma_{xx}^\text{c}=\sigma_{yy}^\text{c}=\sigma_{zz}^\text{c}=-2$ MPa} & \multicolumn{2}{c|}{\begin{tabular}[c]{@{}c@{}}$\sigma_{xx}^\text{c}=-4$ MPa, $\sigma_{yy}^\text{c}=-2$MPa,\\ $\sigma_{zz}^\text{c}=-3$ MPa\end{tabular}} \\ \cline{2-5} 
                                        & $c=3$ MPa                                     & $c=5$ MPa                                    & $c=3$ MPa                                                                   & $c=5$ MPa                                                                   \\ \hline \hline
\multicolumn{1}{|c||}{$\sigma_\text{F}$} & 3.71 MPa                                      & 6.46 MPa                                     & 3.31 MPa                                                                    & 6.06 MPa                                                                    \\ \hline
\multicolumn{5}{|c|}{$\sigma$ relaxation model}                                                                                                                                                                                                                                                    \\ \hline
\multicolumn{1}{|c||}{$\eta$}            & 0.0439 m                                      & 0.0182 m                                     & 0.0520 m                                                                    & 0.0202 m                                                                    \\ \hline
\multicolumn{1}{|c||}{$d_\text{p}$}      & 0.0669 m                                      & 0.0300 m                                     & 0.0779 m                                                                    & 0.0331 m                                                                    \\ \hline
\multicolumn{5}{|c|}{$K$ relaxation model}                                                                                                                                                                                                                                                         \\ \hline
\multicolumn{1}{|c||}{$\eta$}            & 0.0675 m                                      & 0.0223 m                                     & 0.0846 m                                                                    & 0.0253 m                                                                    \\ \hline
\multicolumn{1}{|c||}{$d_\text{p}$}      & 0.1349 m                                      & 0.0445 m                                     & 0.1691m                                                                     & 0.0505 m                                                                    \\ \hline
\end{tabular}
\caption{The crack length extension values, $\eta$, and the sizes of the plastic deformation zone, $d_p$, for the case of non-uniform pressure distribution and different configurations of the confining stress components.}
\label{tabela_d_eta_p_non_un}
\end{center}
\end{table}

The presented data indicates clearly that also in the case of a fracture in a pre-stressed material the trends identified and quantified above are preserved. The $K$ relaxation model overestimates the size of the plastic deformation zone and this overestimation grows with the size of the plastic zone itself. Unlike the $K$ relaxation model, the $\sigma$ relaxation model provides fairly good approximation of the stress in the plane of crack propagation including the region of plastic deformation. Just as in all previously analyzed cases, the new stress redistribution model constitutes a much better and safer choice than the classical variant based on Irwin's single-term approximation.  

The analysis provided above shows that the newly introduced stress relaxation model provides a very good approximation for the $\sigma_{yy}$ stress component in the plane of fracture extension together with the size of the plastic deformation zone. Thus, the underlying assumptions used when deriving the modified crack propagation criterion are satisfied to a large degree. This constitutes a strong argument for the credibility of the new fracture criterion. Obviously, the decisive evidence would require here a detailed computational comparison with other criteria (and/or experiment), which however will be a subject of future studies. For the sake of analysis carried out in the present paper we assume that the new crack propagation condition describes sufficiently well the character and nature of the elasto-plastic fracture especially in the case of small yielding. Thus, respective mechanisms of the elasto-plastic hydraulic fracture can be investigated using the developed numerical tools.

\section{Numerical results: comparison of elastic and elasto-plastic hydraulic fractures}
\label{results}



In the previous section we have shown that newly introduced crack propagation condition in the form \eqref{K_eff}--\eqref{K_I_eff} has capability of correctly describing the nature of elasto-plastic fracture. Now we employ this condition to model a complete hydraulic fracture problem and estimate the effect of plastic deformation.
Below we simulate the hydraulic fracture problem in three variants: 
\begin{itemize}
\item{the classical KGD problem for an elastic solid with the standard LEFM crack propagation condition ($K_I=K_{I\text{c}}$);}
\item{the modified KGD problem for an elastic solid but with the plasticity dependent crack propagation condition \eqref{K_eff};}
\item{the fully elasto-plastic problem (where the elasto-plastic model for the bulk of the fractured material holds) with the crack propagation condition \eqref{K_eff} - in this way the plasticity related effects are embedded in both, the deformation of the solid material and the crack propagation condition.}
\end{itemize}
 Naturally, the last variant utilizes the FEM block to compute the crack opening.  The numerical results obtained for respective variants of the problem are used to investigate the shielding  effect caused by the plastic deformation of the rock formation and whether the simplified second variant of the problem can be a substitute for the full elasto-plastic model.

The computations are carried out for the material parameters and pumping rate value listed in Table \ref{tabela_parametry}. The initial solution comprises an immobile stationary pressurized crack of the aperture:
\[
w(x,0)=w_0\sqrt{a_0^2-x^2}.
\]
The influx magnitude is changed from zero to $q_0$ according to the following function:
\begin{equation}
\label{TP_def}
\bar q_0(t)=
  \begin{cases}
		\left(\frac{3}{t_1^2}t^2-\frac{2}{t_1^3}t^3\right)q_0      &\quad \text{for}  \quad t<t_1,\\
    q_0       & \quad \text{for} \quad t \geq t_1,
  \end{cases}
\end{equation}
which gives a smooth influx transition to the target value over the time span $t \in (0,t_1)$. The value of $t_1$ was set to 0.1 s. The final time instant was $t_\text{end}=20$ s.

\begin{table}[]
\begin{center}
\begin{tabular}{|l|c|c|}
\hline
\multicolumn{3}{|c|}{Elastic constants}                                                                                     \\ \hline
Young modulus, $E$ {[}Gpa{]}                                    & \multicolumn{2}{c|}{14.5}                                 \\ \hline
Poisson's ratio, $\nu$                                          & \multicolumn{2}{c|}{0.24}                                 \\ \hline\hline
\multicolumn{3}{|c|}{Plastic constants (Mohr-Coulomb law)}                                                                  \\ \hline
angle of friction, $\varphi$ {[}deg{]}                             & \multicolumn{2}{c|}{30}                                   \\ \hline
angle of dilation, $\psi$ {[}deg{]}                             & \multicolumn{2}{c|}{30}                                   \\ \hline
\multirow{2}{*}{material cohesion, $c\left(\varepsilon^\text{pl}\right)$ {[}Mpa{]}}               & $\varepsilon^\text{pl}$=0 & $\varepsilon^\text{pl}$=0.011 \\ \cline{2-3} 
                                                                & 5                         & 12                            \\ \hline\hline
\multicolumn{3}{|c|}{Fracturing parameters}                                                                                 \\ \hline
fracture toughness, $K_{I\text{c}}$ {[}MPa$\sqrt{\text{m}}${]}  & \multicolumn{2}{c|}{2}                                    \\ \hline\hline
\multicolumn{3}{|c|}{In-situ stress}                                                                                        \\ \hline
$\sigma_1$ {[}Mpa{]}                                            & \multicolumn{2}{c|}{-14.7}                                \\ \hline
$\sigma_2$ {[}Mpa{]}                                            & \multicolumn{2}{c|}{-3.7}                                 \\ \hline
$\sigma_3$ {[}Mpa{]}                                            & \multicolumn{2}{c|}{-9}                                   \\ \hline\hline
\multicolumn{3}{|c|}{Fluid flow parameters}                                                                                 \\ \hline
fluid viscosity, $\mu$ {[}Pa $\cdot$ s{]}                       & \multicolumn{2}{c|}{0.001}                                \\ \hline
pumping rate, $q_0$ $\left[ \frac{\text{m}^2}{\text{s}}\right]$ & \multicolumn{2}{c|}{0.0005}                               \\ \hline\hline
\multicolumn{3}{|c|}{Initial fracture geometry}                                                                             \\ \hline
initial half-length, $a_0$ {[}m{]}                              & \multicolumn{2}{c|}{0.4}                                  \\ \hline
crack opening multiplier, $w_0$                                 & \multicolumn{2}{c|}{$1.75\cdot 10^{-3}$}                                  \\ \hline
\end{tabular}
\caption{Material and HF process parameters used in numerical simulations. The in-situ stresses are associated with axes $x,y,z$ from Figure \ref{KGD_geom}, respectively.}
\label{tabela_parametry}
\end{center}
\end{table}

\begin{figure}[htb!]
\begin{center}
\includegraphics[scale=0.37]{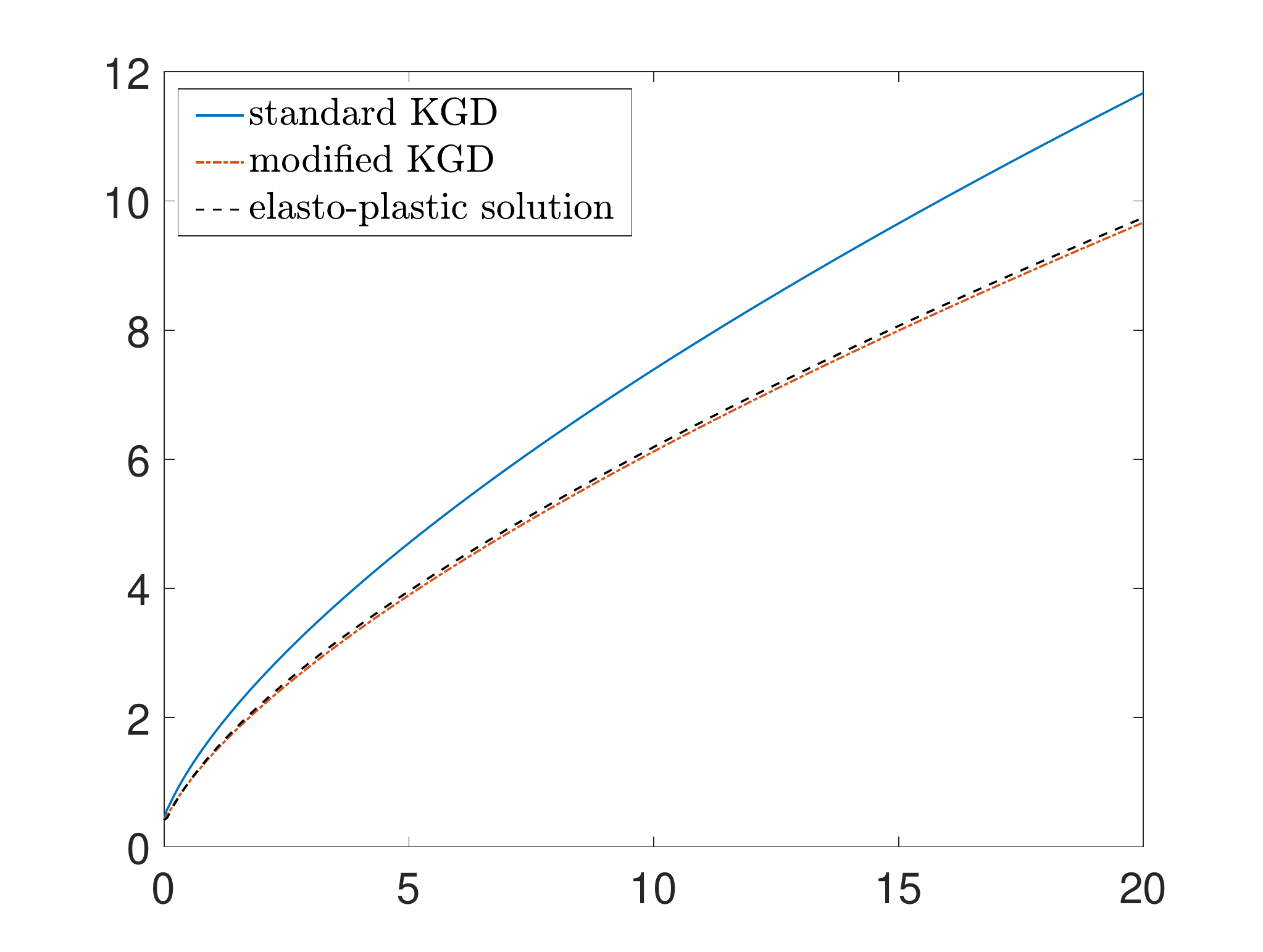}
\hspace{0mm}
\includegraphics[scale=0.37]{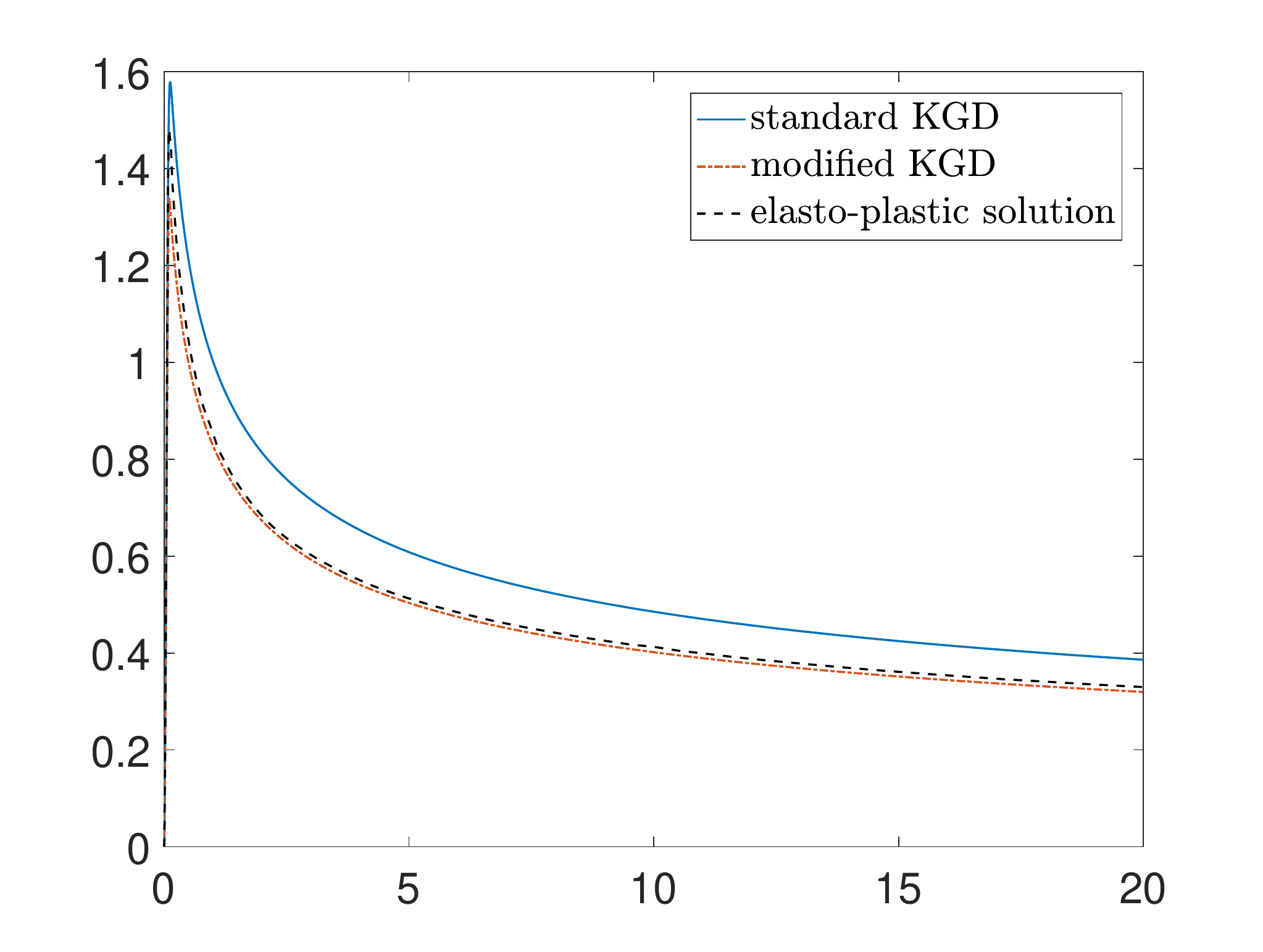}
\put(-335,-5){$t$}
\put(-110,-5){$t$}
\put(-450,155){$\textbf{a)}$}
\put(-225,155){$\textbf{b)}$}
\put(-442,82){$a(t)$}
\put(-225,82){$v_0(t)$}
\caption{Simulation results in terms of: a) the crack length, $a$, b) the crack propagation speed, $v_0$.}
\label{L_v0}
\end{center}
\end{figure}

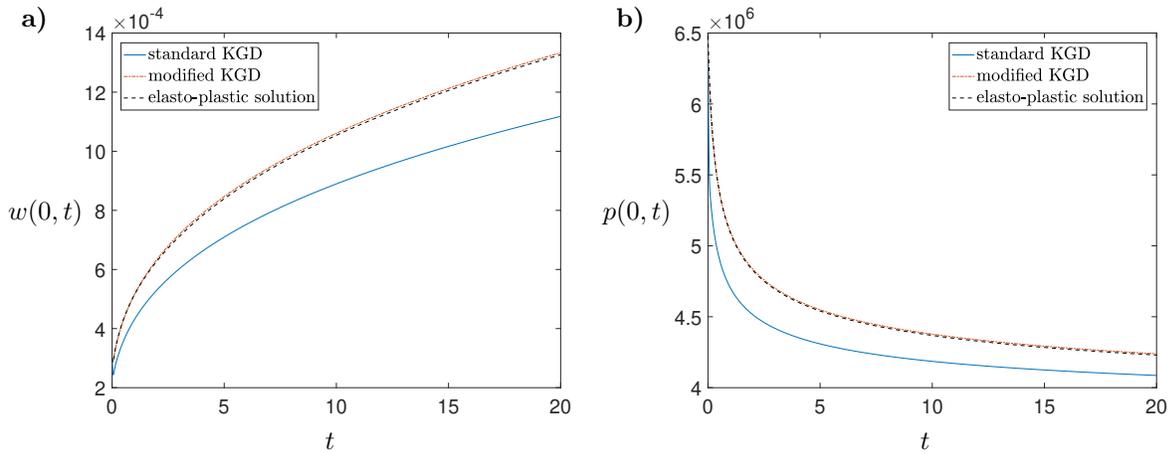
\begin{figure}[htb!]
\begin{center}
\includegraphics[scale=0.37]{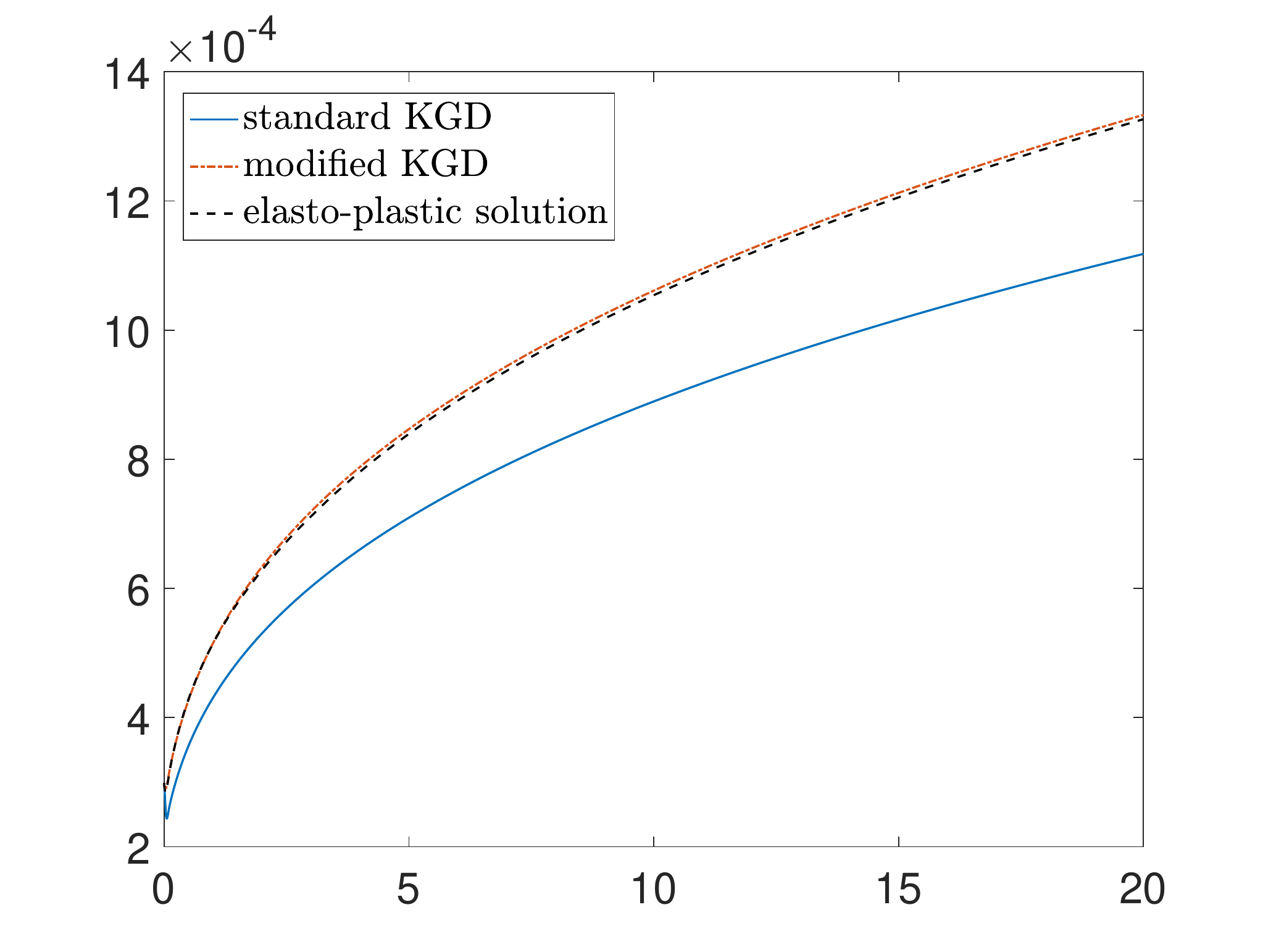}
\hspace{0mm}
\includegraphics[scale=0.37]{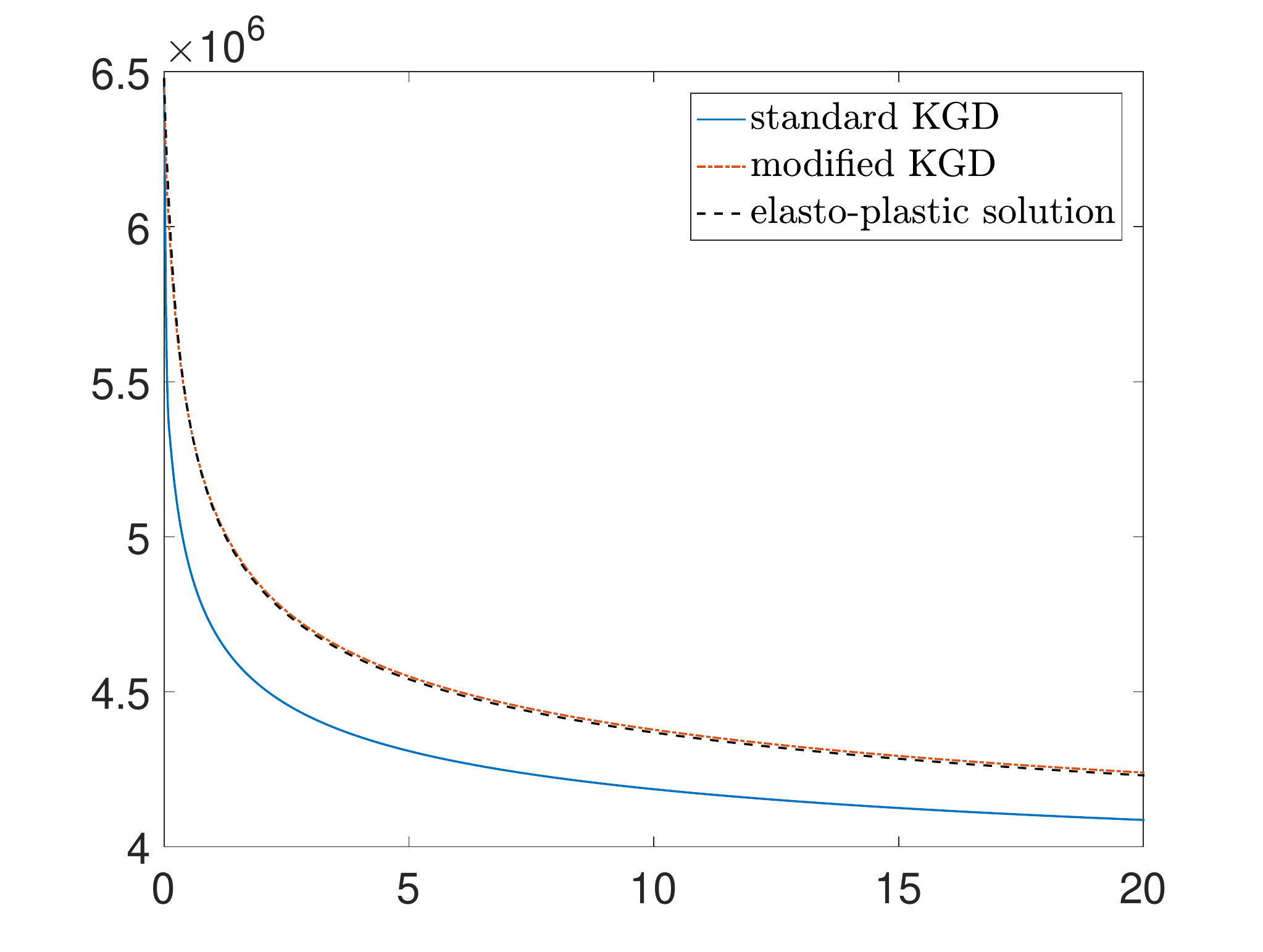}
\put(-335,-5){$t$}
\put(-110,-5){$t$}
\put(-450,155){$\textbf{a)}$}
\put(-225,155){$\textbf{b)}$}
\put(-455,82){$w(0,t)$}
\put(-230,82){$p(0,t)$}
\caption{Simulation results in terms of: a) the fracture opening at the crack mouth, $w(0,t)$, b) the down-hole pressure, $p(0,t)$.}
\label{W_P}
\end{center}
\end{figure}

\begin{figure}[htb!]
\begin{center}
\includegraphics[scale=0.37]{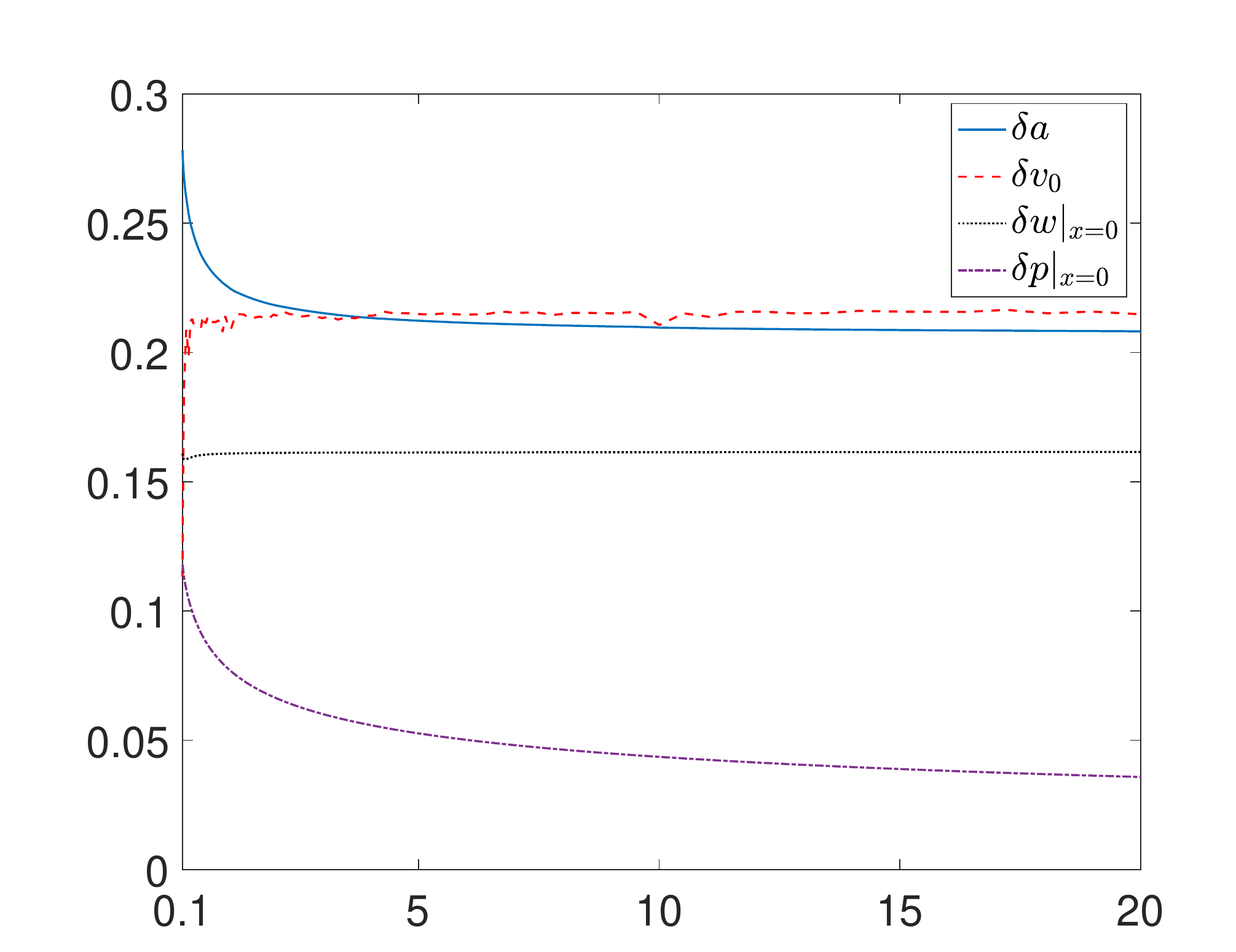}
\put(-102,-5){$t$}
\caption{Relative deviations between the solutions obtained for the standard KGD problem and the elasto-plastic HF model.}
\label{deviations}
\end{center}
\end{figure}

\begin{figure}[htb!]
\begin{center}
\includegraphics[scale=0.37]{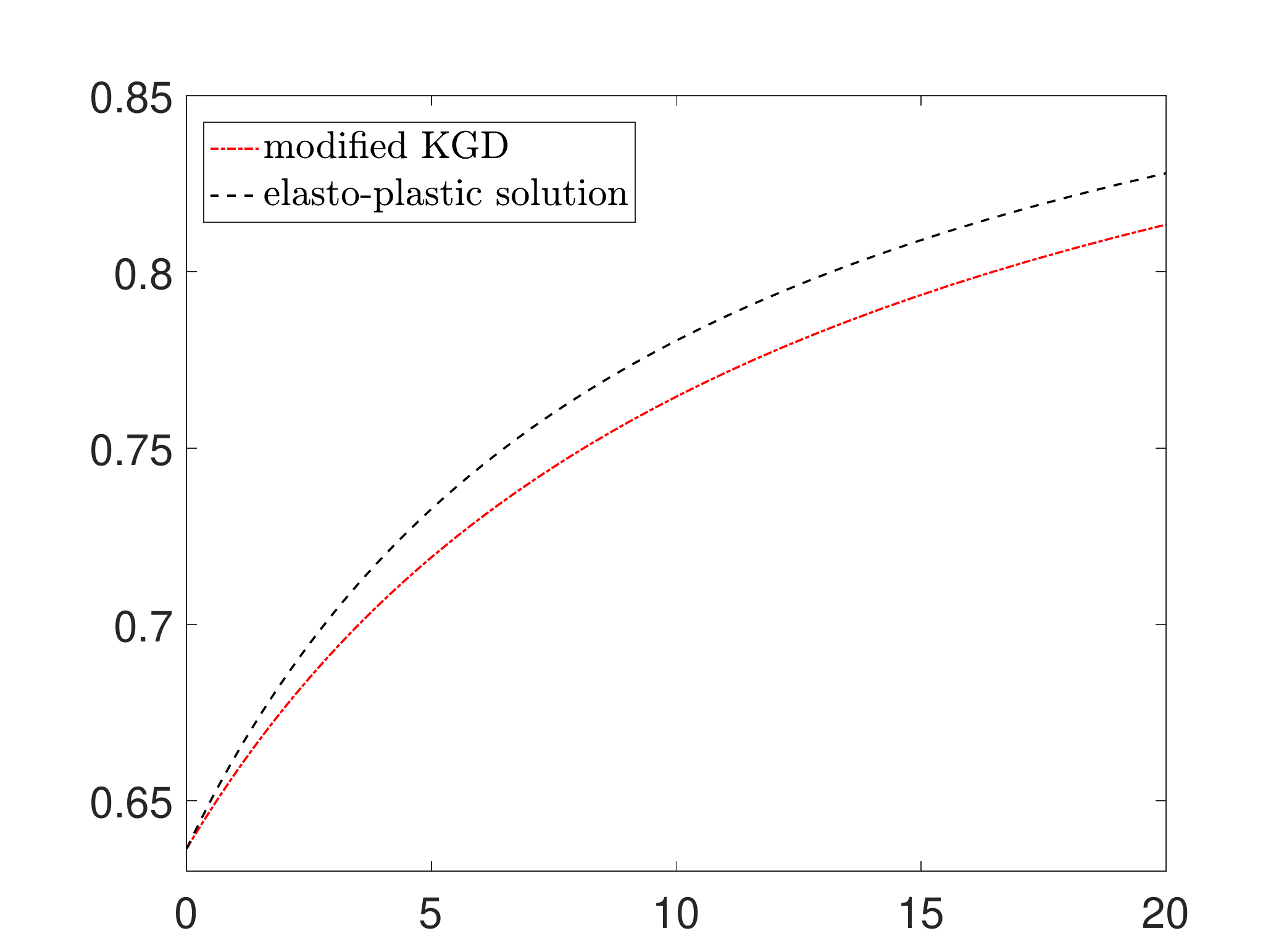}
\hspace{0mm}
\includegraphics[scale=0.37]{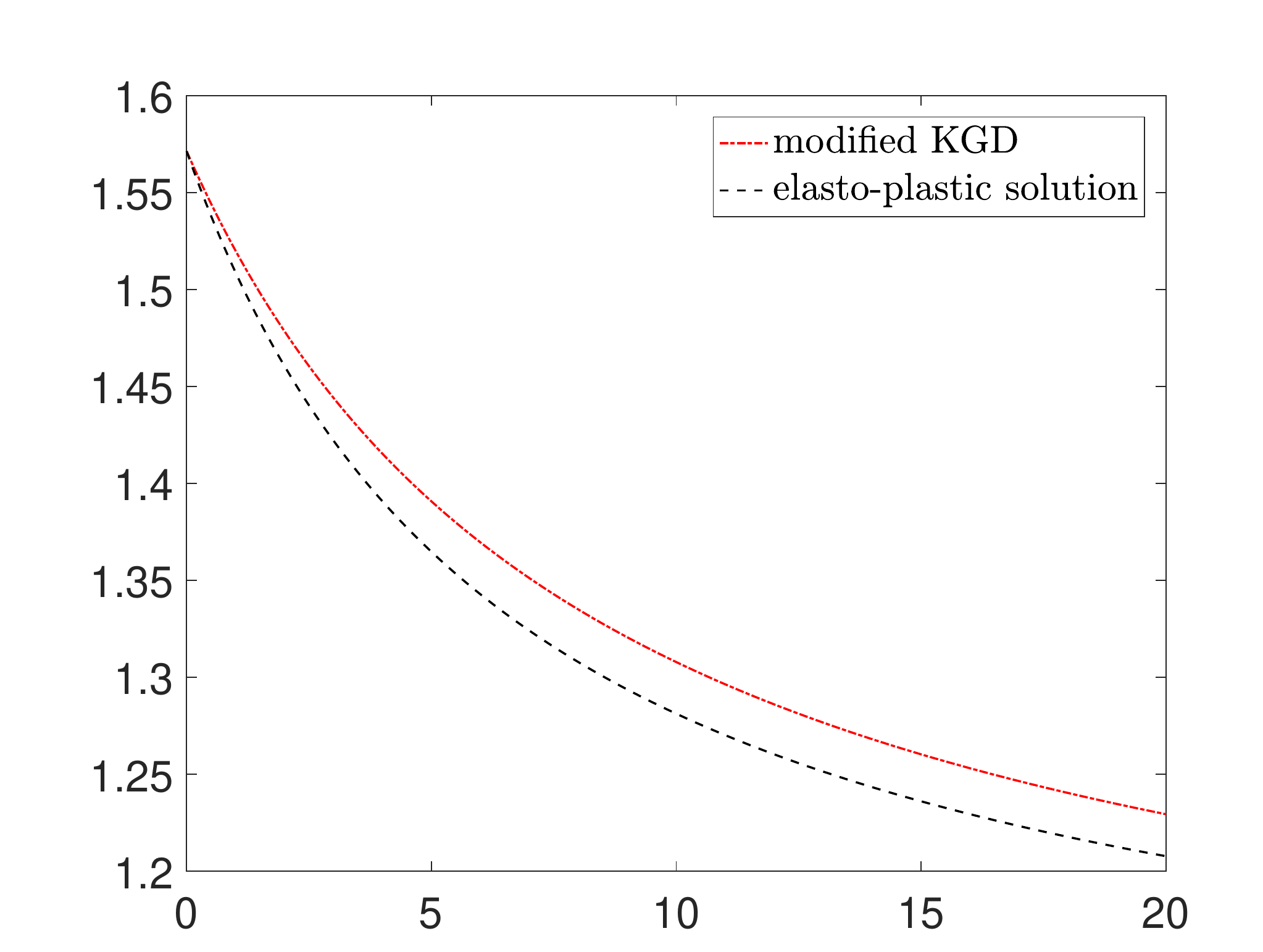}
\put(-330,-5){$t$}
\put(-105,-5){$t$}
\put(-450,155){$\textbf{a)}$}
\put(-225,155){$\textbf{b)}$}
\put(-440,82){$\alpha$}
\put(-225,82){$\frac{K_{I\text{c}}^\text{eff}}{K_{I\text{c}}}$}
\caption{Effective fracture toughness: a) the scaling coefficient $\alpha$, b) the normalized effective toughness, $\frac{K_{I\text{c}}^\text{eff}}{K_{I\text{c}}}$.}
\label{alfa}
\end{center}
\end{figure}
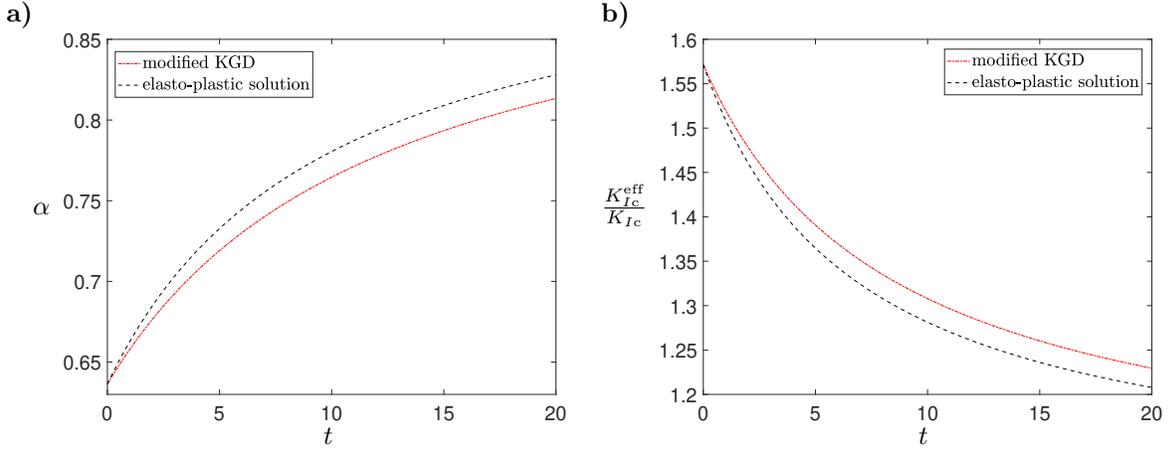

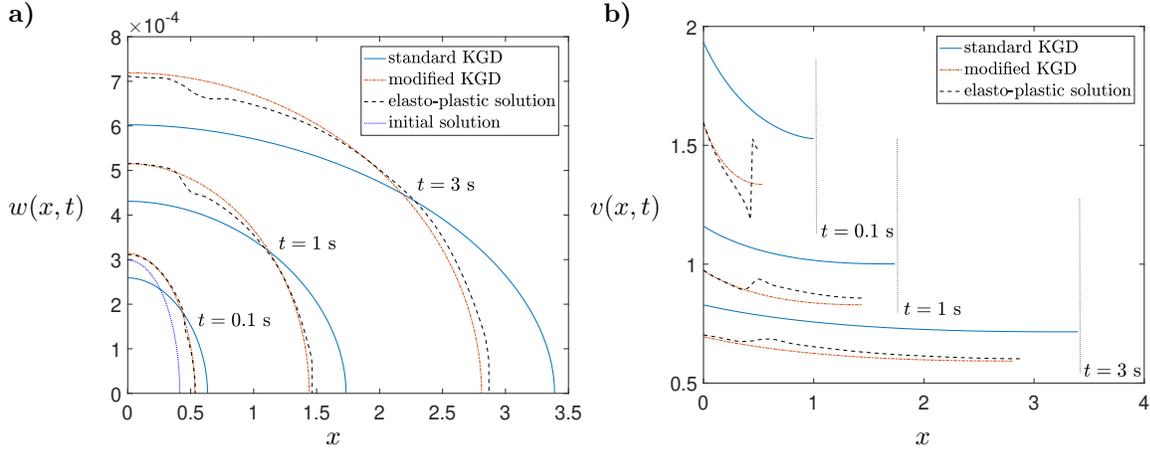
\begin{figure}[htb!]
\begin{center}
\includegraphics[scale=0.37]{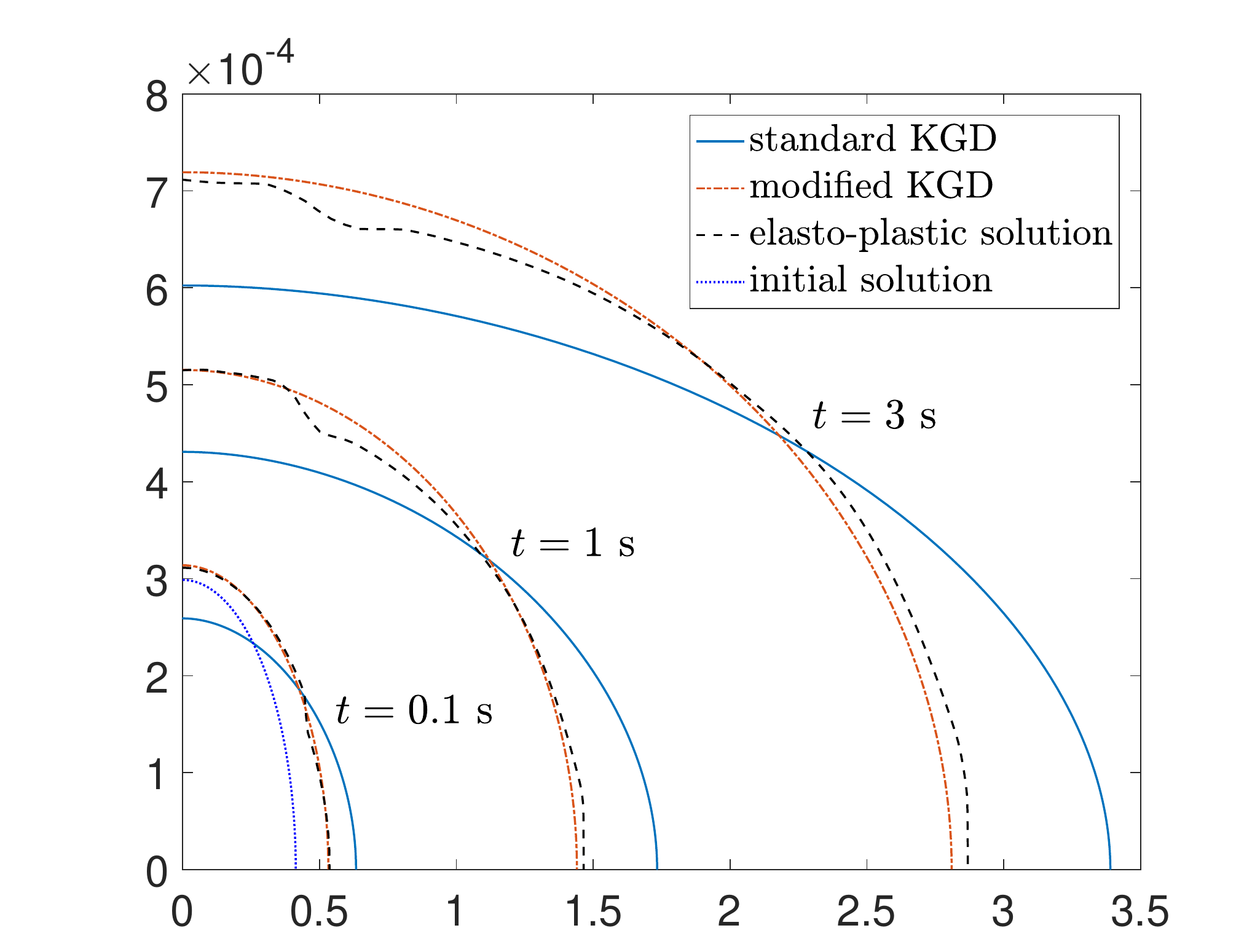}
\hspace{0mm}
\includegraphics[scale=0.37]{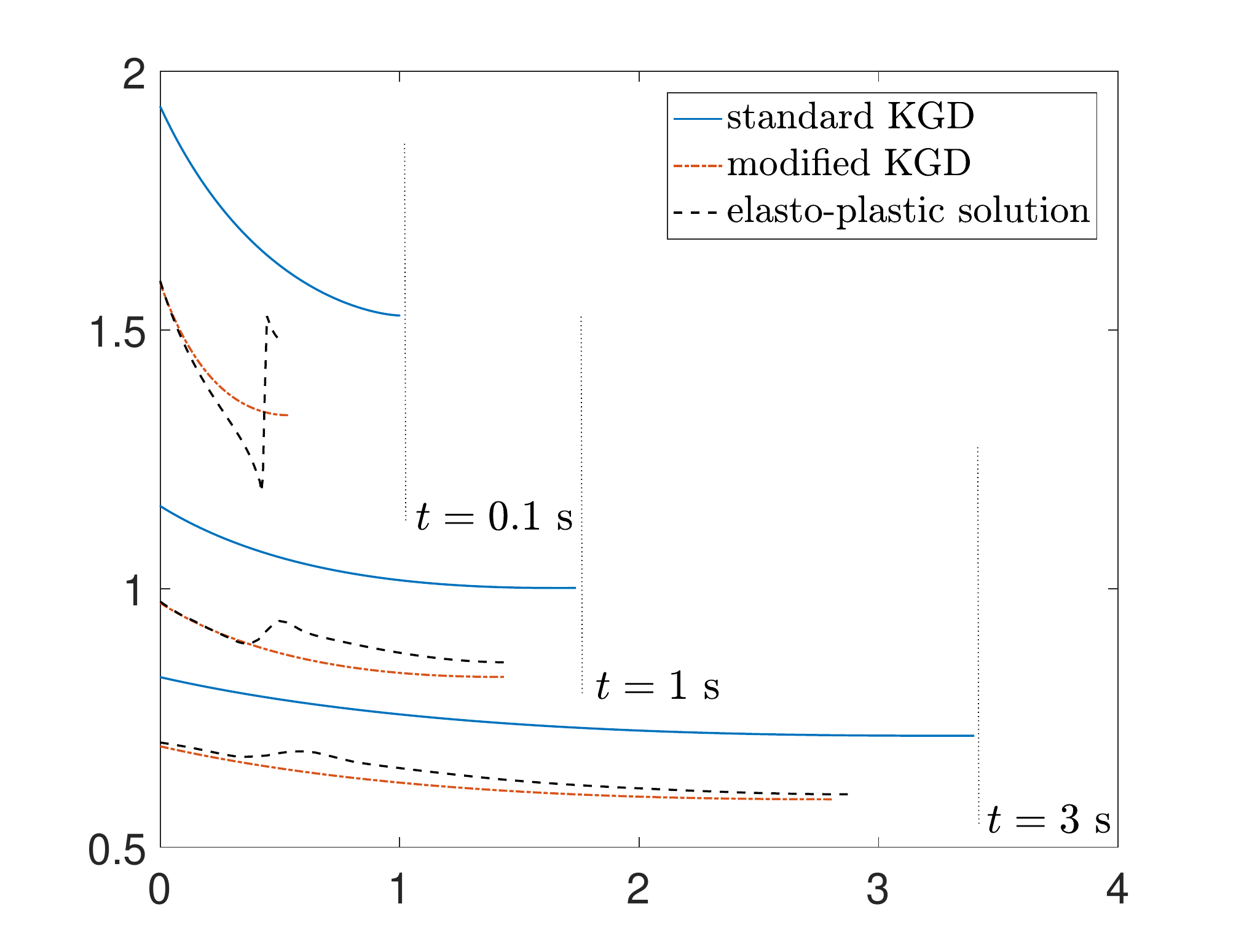}
\put(-330,-5){$x$}
\put(-107,-5){$x$}
\put(-450,155){$\textbf{a)}$}
\put(-225,155){$\textbf{b)}$}
\put(-450,82){$w(x,t)$}
\put(-230,82){$v(x,t)$}
\caption{Spatial distribution of: a) the crack aperture, $w(x,t)$, b) the  fluid velocity, $v(x,t)$, for three time instants: $t=0.1$ s, $t=1$ s, $t=3$ s. The bumps in the elasto-plastic solution are a result of the purely elastic nature of the initial fracture.}
\label{rozwarcia_predkosci}
\end{center}
\end{figure}

The results of computations for the respective variants of the problem are depicted in Figures \ref{L_v0} - \ref{rozwarcia_predkosci}. The general trends regarding the geometry of elasto-plastic hydraulic fracture, already reported in other publications, hold. As a consequence of the shielding effect of the plastic deformations the resulting crack is shorter and wider than its elastic counterpart. It is notable that in the standard KGD model the inlet fracture aperture decreases immediately following the crack initiation, and only starts to grow again after some time (compare Figure \ref{W_P}a) and Figure \ref{rozwarcia_predkosci}a)). This process additionally contributes to the recalled trend.

The relative differences between the solutions obtained for the standard KGD problem and the full elasto-plastic HF model are depicted in Figure \ref{deviations}. Deviations for four dependent variables are presented: the crack length ($\delta a$), the crack propagation speed ($\delta v_0$), the fracture aperture at the crack mouth ($\delta w|_{x=0}$), and the  borehole pressure ($\delta p|_{x=0}$). It shows that for the final time instant one obtains  an approximately $20\%$ longer fracture with the  classical KGD model, however the difference is  larger for smaller times (up to $30\%$). The relative difference between the respective crack propagation speeds retains the level of little more than $20\%$ through almost the entire time of fracture extension. It is only for $t<0.1$  (not shown in the figure in order to make it more legible) that $\delta v_0$ is much higher. The crack aperture at $x=0$ is over $15\%$ greater for the elasto-plastic fracture and this deviation is almost constant during the whole time of the process. Finally, the greatest difference between the respective borehole pressures $\delta p|_{x=0}$ (12 $\%$) is obtained around the time of pumping regime change ($t=0.1$ s). Then $\delta p|_{x=0}$ declines to reach less than $4\%$ at $t=20$ s. As can be concluded from the above description, under the specified conditions the plastic deformation mechanisms have a noticeable effect on the simulation results. 

Meanwhile, the obtained toughness correction coefficients $\alpha$ from equation \eqref{alfa_def} are depicted in Figure \ref{alfa}a). The corresponding values for the second (modified KGD) and third (fully elasto-plastic model) variants of the HF problem are relatively close to one another. They cause the fracture toughness magnification which ranges from a little above 1.55 at the initial time to approximately 1.2 at the final time instant (see Figure \ref{alfa}b)).

Another immediate conclusion from the presented results is that the solution obtained for the modified KGD problem resembles to a very large degree the one produced for the full elasto-plastic HF model (at least in terms of the analyzed above dependent variables). This suggests that the prevailing part of the effects related to the inelastic solid deformation is a result of the crack propagation condition rather than the  deformation of the bulk of the fractured material. In Figure \ref{rozwarcia_predkosci}a) we present a comparison of respective fracture profiles for three moments in time: $t=0.1$ s, $t=1$ s and $t=3$ s. One can see from the first glance that the footprint of the initial fracture is reflected in the elasto-plastic solution in the subsequent time instants (the `bumps' appearing near the fluid inlet at later time-steps). A similar effect was noticed in the paper by \cite{Papanastasiou_1997}. It originates from the fact that the initial solution is defined by an elastic fracture and thus there is no plastic deformation associated with the initial profile. As the plastic zone starts to develop with the fracture growth, the shape peculiarity around the point $x=a_0$ is preserved, however its relative contribution to the crack profile becomes less pronounced with time. Even though the respective characteristics for $a(t)$ and $w(0,t)$ are hardly distinguishable from each other (compare Figure \ref{L_v0}a) and Figure \ref{W_P}a)) the resulting fracture profiles differ noticeably. This is not only due to the aforementioned shape peculiarity but also  because of the more blunt near-tip contour of the fully elasto-plastic fracture. 

 In the paper by \cite{Papanastasiou_1999a} the values of effective fracture toughness computed a posteriori from the elasto-plastic solution were used in the classical KGD model. Then, the resulting fracture profiles were compared with their elasto-plastic counterparts. The comparison revealed good coincidence of respective results but not quite as good as that reported in Figure \ref{rozwarcia_predkosci}a). The explanation of this fact is twofold. Firstly, for the assumed by us material properties and HF process parameters the extent of plastic yielding is smaller than that in the recalled paper. Secondly, the computational algorithm employed in our studies includes (unlike the scheme of   \cite{Papanastasiou_1999a}) the solution dependent effective fracture toughness embedded explicitly in the crack propagation condition. This, together with the property of the algorithm that the global fluid balance condition is identically satisfied (see \cite{Wrobel_2015} for details), contributes to the respective variants of the solution (the one for the modified KGD problem and the full elasto-plastic solution) being largely equivalent to each other.

In Figure \ref{rozwarcia_predkosci}b) the fluid velocity distributions $v(x,t)$ are shown for the same as previously time instants. It is evident that the fracture profile deflection near the point $x=a_0$ substantially disturbs the fluid flow. For $t=0.1$ s one observes virtually a velocity jump around the singular point. Even though the fluid velocity becomes smoother as the fracture evolves, at least two serious issues arise here. The first one is the validity of the lubrication theory and 1D laminar flow model. It seems that under the assumed initial conditions the underlying assumptions for both of them are not satisfied at least in the small time range. The second issue involves accounting for non-Newtonian shear-rate dependent rheologies of fracturing fluids. In such cases the effective local fluid viscosity strongly depends on the variation of velocity profile \citep{Wrobel_Arxiv}. This can considerably affect the regime of flow and thus the propagation of hydraulic fracture (see e.g. the studies by  \cite{Lavrov_2013} and \cite{Felisa_2018}  for the non-Newtonian fluid flow in the rough walled fractures).

Another interesting observation can be made when comparing the crack half-lengths and apertures computed for the modified KGD model and the fully elasto-plastic problem. Namely, it is always the former solution which provides slightly shorter and wider fractures. It is also reflected in the values of the effective fracture toughness where greater toughness magnification is obtained for the modified KGD problem (see Figure \ref{alfa}b)). Although the respective differences are rather small, the trend itself is clear. It suggests that the inelastic deformation of the bulk of the fractured material introduces a screening effect that works against the lateral fracture growth. Thus, the overall fracture geometry results from the interplay between the aforementioned mechanism and the shielding effect pertaining to the plasticity-dependent crack propagation condition. Even though this observation seems trivial, it can be of importance in some variants of the hydraulic fracture problem, e.g. when the interfacial hydrofracturing is analyzed. In such an instance the plastic deformation of rock could increase the fracture length and reduce its aperture, which is contrary to classical results.

\begin{figure}[htb!]
\begin{center}
\includegraphics[scale=0.37]{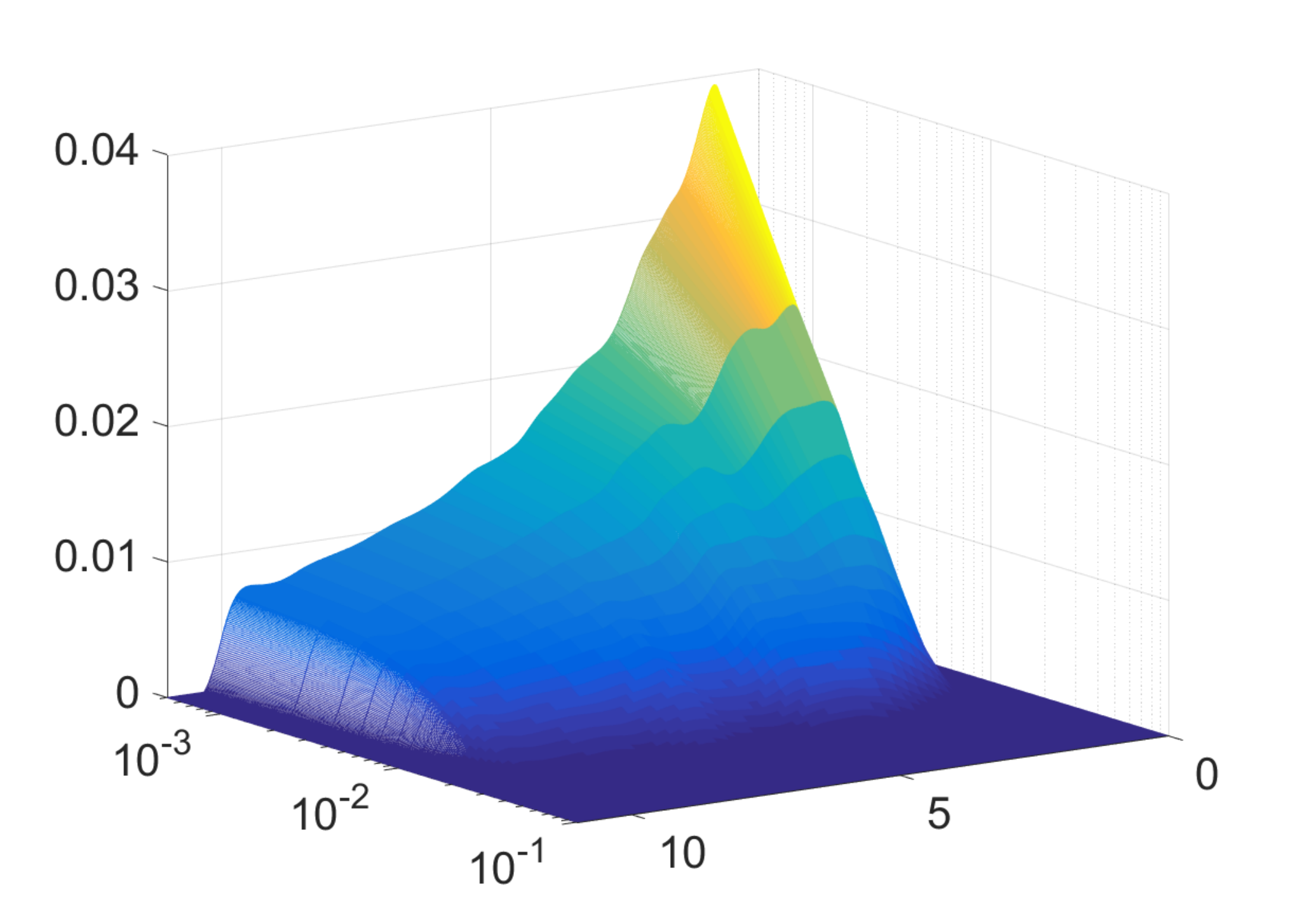}
\hspace{0mm}
\includegraphics[scale=0.37]{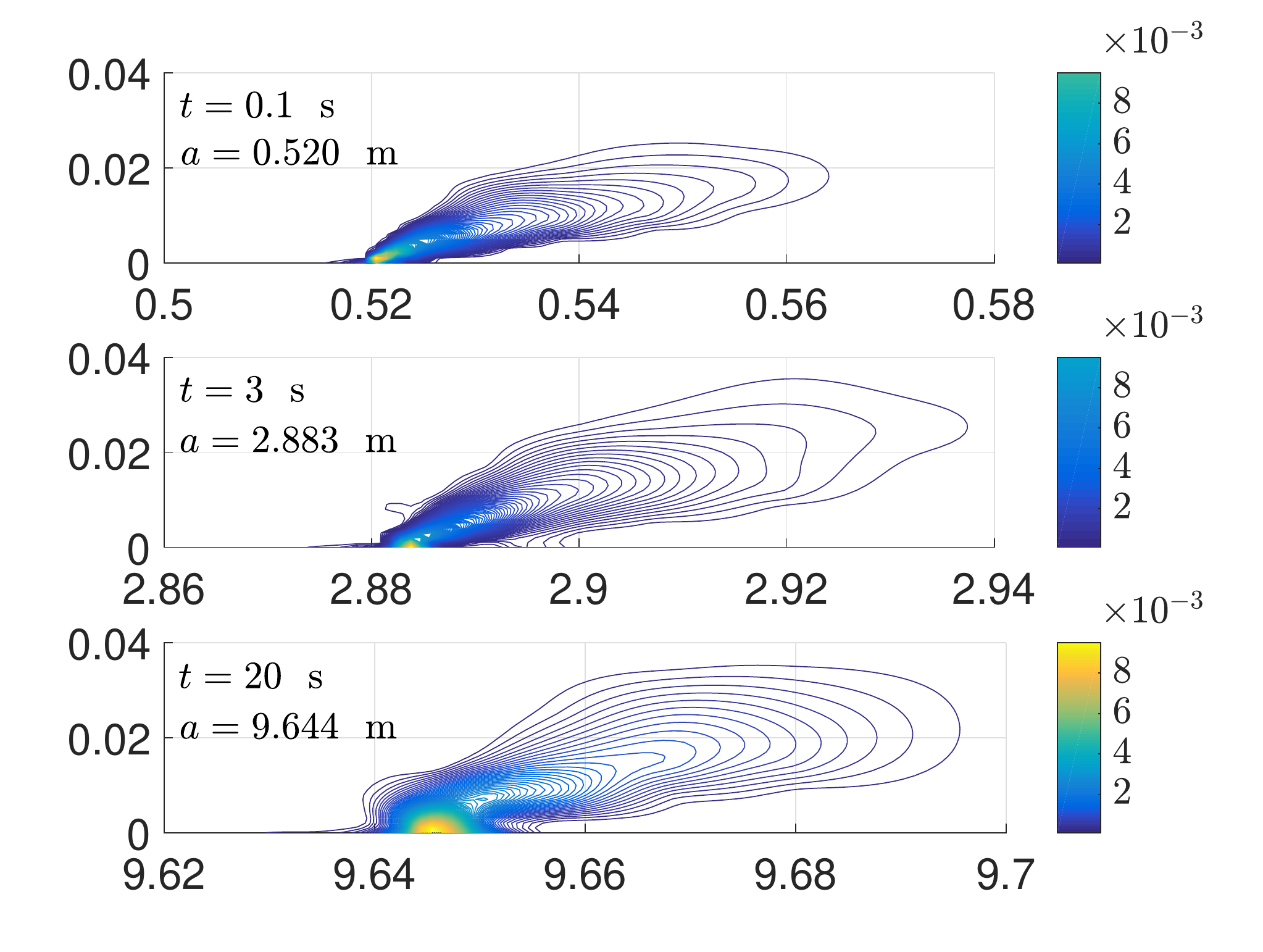}
\put(-285,5){$x$}
\put(-395,5){$y$}
\put(-117,-5){$x$}
\put(-450,155){$\textbf{a)}$}
\put(-225,155){$\textbf{b)}$}
\put(-450,82){$\bar \varepsilon^\text{pl}$}
\put(-220,35){$y$}
\put(-220,83){$y$}
\put(-220,133){$y$}
\put(-18,35){$\bar \varepsilon^\text{pl}$}
\put(-18,83){$\bar \varepsilon^\text{pl}$}
\put(-18,133){$\bar \varepsilon^\text{pl}$}
\caption{Equivalent plastic strain: a) accumulated plastic strain for the time instant $t=20$ s for the elasto-plastic solution, b) the near tip plastic deformation zones for stationary fractures of different lengths (in each subplot the external isoline corresponds to the value $\bar \varepsilon^\text{pl}=10^{-4}$), respective solutions were obtained under the fluid pressures computed for the modified KGD problem.}
\label{odksztalcenia_plastyczne}
\end{center}
\end{figure}
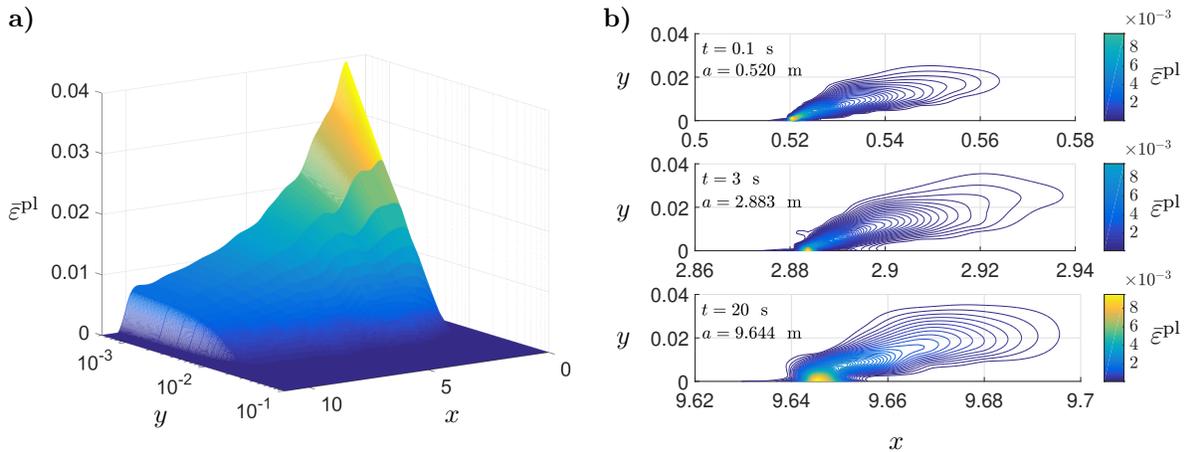

Finally, as the crack propagation condition \eqref{K_eff} was derived on the a priori assumption that the plastic deformation zone is small, let us check to what extent this presumption is satisfied for the respective solutions. For the Mohr-Coulomb plasticity model ABAQUS FEA provides the following measure of the equivalent plastic strain: 
\[
\bar \varepsilon^\text{pl}=\int \frac{1}{c} \boldsymbol{\sigma} :d\varepsilon^\text{pl} , 
\]
with $c$ being the cohesion yield stress.

	In Figure \ref{odksztalcenia_plastyczne}a) we show distribution of $\bar \varepsilon^\text{pl}$ for the elasto-plastic solution in the final time instant ($t=20$ s). The plastic strains accumulated during the whole process of fracture extension are localized in a very narrow strip around the crack profile. In fact, regardless of the position along the crack length, $\bar \varepsilon^\text{pl}$ is well below $1\%$ at a distance of $0.01$ m away from the fracture surface. For comparison we depict in Figure \ref{odksztalcenia_plastyczne}b) the sizes of the near-tip plastic deformation zones obtained by applying the fluid pressures computed for the modified KGD model directly in the FEM module (i.e. no accumulation of plastic deformations in time is present here\footnote{Note that this data was obtained through postpossessing of the results from the modified KGD model and should not be confused with the outcome of a simulation of the complete HF problem. As such it constitutes a partial test where the predefined crack lengths and fluid pressure distributions are used in the $w$ - module of the main algorithm. }). Results for three time instants are shown: $t=0.1$ s, $t=3$ s, and $t=20$ s. Only one of the symmetrical parts is presented ($y>0$). In each of the subplots the most external isoline corresponds to the value $\bar \varepsilon^\text{pl}=10^{-4}$. It shows that in all three moments the external isoline can be well encompassed by a circle of a radius $0.04$ m. The areas over which $\bar \varepsilon^\text{pl}\geq10^{-4}$ are: $5.9\cdot 10^{-4}$ m$^2$ for $t=0.1$ s, $11.3\cdot 10^{-4}$ m$^2$ for $t=1$ s, and $12.9\cdot 10^{-4}$ m$^2$ for $t=20$ s. The corresponding values for $\bar \varepsilon^\text{pl}\geq10^{-3}$ are:  $1.4\cdot 10^{-4}$ m$^2$ for $t=0.1$ s, $2.15\cdot 10^{-4}$ m$^2$ for $t=1$ s, and $2.71\cdot 10^{-4}$ m$^2$ for $t=20$ s.  Considering the above we can confidently conclude that the small yield condition was satisfied in computations.

To complement this subsection we  mention here that the computational cost of using the full elasto-plastic model of HF was much higher than in the case of the simplified variants of the problem. With the employed algorithm, the overall time of computations for a single time step was 70 to 90 times longer for the elasto-plastic model than for the standard or modified KGD problem.

\section{Final conclusions}
\label{conclusions}

In the paper the problem of a hydraulic fracture was considered in three distinct variants: i) the classical KGD model for an elastic solid, ii) the modified KGD problem that, through a dedicated crack propagation condition, accounts for the plastic deformation effects in the near tip zone only, ii) the fully elasto-plastic HF problem in which the elasto-plastic deformation of the bulk of the fractured material is included alongside the plasticity affected crack propagation condition. A new crack propagation condition for the elasto-plastic material, based on the stress relaxation model,  was derived. The underlying assumptions of the employed stress relaxation model were verified in numerical simulations. Computations for the respective variants of the HF problem were performed to examine the shielding effect of the plastic deformations and investigate the possible equivalence of the modified KGD model to the fully elasto-plastic problem.

The following conclusions can be drawn from the conducted analysis:
\begin{itemize}
\item{The underlying assumptions of the newly introduced stress redistribution model and the resulting crack propagation condition are satisfied to a large degree for a range of material and process parameters characteristic for the HF problem.}
\item{At least in the case of small scale yielding the overall effect introduced by the plasticity-dependent crack propagation condition is greater than the one resulting from the plastic deformation of the bulk of the fractured material.}
\item{The modified KGD problem can be a good and credible substitute for the fully elasto-plastic model of hydraulic fracture, at least provided that the inelastic deformations are moderate. The results from the paper by \cite{Papanastasiou_1999a} suggest that this approach could be extended even further with respect to the scale of yielding, which however requires further studies.   }
\item{By using the modified KGD model instead of the full elasto-plastic problem one can immensely reduce the computational cost. With the algorithm employed in this paper  the difference in the overall times of computations amounted to almost two orders of magnitude.}
\end{itemize}

\vspace{5mm}
\noindent
{\bf Funding:} 
This work was funded by European Regional Development Fund and the Republic of Cyprus
through the Research Promotion Foundation (RESTART 2016 - 2020 PROGRAMMES, Excellence Hubs,
Project EXCELLENCE/1216/0481). DP would like to thank the Welsh Government's S\^{e}r Cymru II Research Programme, supported by the European Regional Development Fund.

\vspace{5mm}
\noindent
{\bf Conflicts of interest/Competing interests:} 
The authors have no conflicts of interest to declare that are relevant to the content of this article.

\section*{Acknowledgments}
The authors are thankful to Professor Gennady Mishuris for his useful comments and discussions.


\noindent
 


\begin{thebibliography}{1}

\bibitem[ABAQUS, 2014]{ABAQUS} ABAQUS 6.14 Analysis User's Guid vol III: Materials, \textit{Simulia, Dassault Systems}

\bibitem[Adachi $\&$ Detournay, 2002]{Adachi_Detournay}  Adachi, J., $\&$ Detournay, E. (2002). Self-similar solution of a plane-strain fracture driven by a power-law fluid. \textit{International Journal for Numerical and Analytical Methods in Geomechanics}, 26: 579--604.

\bibitem[Adachi et al., 2007]{Adachi_2007}  Adachi, J., Siebrits, E., Peirce, A., $\&$ Desroches, J. (2007). Computer Simulation of Hydraulic Fractures. \textit{International Journal of Rock Mechanics and Mining Sciences}, 44: 739--757

\bibitem[Atkinson $\&$ Kanninen, 1977]{Atkinson_1977} Atkinson, C., Kanninen, M. (1977) A simple representation of crack tip plasticity: the inclined strip yield superdislocation model. \textit{International Journal of Fracture}, 13(2): 151--163

\bibitem[Bunger, 2013]{Bunger_2013} Bunger, A. (2013) Analysis of the power input needed to propagate multiple hydraulic fractures. \textit{International Journal of Solid and Structures}, 50: 1538 --1549

\bibitem[Detorunay, 2004]{Detournay_2004} Detournay, E. (2004). Propagation regimes of fluid-driven fractures in impermeable rocks. \textit{International Journal of Geomechanics}, 4: 35--45.

\bibitem[Dyskin, 1997]{Dyskin_1997} Dyskin, A. (1997) Crack growth criteria incorporating non-singular stresses: Size effect in apparent fracture toughness. \textit{International Journal of Fracture}, 38(2): 191--206

\bibitem[Felisa et al., 2018]{Felisa_2018} Felisa, G., Lenci, A., Lauriola, I., Longo S., Di Federico, V. (2018) Flow of truncated power-law fluid in fracture channels of variable aperture. \textit{Advances in Water Resources}, 122: 317--327

\bibitem[Garagash, 2009]{Garagash_2009} Garagash, D. (2009). Scaling of physical processes in fluid-driven fracture: perspective from the tip. In \textit{Borodich, F., editor, IUTAM Symposium on Scaling in Solid Mechanics, IUTAM Bookseries}, 10: 91--100, Springer.

\bibitem[Geertsma $\&$ de Klerk, 1969]{Geertsma} Geertsma, J, $\&$ de Klerk, F. (1969). A rapid method of predicting width and extent of hydraulically induced fractures. \textit{Journal of Petroleum Technology}, 21(12): 1571--1581, [SPE 2458].

\bibitem[Inglis, 1913]{Inglis} Inglis, C. (1913) Stresses in Plates Due to the Presence of Cracks and Sharp Corners. \textit{Transactions of the Institute of Naval Architects}, 55: 219--241

\bibitem[Irwin, 1968]{Irwin} Irwin, G. (1968)  Linear fracture mechanics,  fracture transition, and fracture control. \textit{Engineering Fracture Mechanics}, 1: 241--257

\bibitem[Khristianovic  $\&$ Zheltov, 1955]{Khristianovic} Khristianovic, S., $\&$ Zheltov, Y. (1955). Formation of vertical fractures by means of highly viscous liquid. In: \textit{Proceedings of the fourth world petroleum congress, Rome}, 579--586.

\bibitem[Kusmierczyk et al., 2013]{Kusmierczyk} Kusmierczyk, P., Mishuris, G., Wrobel, M. (2013) Remarks on application of different variables for the PKN model of hydrofracturing: various fluid-flow regimes. \textit{International Journal of Fracture}, 184: 185--213

\bibitem[Lavrov, 2013]{Lavrov_2013} Lavrov, A. (2013) Numerical modeling of steady-state flow of a non-Newtonian power-lawfluid in a rough-walled fracture. \textit{Computers and Geotechnics}, 50: 101--109

\bibitem[Liu et al., 2017]{Liu_2017} Liu, F., Gordon, P., Meier, H., Valiveti, D. (2017) A stabilized extended finite element framework for hydraulic fracturing simulations. \textit{International Journal for Numerical and Analytical Methods in Geomechanics}, 41: 654--681.

\bibitem[Menetrey $\&$  William, 1995] {Menetrey_1995} Menetrey, P., William, K. (1995) A triaxial failure criterion for concrete and its generalization. \textit{ACI Structural Journal}, 92: 311--318

\bibitem[Nordgren, 1972]{Nordgren}  Nordgren, R. (1972). Propagation of a Vertical Hydraulic Fracture. \textit{Journal of Petroleum Technology}, 253: 306--314.

\bibitem[Papanastasiou, 1997]{Papanastasiou_1997} Papanastasiou, P. (1997) The influence of plasticity in hydraulic fracturing. \textit{International Journal of Fracture}, 84: 61--79

\bibitem[Papanastasiou, 1999]{Papanastasiou_algorithm} Papanastasiou, P. (1999) An efficient algorithm for propagating fluid driven fractures. \textit{Computational Mechanics}, 24: 258 -- 267

\bibitem[Papanastasiou, 1999a]{Papanastasiou_1999a} Papanastasiou, P. (1999a) The effective fracture toughness in hydraulic fracturing. \textit{International Journal of Fracture}, 96: 127--147

\bibitem[Papanastasiou, 2000]{Papanastasiou_2000} Papanastasiou, P. (2000) Hydraulic fracture closure in a pressure-sensitive elastoplastic medium. \textit{International Journal of Fracture}, 103: 149--161

\bibitem[Papanastasiou $\&$ Atkinson, 2000]{Papanastasiou_Atkinson_2000} Papanastasiou, P., Atkinson C. (2000) Representation of crack-tip plasticity in pressure sensitive geomaterials. \textit{International Journal of Fracture}, 102: 271--286

\bibitem[Papanastasiou $\&$ Atkinson, 2006]{Papanastasiou_Atkinson_2006} Papanastasiou, P., Atkinson C. (2006) Representation of crack-tip plasticity in pressure sensitive geomaterials:large scale yielding. \textit{International Journal of Fracture}, 139: 137--144

\bibitem[Papanastasiou et al., 2016]{Papanastasiou_CO2} Papanastasiou, P., Papamichos, E., Atkinson, C. (2016) On the risk of hydraulic fracturing in CO$_2$ geological storage. \textit{International Journal for Numerical and Analytical Methods in Geomechanics}, 40: 1472--1484

\bibitem[Papazafeiropoulos et al., 2017]{Abaqus2Matlab}  Papazafeiropoulos, G.,  Muniz-Calvente, M., Martinez-Paneda, E. (2017) Abaqus2Matlab: A suitable tool for finite element post-processing. \textit{Advances in Engineering Software}, 105: 9--16

\bibitem[Peck et al., 2018]{Peck_2018_1} Peck, D., Wrobel, M., Perkowska, M., Mishuris, G. (2018) Fluid velocity based simulation of hydraulic fracture: a penny shaped model - part I: the numerical algorithm. \textit{Meccanica}, 53(15): 3615--3635

\bibitem[Peck et al., 2018a]{Peck_2018_2} Peck, D., Wrobel, M., Perkowska, M., Mishuris, G. (2018a) Fluid velocity based simulation of hydraulic fracture - a penny shaped model. Part II: new, accurate semi-analytical benchmarks for an impermeable solid.  \textit{Meccanica}, 53(15): 3637--3650

\bibitem[Perkowska et al., 2016]{Perkowska_2016} Perkowska, M., Wrobel, M., Mishuris, G. (2016) Universal hydrofracturing algorithm for shear--thinning fluids: particle velocity based simulation. \textit{Computers and Geotechnics}, 71: 310--337

\bibitem[Sarris $\&$  Papanastasiou, 2011]{Sarris_2011} Sarris, E., Papanastasiou, P. (2011) The influence of the cohesive process zone in hydraulic fracturing modelling. \textit{International Journal Fracture}, 167: 33--45

\bibitem[Sarris $\&$  Papanastasiou, 2012]{Sarris_2012} Sarris, E., Papanastasiou, P. (2012) Numerical modeling of fluid driven fractures in cohesive poroelastoplastic continuum. \textit{International Journal for Numerical and Analytical Methods in Geomechanics}, 37(12): 1822--1846

\bibitem[Sneddon $\&$ Elliot, 1946]{Sneddon} Sneddon, I., $\&$ Elliot, H. (1946). The opening of a Griffith crack under internal pressure. \textit{Quarterly of Applied Mathematics}, 4: 262--267.

\bibitem[Sun $\&$ Jin, 2012]{Sun_2012} Sun, C., Jin, Z. (2012) Fracture Mechanics. \textit{Academic Press}, ISBN: 978-0-12-385001-0

\bibitem[Wang, 2015]{Wang_2015} Wang, H. (2015) Numerical modeling of non-planar hydraulic fracture propagation in brittle and ductile rocks using XFEM with cohesive zone method. \textit{Journal of Petroleum Science and Engineering}, 135: 127--140

\bibitem[Wang, 2016]{Wang_2016} Wang, H. (2016) Poro-Elasto-Plastic Modeling of Complex Hydraulic Fracture Propagation: Simultaneous Multi-Fracturing and Producing Well Interference. \textit{Acta Mechanica}, 227: 507--525

\bibitem[Wrobel $\&$ Mishuris, 2015]{Wrobel_2015} Wrobel, M., Mishuris, G. (2015) Hydraulic fracture revisited: Particle velocity based simulation. \textit{International Journal of Engineering Science}, 94: 23--58

\bibitem[Wrobel et al., 2017]{Wrobel_2017} Wrobel, M., Mishuris, G., Piccolroaz, A. (2017) Energy Release Rate in hydraulic fracture: can we neglect an impact of the hydraulically induced shear stress? \textit{International Journal of Engineering Science}, 111: 28--51

\bibitem[Wrobel et al., 2018]{Wrobel_2018} Wrobel, M., Mishuris, G., Piccolroaz, A. (2018) On the impact of tangential traction on the crack surfaces induced by fluid in hydraulic fracture: Response to the letter of A.M. Linkov. Int. J. Eng. Sci. (2018) 127, 217--219. \textit{International Journal of Engineering Science}, 127: 220--224

\bibitem[Wrobel, 2020]{Wrobel_Arxiv} Wrobel, M. (2020) An efficient algorithm of solution for the flow of generalized Newtonian fluid in channels of simple geometries. \textit{Rheologica Acta}, 59: 651--663

\bibitem[Wrobel, 2020a]{Wrobel_2020} Wrobel, M. (2020a) On the application of simplified rheological models of fluid in the hydraulic fracture problems. \textit{International Journal of Engineering Science}, 150: 103275

\bibitem[Wrobel et al., 2021]{Wrobel_2021} Wrobel, M., Mishuris, G., Papanastasiou, P. (2021) On the influence of fluid rheology on hydraulic fracture. \textit{International Journal of Engineering Science}, 158: 103426

\bibitem[Wrobel et al., 2021a]{Wrobel_redirection} Wrobel, M., Piccolroaz, A., Papanastasiou, P., Mishuris, G. (2021a) Redirection of a crack driven by viscous fluid taking into account plastic deformation effects in the process zone. \textit{Geomechanics for Energy and Environment}, 26: 100147

\bibitem[Wrobel et al., 2021b]{Wrobel_FEM} Wrobel, M., Papanastasiou, P., Peck, D. (2021b) Numerical simulation of hydraulic fracturing: a hybrid FEM based algorithm. arxiv.org/abs/2108.04608

\bibitem[Wu, 2006]{Wu_phd} Wu, R. (2006) Some fundamental Mechanisms of Hydraulic Fracturing. PhD dissertation, Georgia Institute of Technology

\bibitem[Yao, 2011]{Yao_2011} Yao, Y. (2011) Linear Elastic and Cohesive Fracture Analysis to Model Hydraulic Fracture in Brittle and Ductile Rocks. \textit{Rock Mechanics and Rock Engineering}, 45: 375--387

\bibitem[Zeng et al., 2019]{Zeng_2019} Zeng, Q., Yao, J., Shao, J. (2019) Effect of plastic deformation on hydraulic fracturing with extended element method. \textit{Acta Geotechnica}, 14: 2083--2101



























\end{thebibliography}
\end{document}